# Transport and thermodynamics in quantum junctions: A scattering approach

Alexander Semenov[†] and Abraham Nitzan[†‡1]

[†]*Department of Chemistry, University of Pennsylvania, Philadelphia, Pennsylvania 19104, USA*

[‡]*School of Chemistry, The Sackler Faculty of Science, Tel Aviv University, Tel Aviv 69978, Israel*

Date: March 2nd, 2020

**Abstract**

We present a scattering approach for the study of the transport and thermodynamics of quantum systems strongly coupled to their thermal environment(s). This formalism recovers the standard non-equilibrium Green's function expressions for quantum transport and reproduces recently obtained results for the quantum thermodynamic of slowly driven systems. Using this approach, new results have been obtained. First, we derived of a general explicit expression for non-equilibrium steady state density matrix of a system compromised of multiple infinite baths coupled through a general interaction. Then, we obtained a general expression for the dissipated power for the driven non-interacting resonant level to first order in the driving speeds, where both the dot energy level and its couplings are changing, without invoking the wide band approximation. In addition, we also showed that the symmetric splitting of system bath interaction, employed for the case of a system coupled to one bath to determine the effective system Hamiltonian [Phys. Rev. B **93**, 115318 (2016)] is valid for the multiple baths case as well. Finally, we demonstrated an equivalence of our method to the Landauer-Buttiker formalism and its extension to slowly driven systems developed by von Oppen and co-workers [Phys. Rev. Lett. **120**, 107701 (2018)]. To demonstrate the use of this formalism we analyze the operation a device in which the dot is driven cyclically between two leads under strong coupling conditions. We also generalize the previously obtained expression for entropy production in such driven processes to the many-bath case.

---

[1] Author to whom correspondence should be addressed: anitzan@sas.upenn.edu

**I. Introduction.**

Quantum transport on the nanoscale, e.g. heat and charge transport through molecular junctions, has received a great deal of attention for the past several decades and been extensively studied both theoretically and experimentally[1–5], driven by open fundamental problems, technological promise and continuing progress in nanofabrication. Some of the fundamental problems have led to the emergence of quantum thermodynamics[6,7], which focuses is the interchange of energy and matter between a microscopic system and its environment and its description in terms of thermodynamic quantities such as heat, work, entropy and efficiency, thereby establishing quantum analogues to the three law of thermodynamics that govern energy conversion at the nanoscale.

While a significant progress in the field has been achieved in the limit of weak coupling between system and environment[6–8], the situation of strongly coupled systems, where the total density matrix cannot be, even approximately, represented as a direct tensor product of the densities matrices of the system and the environment (bath), still remains largely unexplored and presents a rich field of active studies[9–14]. On the other hand, the theoretical treatment of quantum transport in the strong coupling regime has been thoroughly established using a variety of methods such as the Landauer - Buttiker scattering description[15–19], the non-equilibrium Green's function (NEGF) formalism[20–23], the numerical renormalization group approach[24] and a multiple time-scale expansion of the total (system plus bath) density matrix[25].

These methods have been recently applied for the development of quantum thermodynamics for non-interacting resonant level connected to one[25–29] or two[30–32] baths, where the system is subject to a slow perturbation which drives it out of equilibrium. These treatments yield similar results when the wide-band approximation is invoked and satisfy the second law up to second order in the driving speed. In their present states, these approaches to the quantum thermodynamics have several weaknesses. First, standard implementations of the NEGF formalism directly address system observables rather than the overall non-equilibrium distribution function. Consequently, an extension of this formalism to the presence of several baths, which requires assessment of the contribution from different baths is less straightforward (see e.g. Eq. (72)). The density matrix expansion can in principle yield the distribution, and has been shown useful for interacting particle models[25,32,33] , however, the construction of the density matrix in the case where the level is coupled to several baths is challenging and has not been yet attempted. The scattering formalism, which treats the central region from an outside perspective[28,34], can be naturally be used in the case of multiple baths. Being based on time independent scattering formalism, it is applicable to steady state fluxes and currents and cannot be easily used, in its present form, for transient response and relaxation processes, and cannot yield cumulative quantities such as total energy and occupations.

Here we propose a scattering approach for the construction of a non-equilibrium steady state (NESS) density matrix and for evaluating quantum thermodynamics of slowly driven systems that are strongly coupled to their thermal environment(s). Within this formalism we reproduce the standard NEGF results for quantum transport and reproduce recently obtained results for the quantum thermodynamic behavior of such system under slow externally controlled driving. Some new results are obtained as well: First, an explicit expression is obtained for the NESS density matrix of a system comprising multiple thermal baths, out of equilibrium between each other, interconnected through a molecular species. This formalism is applied to generalize past work to the systems comprising many baths without invoking the wide band approximation and will be used in future studies of the thermodynamic behavior of such systems. In particular, the power associated with driving the system (resonant level connected to multiple baths and driven by changing both the level energy and its couplings to the baths in a non-interacting particles system) is obtained up to second order in the driving speeds (the lowest order for describing irreversibility). In addition, we show that the symmetric splitting of system bath interaction, employed for the case of a system coupled to one bath to determine the effective system Hamiltonian for calculating the system thermodynamic properties[12,27], also holds for the multiple baths case. Second, we generalize previously obtained expressions for the entropy production and its distribution among the many baths involved. Finally, we explore the implications of driving under coupling to multiple colored baths and show that driving phases (not previously considered in this context) and the baths spectral structures can qualitatively affect the energetic implications of the driving process.

**II. Theory.**

We start with a system of independent baths, described by the Hamiltonian

$$\hat{H}_0 = \sum_\alpha \hat{H}_0^\alpha \quad (1)$$

$\hat{H}_0^\alpha$ is the Hamiltonian of bath $\alpha$. These baths are infinite/semi-finite in size, implying that each $\hat{H}_0^\alpha$ has a continuous unbound spectrum. Each bath is assumed to be in its own thermal equilibrium,

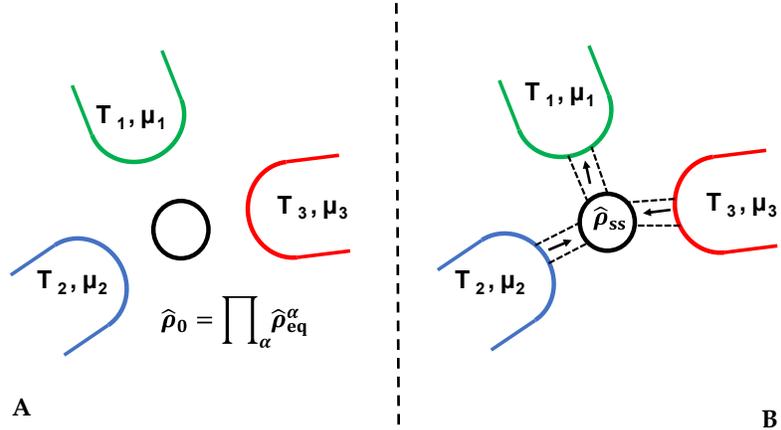

**Figure 1**. A sketch of the problem: A) infinite baths are initially uncoupled B) the same baths are coupled through the central region and the system is in a non-equilibrium steady state. A transition from A to B is done by turning the interaction adiabatically.

characterized by an inversed temperature $\beta_\alpha$ and a chemical potential $\mu_\alpha$, so the density operator of this system is $\hat{\rho}_0 = \prod_\alpha \hat{\rho}_0^\alpha$, $\hat{\rho}_0^\alpha = \exp\{-\beta_\alpha(\hat{H}_0^\alpha - \mu_\alpha \hat{N}_0^\alpha)\}/Z_\alpha$

For definiteness, we take the baths to by infinite systems of non-interacting particles or quasiparticles described by the Hamiltonian

$$\hat{H}_0^\alpha = \sum_k \varepsilon_{k\alpha} \hat{c}_{k\alpha}^\dagger \hat{c}_{k\alpha} \tag{2}$$

where $k$ stands for an eigen level within a bath and $\hat{c}_{k\alpha}^\dagger (\hat{c}_{k\alpha})$ are the corresponding creation/annihilation operators. Thus

$$\hat{\rho}_0^\alpha = \frac{1}{Z_\alpha} \exp\{-\beta_\alpha(\varepsilon_{k\alpha} - \mu_\alpha)\hat{c}_{k\alpha}^\dagger \hat{c}_{k\alpha}\}; \quad \hat{\rho}_0 = \frac{1}{Z} \prod_{k\alpha} \exp\{-\beta_\alpha(\varepsilon_{k\alpha} - \mu_\alpha)\hat{c}_{k\alpha}^\dagger \hat{c}_{k\alpha}\} \tag{3}$$

The density operators (3) satisfy the equilibrium Liouville equations:

$$\partial_t \hat{\rho}_0^\alpha = -\frac{i}{\hbar}[\hat{H}_0^\alpha, \hat{\rho}_0^\alpha] = 0 \tag{4}$$

$$\partial_t \hat{\rho}_0 = -\frac{i}{\hbar}[\hat{H}_0, \hat{\rho}_0] = 0 \tag{5}$$

Next, coupling between the baths, $\hat{V}$, is switched on adiabatically according to:

$$\hat{H}(t) = \hat{H}_0 + \exp\{\eta t \theta(-t)\}\hat{V} \tag{6}$$

where $\eta$ is an infinitesimally small positive number and $\theta(-t)$ is a step function. Eq. (6) describes an adiabatic buildup of the interaction and a corresponding change in the density operator $\hat{\rho}(t)$ according to

$$\partial_t \hat{\rho}(t) = -\frac{i}{\hbar}[\hat{H}(t), \hat{\rho}(t)] \tag{7}$$

with the boundary conditions $\hat{\rho}(t=-\infty) = \hat{\rho}_0$ and $\hat{H}(t=-\infty) = \hat{H}_0$, $\hat{H}(t \geq 0) = \hat{H}$. This adiabatic turn-on of the coupling between baths leads, for $t > 0$, to the steady state associated with the boundary conditions imposed by the baths. Indeed, in Appendix A we show that for $t > 0$ the state

$$\hat{\rho}_{ss} = \hat{\Omega}_+ \hat{\rho}_0 \hat{\Omega}_+^\dagger \tag{8}$$

where $\hat{\Omega}_+$ is a Moller (wave) scattering operator:

$$\hat{\Omega}_+ = \lim_{t\to\infty} \exp(-i\hat{H}t)\exp(i\hat{H}_0 t) \tag{9}$$

is a solution of the corresponding Liouville equation[35]

$$\partial_t \hat{\rho}_{ss} = -\frac{i}{\hbar}[\hat{H},\hat{\rho}_{ss}] = 0. \tag{10}$$

In Appendix B we show that Expression (8) equivalent to both McLennan-Zubarev[36,37] and Hershfield[38] non-equilibrium steady state density matrices.

Note that Eqs. (8)-(10) are quite general and can be applied to both bosonic and fermionic baths and different scenarios for inter-bath coupling. In the resonance level model considered in the next Section, the inter-bath coupling is mediated by a single 'dot' level (or, for a boson model, a single boson). Accordingly, the coupling $\hat{V}$ between the baths, Eq. (19) below, includes the Hamiltonian of this dot. Also note that the transformation (8) that yields this non-equilibrium steady state density matrix is unitary. This seemingly contradicts the fact that the evolution from scenario (a) to (b) in Fig. 1 is a relaxation process. To resolve this apparent contradiction, one needs to keep in mind that the baths are infinite. Thus, if we apply this mathematical description to two *finite* isolated leads connected through a quantum dot, then after the interaction is turned on, the inter-bath current through the junction will first increase then reach a plateau, but on longer timescale will oscillate between the finite leads. Increasing the lead sizes will lead to the extension of the plateau region and in the limit of an infinite size this plateau becomes infinite which in turn, corresponds to a steady state.

Using Eq. (8) and the unitarity of Moller operators, the steady state density operator takes the form

$$\hat{\rho}_{ss} = \frac{1}{Z}\prod_{k\alpha}\exp\left\{-\beta_\alpha(\varepsilon_{k\alpha}-\mu_\alpha)\hat{\Omega}_+\hat{c}^\dagger_{k\alpha}\hat{\Omega}^\dagger_+\hat{\Omega}_+\hat{c}_{k\alpha}\hat{\Omega}^\dagger_+\right\} \tag{11}$$

Introducing the new asymptotic operators:

$$\hat{\psi}^\dagger_{k\alpha} \equiv \hat{\Omega}_+ \hat{c}^\dagger_{k\alpha} \hat{\Omega}^\dagger_+ \qquad (12)$$

Eq. (11) becomes

$$\hat{\rho}_{ss} = \frac{1}{Z}\prod_{k\alpha}\exp\left\{-\beta_\alpha(\varepsilon_{k\alpha}-\mu_\alpha)\hat{\psi}^\dagger_{k\alpha}\hat{\psi}_{k\alpha}\right\} \qquad (13)$$

The significance of the form (13) can be seen from the following observations: First note that

$$\partial_t \hat{c}^\dagger_{k\alpha} = \frac{i}{\hbar}[\hat{H}_0, \hat{c}^\dagger_{k\alpha}] = \frac{i}{\hbar}\varepsilon_{k\alpha}\hat{c}^\dagger_{k\alpha} \qquad (14)$$

which is valid for both bosons and fermions. For the asymptotic operator (12) we have:

$$\partial_t\hat{\psi}^\dagger_{k\alpha} = \frac{i}{\hbar}[\hat{H},\hat{\psi}^\dagger_{k\alpha}] = \frac{i}{\hbar}[\hat{H}\hat{\Omega}_+\hat{c}^\dagger_{k\alpha}\hat{\Omega}^\dagger_+ - \hat{\Omega}_+\hat{c}^\dagger_{k\alpha}\hat{\Omega}^\dagger_+\hat{H}]$$
$$= \frac{i}{\hbar}[\hat{\Omega}_+\hat{H}_0\hat{c}^\dagger_{k\alpha}\hat{\Omega}^\dagger_+ - \hat{\Omega}_+\hat{c}^\dagger_{k\alpha}\hat{H}_0\hat{\Omega}^\dagger_+] = \frac{i}{\hbar}\varepsilon_{k\alpha}\hat{\psi}^\dagger_{k\alpha} \qquad (15)$$

where we used the intertwining relation $\hat{H}\hat{\Omega}_+ = \hat{\Omega}_+\hat{H}_0$ (see Appendix A, Eq. (A19)). Furthermore the $\hat{\psi}^\dagger_{k\alpha}$ operators satisfy the standard boson/fermions commutation relations:

$$[\hat{\psi}^\dagger_{k\alpha},\hat{\psi}^\dagger_{n\beta}] = \hat{\Omega}_+[\hat{c}^\dagger_{k\alpha},\hat{c}^\dagger_{n\beta}]\hat{\Omega}^\dagger_+ = 0 \qquad (16)\text{a}$$

$$[\hat{\psi}_{k\alpha},\hat{\psi}^\dagger_{n\beta}] = \hat{\Omega}_+[\hat{c}_{k\alpha},\hat{c}^\dagger_{n\beta}]\hat{\Omega}^\dagger_+ = \delta_{\alpha\beta}\delta_{nk} \qquad \text{(for bosons)} \qquad (16)\text{b}$$

$$[\hat{\psi}^\dagger_{k\alpha},\hat{\psi}_{n\beta}]_+ = \hat{\Omega}_+[\hat{c}^\dagger_{k\alpha},\hat{c}_{n\beta}]_+\hat{\Omega}^\dagger_+ = \delta_{\alpha\beta}\delta_{nk} \qquad \text{(for fermions)} \qquad (16)\text{c}$$

Eqs. (15)-(16) imply that the Moller operators preserve the spectra as well as the commutation properties of the fermion/boson operators. It should also be noted that the expression (13) is quite general and emphasizes the fact that a non-equilibrium steady state density matrix can be seen as a direct product of equilibrium density matrices. Finally, we note that operators $\hat{\psi}^\dagger_{k\alpha}(\hat{\psi}_{n\beta})$ describe scattering states. Bound states belong to the kernel (null space) of the Moller operator [39](i.e. the series (A9) for the Moller operator do not converge on the subspace of bound states of $\hat{H}$ ). In this case one can in principle use the same adiabatic procedure given by Eq.(6) and employ the Gell-Mann and Low theorem[40] to obtain the bound states solution of $\hat{H}$ after the steady state is reached. However, if $\hat{H}$ does not contain bound states, Eq. (8) remains valid.[41]

It is useful to introduce single excitation states:

$$\hat{\psi}^\dagger_{k\beta}|0\rangle = |\psi_{k\beta}\rangle \tag{17a}$$

$$\hat{c}^\dagger_{k\beta}|0\rangle = |c_{k\beta}\rangle \tag{17b}$$

where $|0\rangle$ stands for the ground state of the system. The states (17) are connected through so-called Lippmann-Schwinger equation:

$$|\psi_{k\beta}\rangle = \left(\hat{I} + \hat{G}^r(\varepsilon_{k\beta})\hat{V}\right)|c_{k\beta}\rangle \tag{18}$$

where $\hat{G}^{r/a}(\varepsilon) = \dfrac{1}{\varepsilon - \hat{H} \pm i\eta}$ is the Green function. Eq. (18) is obtained in Appendix C.

**III. The Fermionic Resonant Level Model – Steady State**

In this section we apply the formalism developed in the previous section to an electron transport system represented by the non-interacting fermionic resonant level model. In this model the interaction has the form:

$$\hat{V} = \varepsilon_d \hat{d}^\dagger \hat{d} + \sum_{k\alpha}\left(V_{k\alpha}\hat{c}^\dagger_{k\alpha}\hat{d} + V^*_{k\alpha}\hat{d}^\dagger \hat{c}_{k\alpha}\right) \tag{19}$$

First, explicit forms are obtained for the asymptotic field operators. In Appendix C the following expressions for the scattering operators are derived

$$\hat{\psi}^\dagger_{k\beta} = V^*_{k\beta}G^r_{dd}(\varepsilon_{k\beta})\hat{d}^\dagger + \sum_{n\alpha}\left\{\delta_{k\beta n\alpha} + V_{n\alpha}\frac{V^*_{k\beta}G^r_{dd}(\varepsilon_{k\beta})}{\varepsilon_{k\beta} - \varepsilon_{n\alpha} + i\eta}\right\}\hat{c}^\dagger_{n\alpha} \tag{20a}$$

$$\hat{\psi}_{k\beta} = V_{k\beta}G^a_{dd}(\varepsilon_{k\beta})\hat{d} + \sum_{n\alpha}\left\{\delta_{k\beta n\alpha} + V^*_{n\alpha}\frac{V_{k\beta}G^a_{dd}(\varepsilon_{k\beta})}{\varepsilon_{k\beta} - \varepsilon_{n\alpha} - i\eta}\right\}\hat{c}_{n\alpha} \tag{20b}$$

Here $\hat{\psi}^{\dagger}_{k\beta}$ and $\hat{\psi}_{k\beta}$ are, respectively, creation and annihilation operators for a particle in the scattering state that correspond to an incoming particle in state $k$ of bath (or lead) $\beta$.

The corresponding inverted expressions are obtained in the forms

$$\hat{d}^{\dagger} = \sum_{k\beta} G^{a}_{dd}(\varepsilon_{k\beta}) V_{k\beta} \hat{\psi}^{\dagger}_{k\beta} \tag{21a}$$

$$\hat{d} = \sum_{k\beta} G^{r}_{dd}(\varepsilon_{k\beta}) V^{*}_{k\beta} \hat{\psi}_{k\beta} \tag{21b}$$

$$\hat{c}^{\dagger}_{n\alpha} = \sum_{k\beta} \left( \delta_{k\beta n\alpha} + V^{*}_{n\alpha} \frac{V_{k\beta} G^{a}_{dd}(\varepsilon_{k\beta})}{\varepsilon_{k\beta} - \varepsilon_{n\alpha} - i\eta} \right) \hat{\psi}^{\dagger}_{k\beta} \tag{22a}$$

$$\hat{c}_{n\alpha} = \sum_{k\beta} \left\{ \delta_{k\beta n\alpha} + V_{n\alpha} \frac{V^{*}_{k\beta} G^{r}_{dd}(\varepsilon_{k\beta})}{\varepsilon_{k\beta} - \varepsilon_{n\alpha} + i\eta} \right\} \hat{\psi}_{k\beta} \tag{22b}$$

where $\delta_{k\beta n\alpha} = \delta_{kn}\delta_{\beta\alpha}$ and $G^{r/a}_{dd}$ are the retarded/advanced Green functions of the dot level

$$G^{r}_{dd}(\varepsilon) = G^{a}_{dd}{}^{*}(\varepsilon) = \frac{1}{\varepsilon - \varepsilon_d - \Sigma^{r}_{dd}(\varepsilon)} \tag{23}$$

$$\Sigma^{r}_{dd}(\varepsilon) = \Sigma^{a}_{dd}(\varepsilon)^{*} = \sum_{n\alpha} |V_{n\alpha}|^2 \frac{1}{\varepsilon - \varepsilon_{n\alpha} + i\eta} = \Lambda(\varepsilon) - i\Gamma(\varepsilon)/2 \tag{24}$$

$$\Gamma(\varepsilon) = \sum_{\alpha} \Gamma_{\alpha}(\varepsilon) \,;\quad \Gamma_{\alpha}(\varepsilon) = \sum_{n} 2\pi |V_{n\alpha}|^2 \,\delta(\varepsilon - \varepsilon_{n\alpha}) \tag{25}$$

$$\Lambda(\varepsilon) = \sum_{\alpha} \Lambda_{\alpha}(\varepsilon);\quad \Lambda_{\alpha}(\varepsilon) = \sum_{n} |V_{n\alpha}|^2 \, \text{PP} \frac{1}{\varepsilon - \varepsilon_{n\alpha}} \tag{26}$$

In Eqs. (20)-(22) the limit $\eta \to +0$ is implied. We further show (SI, Section A) that the Hamiltonian and number operators assume their standard forms when expressed in terms of the local creation and annihilation operators:

$$\hat{H} = \sum_{k\alpha} \varepsilon_{k\alpha} \hat{\psi}^{\dagger}_{k\alpha} \hat{\psi}_{k\alpha} = \varepsilon_d \hat{d}^{\dagger} \hat{d} + \sum_{k\alpha} \left( V_{k\alpha} \hat{c}^{\dagger}_{k\alpha} \hat{d} + V^{*}_{k\alpha} \hat{d}^{\dagger} \hat{c}_{k\alpha} \right) + \sum_{k\alpha} \varepsilon_{k\alpha} \hat{c}^{\dagger}_{k\alpha} \hat{c}_{k\alpha} \tag{27a}$$

$$\hat{N} = \sum_{k\alpha} \hat{\psi}^{\dagger}_{k\alpha} \hat{\psi}_{k\alpha} = \hat{d}^{\dagger} \hat{d} + \sum_{k\alpha} \hat{c}^{\dagger}_{k\alpha} \hat{c}_{k\alpha} \tag{27b}$$

Eqs. (27) imply that the total energy and the total number of particles are conserved proving the completeness of the scattering states basis.

In what follows we employ Eqs.(20)-(22) to calculate various transport and thermodynamic quantities of the static resonance level model as well as well as its extension to the case in which one or more parameters in the Hamiltonian $\hat{H} = \hat{H}_0 + \hat{V}$ ( $\hat{H}_0$ and $\hat{V}$ are given by Eqs. (2) and (19)) are slowly driven.

### IIIa. Steady state observables

The key point in the calculation is to express any single particle operator $\hat{A}$ by the asymptotic field operators $\hat{\psi}^\dagger_{k\beta}(\hat{\psi}_{n\alpha})$:

$$\hat{A} = \sum_{k\beta n\alpha} \gamma_{k\beta n\alpha} \hat{\psi}^\dagger_{k\beta} \hat{\psi}_{n\alpha} \tag{28}$$

Once this is done, the steady-state expectation value of $\hat{A}$ is obtained from (13)

$$\langle \hat{A} \rangle = \text{Tr}\{\hat{\rho}_{ss}\hat{A}\} = \sum_{n\alpha} \gamma_{n\alpha n\alpha} f_\alpha(\varepsilon_{n\alpha}) \tag{29}$$

which is a direct consequence of the form (13) of the steady-state density operator.

As a simple example consider the dot population. We use Eqs (21) and (13) to get:

$$\begin{aligned} N_d &= \text{Tr}\{\hat{d}^\dagger \hat{d} \hat{\rho}_{ss}\} = \sum_{k\beta} G^a_{dd}(\varepsilon_{k\beta}) V_{k\beta} \sum_{n\alpha} G^r_{dd}(\varepsilon_{n\alpha}) V^*_{n\alpha} \text{Tr}\{\hat{\psi}^\dagger_{k\beta} \hat{\psi}_{n\alpha} \hat{\rho}_{ss}\} \\ &= \sum_{n\alpha} G^r_{dd}(\varepsilon_{n\alpha}) G^a_{dd}(\varepsilon_{n\alpha}) |V_{n\alpha}|^2 f_\alpha(\varepsilon_{n\alpha}) \end{aligned} \tag{30}$$

Using $\sum_{n\alpha} f(\varepsilon_{n\alpha})|V_{n\alpha}|^2 = (2\pi)^{-1} \int d\varepsilon f(\varepsilon) \Gamma_\alpha(\varepsilon)$ and $G^r_{dd}(\varepsilon) G^a_{dd}(\varepsilon) = A_{dd}(\varepsilon)/\Gamma(\varepsilon)$ where $\Gamma(\varepsilon) = \sum_\alpha \Gamma_\alpha(\varepsilon)$ and $A_{dd}(\varepsilon)$ is the spectral density associated with the dot level, Eq. (30) may be cast in the more familiar form for the dot population

$$N_d = \frac{1}{2\pi} \int_{-\infty}^{\infty} A_{dd}(\varepsilon) \sum_\alpha f_\alpha(\varepsilon) \frac{\Gamma_\alpha(\varepsilon)}{\Gamma(\varepsilon)} d\varepsilon \tag{31}$$

As another example we next show that the present procedure leads to the Landauer expression for the junction current, given for a two-lead model by Eqs. (34) and (35) below. We start with the expression for the current associated with bath α

$$J_\alpha = \frac{d\langle N_\alpha \rangle}{dt} = i\mathrm{Tr}\{\hat{\rho}[\hat{V}_\alpha, \hat{N}_\alpha]\} = i\sum_n \mathrm{Tr}\{\hat{\rho} V_{n\alpha}\hat{c}^\dagger_{n\alpha}\hat{d} - \rho V^*_{n\alpha}\hat{d}^\dagger \hat{c}_{n\alpha}\} \tag{32}$$

which, using Eqs. (21-22) takes the form

$$\begin{aligned}
J_\alpha = &i\sum_n \mathrm{Tr}\left[\rho V_{n\alpha} \sum_{k\beta}\left\{\delta_{k\beta n\alpha} + V^*_{n\alpha}\frac{V_{k\beta}G^a_{dd}(\varepsilon_{k\beta})}{\varepsilon_{k\beta} - \varepsilon_{n\alpha} - i\eta}\right\}\hat{\psi}^\dagger_{k\beta}\sum_{m\gamma}G^r_{dd}(\varepsilon_{m\gamma})V^*_{m\gamma}\hat{\psi}_{m\gamma}\right] \\
&-i\sum_n \mathrm{Tr}\left[\rho V^*_{n\alpha} \sum_{m\gamma}G^a_{dd}(\varepsilon_{m\gamma})V_{m\gamma}\hat{\psi}^\dagger_{m\gamma}\sum_{k\beta}\left\{\delta_{k\beta n\alpha} + V_{n\alpha}\frac{V^*_{k\beta}G^r_{dd}(\varepsilon_{k\beta})}{\varepsilon_{k\beta} - \varepsilon_{n\alpha} + i\eta}\right\}\hat{\psi}_{k\beta}\right]
\end{aligned} \tag{33}$$

This has the general form of Eqs. (28), (29) and can be evaluated along similar lines as above (see SI, Section B). For a two terminal (α = L,R) junction this leads to

$$J = \frac{1}{2\pi}\int T(\varepsilon)\{f_L(\varepsilon) - f_R(\varepsilon)\}d\varepsilon \tag{34}$$

$$T(\varepsilon) = \Gamma_R(\varepsilon)\Gamma_L(\varepsilon)G^r_{dd}(\varepsilon)G^a_{dd}(\varepsilon) \tag{35}$$

To end this subsection we note that one could also construct (SI, Section C), starting from the present formalism, the full S-matrix theory of junction scattering (generalized to the many-baths model) which is the basis for the Landauer-Buttiker theory of junction transport.

**IIIb.** *Symmetric Splitting*

In Refs.[12,27,30] it was shown that for the model under discussion the $\varepsilon_d$ dependence of the total energy, expressed by the derivative $\partial\left(\mathrm{Tr}\left(\hat{\rho}_{eq}\hat{H}\right)\right)/\partial\varepsilon_d$, is completely captured by a similar expression, $\partial\left(\mathrm{Tr}\left(\hat{\rho}_{eq}\hat{H}_{eff}\right)\right)/\partial\varepsilon_d$, where

$$\hat{H}_{eff} \equiv \varepsilon_d \hat{d}^\dagger\hat{d} + \frac{1}{2}\left\{\sum_k V_k \hat{c}^\dagger_k\hat{d} + V^*_k \hat{d}^\dagger \hat{c}_k\right\} \tag{36}$$

may be considered as an effective "dot Hamiltonian" defined by splitting the dot-baths interaction evenly between the dot and the baths[42]. This symmetric splitting of the interaction[12,26,27], while sometimes used as an assumption of practical consequences is by no means a general principle, and can be justified only for the average energy in non-interacting particles models. It is nevertheless useful for addressing subsystem thermodynamic properties in such systems.

Here we show that this symmetric splitting remains valid (in the sense above) also for non-equilibrium steady states involving multiple baths, at least under the wide-band approximation. In this approximation, the $\varepsilon_d$- dependent part of the total density of states is given by

$$D(\varepsilon) = \frac{1}{\pi} \text{Im}\{G_{dd}^r(\varepsilon)\} = \sum_{\beta} D_\beta(\varepsilon) \qquad (37)$$

In the second equality of (37) we have written $D$ as a sum over contributions from the different leads. In SI, Section D we show that

$$D_\beta(\varepsilon) = \frac{\Gamma_\beta}{\pi \Gamma} \text{Im}\{G_{dd}^r(\varepsilon)\} = \frac{\Gamma_\beta}{2\pi \Gamma} A_{dd}(\varepsilon) \qquad (38)$$

where $A_{dd} = \langle d|\hat{A}|d \rangle$ is the spectral function, $\hat{A} = i(\hat{G}^r - \hat{G}^a)$. The $\varepsilon_d$-dependent part of the total system energy, denoted by $\langle \hat{H} \rangle^{(d)}$, is consequently given by[43]

$$\langle \hat{H} \rangle^{(d)} = \sum_\beta \int_{-\infty}^{\infty} \varepsilon D_\beta(\varepsilon) f_\beta(\varepsilon) d\varepsilon = \sum_\beta \int_{-\infty}^{\infty} \varepsilon \frac{\Gamma_\beta}{2\pi \Gamma} A_{dd}(\varepsilon) f_\beta(\varepsilon) d\varepsilon \qquad (39)$$

Next consider the Hamiltonian (36): from the SI, (Section A, Eqs. (SA9) and (SA17)) it follows that

$$\hat{H}_{eff} = \varepsilon_d \hat{d}^\dagger \hat{d} + \frac{1}{2} \left\{ \sum_{k\alpha} V_{k\alpha} \hat{c}_{k\alpha}^\dagger \hat{d} + V_{k\alpha}^* \hat{d}^\dagger \hat{c}_{k\alpha} \right\} = \sum_{k\alpha} \sum_{n\beta} G_{dd}^r(\varepsilon_k) G_{dd}^a(\varepsilon_n) V_{n\beta} V_{k\alpha}^* \hat{\psi}_{n\beta}^\dagger \hat{\psi}_{k\alpha} \{\varepsilon_k + \varepsilon_n\}/2 \quad (40)$$

Thus, using Eq. (29), we obtain

$$\langle \hat{H}_{eff} \rangle = \sum_{n\alpha} G_{dd}^r(\varepsilon_n) G_{dd}^a(\varepsilon_n) |V_{n\alpha}|^2 \varepsilon_n f_\alpha(\varepsilon_n) \qquad (41)$$

Using Eq.(25) and introducing the integral $\int_{-\infty}^{\infty} \delta(\varepsilon - \varepsilon_n) d\varepsilon$ Eq. (41) may be rewritten in the form

$$\langle \hat{H}_{eff} \rangle = \frac{1}{2\pi} \int_{-\infty}^{\infty} \varepsilon G_{dd}^r(\varepsilon) G_{dd}^a(\varepsilon) \sum_\alpha f_\alpha(\varepsilon) \Gamma_\alpha(\varepsilon) d\varepsilon$$
$$= \frac{1}{2\pi} \int_{-\infty}^{\infty} \varepsilon A_{dd}(\varepsilon) \sum_\alpha f_\alpha(\varepsilon) \frac{\Gamma_\alpha(\varepsilon)}{\Gamma(\varepsilon)} d\varepsilon \qquad (42)$$

We see that (42) coincides with (39), thus we can conclude that $\hat{H}_{eff}$ indeed contains all the $\varepsilon_d$ dependence of the total Hamiltonian.

## IV. Externally Imposed Driving

Next, consider the case where the total Hamiltonian $\hat{H}$ parametrically depends on one or more parameters $R^\nu$ that undergo slow externally controlled driving. The following derivation is valid for both fermions and bosons. In the adiabatic approximation the non-equilibrium density matrix is given by Eq.(13) $\hat{\rho}_{ss}(R^\nu) = \frac{1}{Z} \prod_{k\alpha} \exp\{-\beta_\alpha(\varepsilon_{k\alpha} - \mu_\alpha) \hat{\psi}_{k\alpha}^\dagger(R^\nu) \hat{\psi}_{k\alpha}(R^\nu)\}$ where the field operators correspond to the instantaneous Hamiltonian $\hat{H}(R^\nu)$. A non-adiabatic correction, $\Delta\hat{\rho}(t) = \hat{\rho}(t) - \hat{\rho}_{ss}(R^\nu(t))$, to the density matrix due to a finite driving speed $\dot{R}^\nu$ can be obtained from the Liouville equation:

$$\partial_t \{\Delta\hat{\rho}(t) + \hat{\rho}_{ss}(R^\nu(t))\} = -\frac{i}{\hbar} \left[ \hat{H}(R^\nu(t)), \Delta\hat{\rho}(t) + \hat{\rho}_{ss}(R^\nu(t)) \right] \qquad (43)$$

Since

$$[\hat{H}(R^\nu), \hat{\rho}_{ss}(R^\nu)] = 0 \; , \; d\hat{\rho}_{ss}(R^\nu(t))/dt = \sum_\nu \dot{R}^\nu \partial_{R^\nu} \hat{\rho}_{ss}(R^\nu(t)) \qquad (44)$$

we have

$$\partial_t \Delta\hat{\rho} = -\frac{i}{\hbar}[\hat{H}, \Delta\hat{\rho}] - \sum_\nu \dot{R}^\nu \partial_{R^\nu} \hat{\rho}_{ss} \qquad (45)$$

Note that Eq. (45) is an exact equation for the non-adiabatic correction. Its solution

$$\Delta\hat{\rho}(t) = \Delta\hat{\rho}(T) - \sum_{v}\dot{R}^{v}\int_{T}^{t}\hat{U}(t,\tau)\left\{\partial_{R^{v}}\hat{\rho}_{ss}\left(R^{v}(\tau)\right)\right\}\hat{U}^{\dagger}(t,\tau)d\tau \quad (46)$$

is an exact formal solution of Eq. (45). To guarantee that the integral in Eq. (46) converges uniformly in the limit $T \to -\infty$ we re-write it in the form

$$\Delta\hat{\rho}(t) = \Delta\hat{\rho}(T) - \lim_{\eta \to +0}\sum_{v}\dot{R}^{v}\int_{T}^{t}\exp(\eta(\tau-t)/\hbar)\hat{U}(t,\tau)\left\{\partial_{R^{v}}\hat{\rho}_{ss}\left(R^{v}(\tau)\right)\right\}\hat{U}^{\dagger}(t,\tau)d\tau \quad (47)$$

Introducing an adiabatic approximation $\hat{U}(t,\tau) \approx \exp\left(-i\hat{H}\left(R^{v}(t)\right)(t-\tau)/\hbar\right)$, $\partial_{R^{v}}\hat{\rho}_{ss}\left(R^{v}(\tau)\right) \approx \partial_{R^{v}}\hat{\rho}_{ss}\left(R^{v}(t)\right)$ and setting the boundary condition $\hat{\rho}^{(1)}(-\infty) = 0$ we have, now to first order in $\dot{R}^{v}$ [44]

$$\hat{\rho}_{ss}^{(1)}(t) = -\lim_{\eta \to +0}\sum_{v}\dot{R}^{v}\int_{-\infty}^{0}\exp(\eta\tau/\hbar)\exp(i\hat{H}(R^{v}(t))\tau/\hbar)\left\{\partial_{R^{v}}\hat{\rho}_{ss}\left(R^{v}(t)\right)\right\}\exp\left(-i\hat{H}\left(R^{v}(t)\right)\tau/\hbar\right)d\tau \quad (48)$$

where we made the change of variables $\tau \equiv \tau - t$. It is easy directly to verify that $\hat{\rho}_{ss}^{(1)}$ in Eq.(48) is Hermitian and $\text{Tr}\{\hat{\rho}_{ss}^{(1)}\} = 0$.

Consider an operator $\hat{A}$, written in terms of the adiabatic scattering operators as in Eq.(28), namely

$$\hat{A}(R^{v}) = \sum_{k\beta n\alpha}\gamma_{k\beta n\alpha}(R^{v})\hat{\psi}_{k\beta}^{\dagger}(R^{v})\hat{\psi}_{n\alpha}(R^{v}) \quad (49)$$

The adiabatic expectation value of this operator is obtained from Eq. (29) for the instantaneous value of $R^{v}$. To obtain the non-adiabatic correction to this expectation value we can use the non-adiabatic correction to the density operator, Eq. (48), in evaluating $\langle\hat{A}(t)\rangle = \text{Tr}\left(\hat{\rho}(t)\hat{A}(R^{v})\right)$. This leads to (see SI, Section E):

$$\langle\hat{A}\rangle^{(1)} = \hbar\lim_{\eta \to +0}\sum_{v}\dot{R}^{v}\sum_{k\beta n\alpha}\left(\frac{\eta}{(\varepsilon_{k\beta}-\varepsilon_{n\alpha})^{2}+\eta^{2}} + i\frac{\varepsilon_{k\beta}-\varepsilon_{n\alpha}}{(\varepsilon_{k\beta}-\varepsilon_{n\alpha})^{2}+\eta^{2}}\right)\gamma_{k\beta n\alpha}\text{Tr}\left[\hat{\rho}_{ss}\partial_{R^{v}}\left(\hat{\psi}_{k\beta}^{\dagger}\hat{\psi}_{n\alpha}\right)\right] \quad (50)$$

An alternative but equivalent procedure that should (and does: SI, Section E) yield the same result is to evaluate the non-adiabatic corrections to the Heisenberg representation, $\hat{A}_H(t) = \hat{A}_H(\{R(t)\}) + \hat{A}_H^{(1)}(t)$, and use it with the adiabatic density operator, $\langle \hat{A} \rangle^{(1)}(t) = \text{Tr}\left(\hat{\rho}_{ss}(\{R(t)\})\hat{A}_H^{(1)}(t)\right)$.

*Driving the dot level.* If the driving is done by a process that changes $\varepsilon_d$, e.g., by varying a gate potential, we can further use the identity (SI, Section F)

$$\partial_{\varepsilon_d}\left(\hat{\psi}_{k\beta}^\dagger \hat{\psi}_{n\alpha}\right) = -\sum_{m\gamma} V_{k\beta}^* \frac{V_{m\gamma} G_{dd}^r(\varepsilon_{k\beta}) G_{dd}^a(\varepsilon_{m\gamma})}{\varepsilon_{m\gamma} - \varepsilon_{k\beta} - i\eta_1} \hat{\psi}_{m\gamma}^\dagger \hat{\psi}_{n\alpha} + \sum_{m\gamma} V_{n\alpha} \frac{V_{m\gamma}^* G_{dd}^a(\varepsilon_{n\alpha}) G_{dd}^r(\varepsilon_{m\gamma})}{\varepsilon_{m\gamma} - \varepsilon_{n\alpha} + i\eta_1} \hat{\psi}_{k\beta}^\dagger \hat{\psi}_{m\gamma}$$

(51)

to get

$$\text{Tr}\left\{\hat{\rho}_{ss}\partial_{\varepsilon_d}\left(\hat{\psi}_{k\beta}^\dagger \hat{\psi}_{n\alpha}\right)\right\} = -V_{k\beta}^* \frac{V_{n\alpha} G_{dd}^r(\varepsilon_{k\beta}) G_{dd}^a(\varepsilon_{n\alpha})}{\varepsilon_{n\alpha} - \varepsilon_{k\beta} - i\eta_1}\left\{f_\alpha(\varepsilon_{n\alpha}) - f_\beta(\varepsilon_{k\beta})\right\}$$

(52)

In (51) and (52), the limit $\eta_1 \to +0$ is implied. We can now test this formalism against previously obtained results for a for the single lead case. Substituting Eq. (52) into Eq. (50) for the single lead case and for $\hat{A} = \hat{d}^\dagger \hat{d}$, using the identity

$$\int_{-\infty}^{\infty} \delta(\varepsilon_n - \varepsilon) \frac{F(\varepsilon_n) - F(\varepsilon)}{\varepsilon_n - \varepsilon} d\varepsilon = \partial_\varepsilon F(\varepsilon_n)$$ where $F(\varepsilon_n)$ is an analytical function, and

transforming the double sum into an integral leads to (SI, Section G) the following expression for the lowest non-adiabatic correction to the particle number[45]:

$$N^{(1)} = \frac{-\dot{\varepsilon}_d \hbar}{4\pi} \int d\varepsilon A_{dd}^2(\varepsilon) \partial_\varepsilon f$$

(53)

The result (53) leads to the 2nd order expression for the power:

$$\dot{W}^{(2)} = \dot{\varepsilon}_d \left\langle \frac{\partial \hat{H}}{\partial \varepsilon_d} \right\rangle^{(1)} = \dot{\varepsilon}_d N^{(1)} = \frac{-(\dot{\varepsilon}_d)^2 \hbar}{4\pi} \int d\varepsilon A_{dd}^2(\varepsilon) \partial_\varepsilon f$$

(54)

This single lead result was obtained earlier[27,28]. The equivalent result for a multi-terminal junction is (see SI, Section G) is:

$$\dot{W}^{(2)} = \frac{-(\dot{\varepsilon}_d)^2 \hbar}{4\pi} \int d\varepsilon \, A_{dd}^2(\varepsilon) \partial_\varepsilon \tilde{f} \tag{55}$$

where we have introduced a weighted distribution function

$$\tilde{f}(\varepsilon) = \sum_\alpha \frac{\Gamma_\alpha(\varepsilon) f_\alpha(\varepsilon)}{\Gamma(\varepsilon)} \tag{56}$$

In the wide-band limit ($\Gamma$, $\Gamma_\alpha$ constant), Eq. (55) coincides with the result of Ref [34].

*Driving both the dot level and the dot-lead coupling.* Next, let both the dot $\varepsilon_d$ and the couplings $V_{k\alpha} = |V_{k\alpha}| \exp(-i\Phi_{k\alpha})$ be subjects of slow driving, characterized by the driving parameters:

$$\dot{\varepsilon}_d = \dot{R} \frac{d\varepsilon_d}{dR} \equiv \dot{R} K_d \tag{57a}$$

$$\dot{\Gamma}_\alpha = \dot{R} \frac{d\Gamma_\alpha}{dR} \equiv \dot{R} K_{\Gamma\alpha} \tag{57b}$$

$$\dot{\Lambda} = \dot{R} \frac{d\Lambda}{dR} \equiv \dot{R} K_\Lambda \tag{57c}$$

$$\dot{\Phi}_\alpha = \dot{R} \frac{d\Phi_\alpha}{dR} \equiv \dot{R} K_{\Phi\alpha} \tag{57d}$$

Note that $K_\Lambda = \sum_\alpha K_{\Lambda\alpha}$ and $K_{\Gamma\alpha}$ are not independent both reflect the energy dependence of the system-bath couplings $V$ in Eqs. (25) and (26). Also note that their dependence on the energy $\varepsilon$ was suppressed just in order to shorten the notation. The result for the non-adiabatic correction to the power is (see SI, Section H)

$$\dot{W}^{(2)} = \dot{W}_I^{(2)} + \dot{W}_{II}^{(2)} \tag{58}$$

where

$$\dot{W}_I^{(2)} = -\frac{\hbar(\dot{R})^2}{4\pi} \int_{-\infty}^{\infty} d\varepsilon \left(\frac{A_{dd}}{\Gamma}\right)^2 \sum_\alpha (\partial_\varepsilon f_\alpha)\Gamma_\alpha \sum_\beta \Gamma_\beta$$

$$\times \left\{ K_\Lambda + K_d + \frac{1}{2}\sum_\gamma \left\{(2K_{\Phi\gamma} - K_{\Phi\alpha} - K_{\Phi\beta})\Gamma_\gamma\right\} + (\varepsilon - \varepsilon_d - \Lambda)\left(\frac{K_{\Gamma\alpha}}{2\Gamma_\alpha} + \frac{K_{\Gamma\beta}}{2\Gamma_\beta}\right)\right\}^2 \quad (59)$$

$$\dot{W}_{II}^{(2)} = -\frac{\hbar(\dot{R})^2}{8\pi} \int_{-\infty}^{\infty} d\varepsilon \left(\frac{A_{dd}}{\Gamma}\right)^2 \sum_\alpha \sum_\beta (f_\alpha - f_\beta)$$

$$\times \left[\left\{(\partial_\varepsilon \Gamma_\alpha)\Gamma_\beta - (\partial_\varepsilon \Gamma_\beta)\Gamma_\alpha\right\}\left[K_\Lambda + K_d + \frac{1}{2}\sum_\gamma\left\{(2K_{\Phi\gamma} - K_{\Phi\alpha} - K_{\Phi\beta})\Gamma_\gamma\right\} + (\varepsilon - \varepsilon_d - \Lambda)\left(\frac{K_{\Gamma\alpha}}{2\Gamma_\alpha} + \frac{K_{\Gamma\beta}}{2\Gamma_\beta}\right)\right]^2\right.$$

$$+\Gamma_\alpha \Gamma_\beta \left[\left\{2(K_\Lambda + K_d) + \sum_\gamma\left\{(2K_{\Phi\gamma} - K_{\Phi\alpha} - K_{\Phi\beta})\Gamma_\gamma\right\} + (\varepsilon - \varepsilon_d - \Lambda)\left(\frac{K_{\Gamma\alpha}}{\Gamma_\alpha} + \frac{K_{\Gamma\beta}}{\Gamma_\beta}\right)\right\}\right.$$

$$\times \left\{\frac{1}{2}\sum_\gamma \partial_\varepsilon\left\{(K_{\Phi\beta} - K_{\Phi\alpha})\Gamma_\gamma\right\} + (1 - \partial_\varepsilon \Lambda)\left(\frac{K_{\Gamma\alpha}}{2\Gamma_\alpha} - \frac{K_{\Gamma\beta}}{2\Gamma_\beta}\right) + (\varepsilon - \varepsilon_d - \Lambda)\partial_\varepsilon\left(\frac{K_{\Gamma\alpha}}{2\Gamma_\alpha} - \frac{K_{\Gamma\beta}}{2\Gamma_\beta}\right)\right\}$$

$$+\left\{2(\varepsilon - \varepsilon_d - \Lambda)(K_{\Phi\alpha} - K_{\Phi\beta}) + \Gamma\left(\frac{K_{\Gamma\alpha}}{2\Gamma_\alpha} - \frac{K_{\Gamma\beta}}{2\Gamma_\beta}\right)\right\}$$

$$\times\left\{-\frac{1}{2}\partial_\varepsilon K_\Gamma + \partial_\varepsilon\left\{\text{PP}\int_{-\infty}^{\infty} d\varepsilon' \sum_\gamma \Gamma_\gamma(\varepsilon')\frac{(2K_{\Phi\gamma}(\varepsilon') - K_{\Phi\alpha} - K_{\Phi\beta})}{2\pi(\varepsilon - \varepsilon')}\right\} + K_{\Phi\alpha} + K_{\Phi\beta}\right.$$

$$\left.\left.+(\varepsilon - \varepsilon_d)\partial_\varepsilon(K_{\Phi\alpha} + K_{\Phi\beta}) + \partial_\varepsilon\left(\left\{\frac{K_{\Gamma\alpha}}{\Gamma_\alpha} + \frac{K_{\Gamma\beta}}{\Gamma_\beta}\right\}\frac{\Gamma}{4}\right)\right\}\right]$$

(60)

with $K_\Gamma = \sum_\alpha K_{\Gamma\alpha}$. The contribution $\dot{W}_{II}^{(2)}$ arises solely from the time and/or energy dependence of the self-energy and vanishes in the wide band limit if the dot-baths coupling is not driven.

The following observations are noteworthy:

(a) In the absence of voltage and/or thermal bias, i.e., when the dot effectively interacts with a single bath, $\dot{W}_{II}^{(2)}$ vanishes. In this case Eqs. (58), (59) reduce to:

$$\dot{W}^{(2)} = -\frac{\hbar}{4\pi}(\dot{R})^2 \int_{-\infty}^{\infty} d\varepsilon (\partial_\varepsilon f)(\Gamma(\varepsilon)A_{dd}(\varepsilon))^2 \left\{\partial_R\left(\frac{\varepsilon-\varepsilon_d-\Lambda(\varepsilon)}{\Gamma(\varepsilon)}\right)\right\}^2 \qquad (61)$$

If only $\varepsilon_d$ is driven (see SI, Sect. H for details) this result is equivalent to Eq. (54),. In this case $\dot{W}^{(2)}$ can be identified with the excess work against friction resulting from driving at finite speed.

(b) The wide band limit of Eq. (61)

$$\begin{aligned}\dot{W}^{(2)} &= -\frac{\hbar}{4\pi}(\dot{R})^2 \Gamma^2 \int_{-\infty}^{\infty} d\varepsilon (\partial_\varepsilon f) A_{dd}^2(\varepsilon) \left\{\partial_R\left(\frac{\varepsilon-\varepsilon_d}{\Gamma}\right)\right\}^2 \\ &= -\frac{\hbar}{4\pi}(\dot{R})^2 \Gamma^2 \int_{-\infty}^{\infty} d\varepsilon (\partial_\varepsilon f) A_{dd}^2(\varepsilon) \left\{\frac{K_d}{\Gamma} - \frac{\varepsilon-\varepsilon_d}{\Gamma^2} K_\Gamma\right\}^2\end{aligned} \qquad (62)$$

describes the case where the dot energy and the dot-coupling are driven in the single lead case. This result coincides with that obtained for this limit by Haughian and co-workers[46] (see Eq. (44) in this reference).

(c) It is easily seen that $\dot{W}_I^{(2)}$ is always positive, namely representing work done on the system while driving it. On the other hand, $\dot{W}_{II}^{(2)}$ can be positive or negative. Moreover, their sum $\dot{W}^{(2)}$ can be negative, representing conversion of voltage bias into potentially useful work. Note that a positive $\dot{W}^{(2)}$ in a biased junction represent not only excess work against friction but possibly also work spent to drive electronic current against bias voltage. We show below that these scenarios can translate into a pump or motor action of this model device under cyclic driving.

(d) While the attribute discussed above exist also in the absence of phase driving, a novel aspect of the results (58)-(60) is the contribution from driving the phase of the system-baths coupling. Physically, the driven phase may reflect driving by an external electromagnetic field and the power may be generated due to the Lorentz force[34,47] between the excess current and the field. A connection between the driven phases, the excess current and power production is shown in the SI, Section L. Also, if only the phases are driven and $K_{\Phi\beta} \neq K_{\Phi\alpha}$, the excess current is produced by the interference of the waves coming from different baths as we show in the SI, Section I.

(e) $\dot{W}_{II}^{(2)}$ can become negative (implying the possibility to extract work from a voltage bias) even when only the $\varepsilon_d$ is driven if beyond the wide-band limit. One possible scenario for such colored bath is to have the driven dot level coupled to wide-band baths through one or more static levels. An example of such case is presented below.

The physical behavior stemming from the colored character of the leads is demonstrated using the model displayed in the inset to Fig. 2: the energy $\varepsilon_d$ of the dot level is driven between two leads whose spectral densities translate into the self-energy functions are given by $\Sigma_{L/R}(\varepsilon) = \Lambda_{L/R}(\varepsilon) - (i/2)\Gamma_{L/R}(\varepsilon)$ with

$$\Gamma_{L/R} = \Gamma \frac{\delta^2}{(\varepsilon \pm E_0)^2 + \delta^2}; \qquad \Lambda_{L/R} = \frac{\Gamma}{2} \frac{\delta(\varepsilon \pm E_0)}{(\varepsilon \pm E_0)^2 + \delta^2} \qquad (63)$$

with the + and − signs correspond to the left and right sides, respectively. These contributions to the dot self-energy correspond to a system in which the dot level is coupled to static levels of unperturbed energies $-\varepsilon_0$ and $+\varepsilon_0$ that are in turn coupled to a wide band bath on the left and right, respectively. (Note that the functions $\Gamma(\varepsilon)$ and $\Lambda(\varepsilon)$ are related to each other by the standard Kramers-Kroning relations). Figure 2 shows the integrated second order work functions, $\oint \dot{W}_{I}^{(2)}$, $\oint \dot{W}_{II}^{(2)}$ and $\oint \dot{W}^{(2)} = \oint \dot{W}_{I}^{(2)} + \oint \dot{W}_{II}^{(2)}$ calculated for periodic driving of $\varepsilon_d$ (see caption to Fig. 2 for choice of parameters), plotted against the bias voltage where positive bias corresponds to the situation seen in the inset (electronic chemical potential higher on the right). It is seen that the excess work $\oint \dot{W}_{II}^{(2)}$ changes from positive at small bias to negative at high bias and moreover dominates the sign of the total excess work, $\oint \dot{W}^{(2)}$, when the bias voltage becomes comparable to and larger than $2\varepsilon_0$. Indeed, it is easily realized that for such bias the cyclic driving moves the dot level up when it is mostly empty, and down when it is mostly occupied, extracting a net mechanical power from the voltage bias. This gain can be substantially larger than friction losses that dominate the $\oint \dot{W}_I^{(2)}$ contribution. In particular, the latter become small when $\Delta V$ is large because at large bias the dot level does not oscillates near the high friction domains near the left or right Fermi energies.

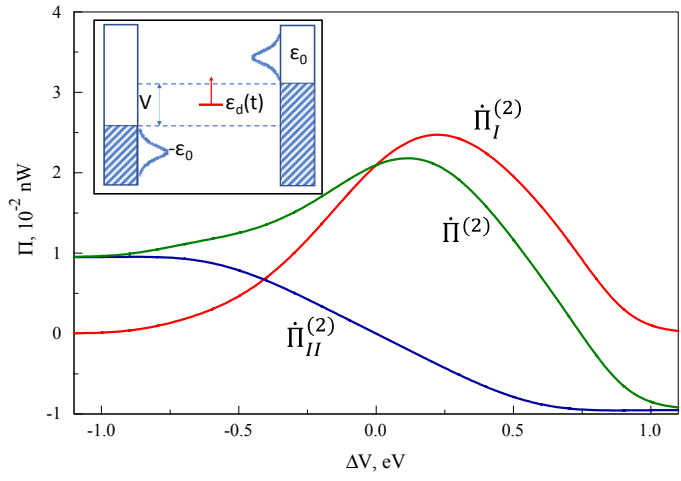

Figure 2. The excess work per period $\Pi = \tau^{-1}W^{(2)} = \tau^{-1}\oint \dot{W}^{(2)}$ and its components $\Pi_I = \tau^{-1}W_I^{(2)}$ and $\Pi_{II} = \tau^{-1}W_{II}^{(2)}$ calculated for the model where a single dot level is driven periodically according to $\varepsilon_d(t) = E_0 \sin(2\pi \frac{t}{\tau})$ between leads whose interaction with the dot is characterized by the self-energies (63), plotted against the bias potential ΔV. The latter is divided evenly relative to the unbiased junction between the leads (see the insert). Here Γ = 0.1eV, δ = 0.3eV, E₀=0.4eV, ε₀ = 0.4eV, τ = 1ps.

Finally, it is of interest to examine the connection of the present formalism to the extension, developed by von Oppen and co-workers for slowly driven systems[28,34,47], of the Landauer-Buttiker S-matrix formalism[15,17]. In Refs. 28,34,47 driving induced corrections to the scattering matrix were obtained using the NEGF formalism. Here we obtained the same results by calculating the first order correction to the net flux into a given bath β using the first order corrections due to driving to the density matrix $\hat{\rho}_{ss}^{(1)}$, and to the outgoing waves, $\left(\hat{\psi}^\dagger_{k\alpha,-}\hat{\psi}_{n\beta,-}\right)^{(1)}$. To facilitate comparison with results of Refs. [28,34,47] we specify in what follows to one lead (denoted β) and to the case where only the dot level energy is driven. The net flux into the lead per unit energy at steady state is given by

$$j_\beta(\varepsilon) = j_{\beta,out}(\varepsilon) - j_{\beta,in}(\varepsilon) \tag{64}$$

where[48]

$$j_{\beta,in}(\varepsilon) = \frac{1}{2\pi\hbar}\left(\text{Tr}\left[\hat{\rho}_{ss}\hat{\psi}^\dagger_{k\beta,+}\hat{\psi}_{k\beta,+}\right]\right)_{\varepsilon_k=\varepsilon} = \frac{f(\varepsilon)}{2\pi\hbar}$$

$$j_{\beta,out}(\varepsilon) = \frac{1}{2\pi\hbar}\left(\text{Tr}\left[\hat{\rho}_{ss}\hat{\psi}^\dagger_{k\beta,-}\hat{\psi}_{k\beta,-}\right]\right)_{\varepsilon_k=\varepsilon} = \frac{f(\varepsilon)+\phi(\varepsilon)}{2\pi\hbar} \tag{65}$$

and where have denoted

$$\phi(\varepsilon) = \left(\text{Tr}\left[\hat{\rho}_{ss}\left(\hat{\psi}^\dagger_{k\beta,-}\hat{\psi}_{k\beta,-} - \hat{\psi}^\dagger_{k\beta,+}\hat{\psi}_{k\beta,+}\right)\right]\right)_{\varepsilon_k=\varepsilon} \tag{66}$$

We show in SI, Section J that to first order in $\dot{\varepsilon}_d$ $\phi(\varepsilon)$ is given by

$$\phi^{(1)}(\varepsilon) = -\hbar\dot{\varepsilon}_d A_{dd}(\varepsilon)\partial_\varepsilon f(\varepsilon) \tag{67}$$

which coincides with the correction given by Eq. (26) of Ref. 28. The dissipated power can be then derived from the correction (67) ( see Eq. (20) in Ref. 28)

$$\dot{W}^{(2)} = -\frac{1}{4\pi\hbar}\int_{-\infty}^{\infty}d\varepsilon\frac{1}{\partial_\varepsilon f}\left(\phi^{(1)}(\varepsilon)\right)^2 = \frac{-(\dot{\varepsilon}_d)^2\hbar}{4\pi}\int_{-\infty}^{\infty}d\varepsilon A_{dd}^2(\varepsilon)(\partial_\varepsilon f) \tag{68}$$

which coincides with (54).

For the same resonance level/one lead model, if both the dot energy and dot-lead coupling are driven as a function of some parameter R, the correction to the distribution for a single lead is obtained in the form (Supplementary Information, Section J)

$$\phi^{(1)}(\varepsilon) = \hbar\dot{R}A_{dd}(\varepsilon)\Gamma(\varepsilon)\left(\partial_\varepsilon f(\varepsilon)\right)\partial_R\left(\frac{\varepsilon-\varepsilon_d-\Lambda(\varepsilon)}{\Gamma(\varepsilon)}\right). \tag{69}$$

Substituting (69) into (68) recovers the result (61). Thus, using our scattering approach we were able to rigorously generalize the extension by von Oppen and coworkers of the Landauer-Buttiker S-matrix theory to driven systems beyond the wide-band approximation.

**V. Thermodynamic fluxes and the second law**

Using the results of the previous sections, we can now establish the thermodynamics of the dot coupled to multiple baths. Here we limit ourselves to the special case where the wide-band limit is assumed and only the dot energy level is driven. The results obtained below generalize the single-bath results obtained in References 27-28.

The fact that the total non-equilibrium distribution is a product of equilibrium distributions in the basis of scattering states implies the following expression for the total system entropy:

$$S^{(0)} = \sum_\beta S^{(0)}_\beta = k_B \sum_\beta \int D_\beta(\varepsilon) \sigma(f_\beta(\varepsilon)) d\varepsilon \tag{70}$$

where $D_\beta(\varepsilon) = \frac{1}{2\pi} \frac{\Gamma_\beta}{\Gamma} A_{dd}(\varepsilon)$ (see Eq. (38)) is a partial density of states of lead β obtained from the scattering formalism in SI, Section G and where
$$\sigma(x) \equiv -x \ln x - (1-x) \ln(1-x) \tag{71}$$
The first order correction to the entropy due to the driving takes the form

$$S^{(1)} = k_B \sum_\beta \int D_\beta(\varepsilon) \frac{\partial \sigma(x)}{\partial x}\Big|_{x=f_\beta(\varepsilon)} \delta f_\beta(\varepsilon) d\varepsilon = \sum_\beta \int D_\beta(\varepsilon) \frac{\varepsilon - \mu_\beta}{T_\beta} \delta f_\beta(\varepsilon) d\varepsilon \tag{72}$$

where $\delta f_\beta(\varepsilon)$ is a correction to the Fermi distribution. This correction is derived in the SI, Section K, and is given by

$$\delta f_\beta(\varepsilon) = -\hbar \frac{\dot{\varepsilon}_d}{2} A_{dd}(\varepsilon) \partial f_\beta(\varepsilon) \tag{73}$$

(note that this expression is identical to that in the single bath treatments of Refs. 27 and 28, originally derived by Kita[49]). This leads to:

$$S^{(1)} = -\frac{\dot{\varepsilon}_d}{2} \hbar \sum_\beta \int \frac{\Gamma_\beta}{2\pi \Gamma} \frac{\varepsilon - \mu_\beta}{T_\beta} A^2_{dd}(\varepsilon) \partial(f_\beta(\varepsilon)) d\varepsilon \tag{74}$$

The second order correction to the entropy is obtained from (74)

$$\dot{S}^{(2)} = \dot{\varepsilon}_d \partial_{\varepsilon_d}(S^{(1)}) = \frac{(\dot{\varepsilon}_d)^2}{2} \sum_\beta \int \frac{\Gamma_\beta}{2\pi \Gamma} \frac{\varepsilon - \mu_\beta}{T_\beta} \partial(A^2_{dd}(\varepsilon)) \partial(f_\beta(\varepsilon)) d\varepsilon \tag{75}$$

Note that the individual contributions to this excess entropy are the net entropy fluxes associated with the different baths and are identical in form to the flux calculated by Bruch et al[27,28] in the single bath case.

Next, consider the energy, which can again be written as a sum of the terms associated with the different baths (leads):

$$E_\beta^{(0)} = \int \varepsilon D_\beta(\varepsilon) f_\beta(\varepsilon) d\varepsilon \tag{76}$$

The corresponding correction due to the dot driving may be again expanded in powers of $\dot{\varepsilon}_d$. To first order we get

$$E_\beta^{(1)} = \int \varepsilon D_\beta(\varepsilon) \delta f_\beta(\varepsilon) d\varepsilon = -\frac{\dot{\varepsilon}_d}{2} \hbar \int \frac{\Gamma_\beta}{2\pi\Gamma} A_{dd}^2(\varepsilon) \partial(f_\beta(\varepsilon)) d\varepsilon \tag{77}$$

As a consistency check we note that this result can be also obtained (SI, Section K) using $E^{(1)} = \text{Tr}\{\hat{\rho}^{(1)} \hat{H}_{eff}\}$ where $\hat{H}_{eff}$ is the split Hamiltonian given by Eq. (40)

The quasistatic power production associated with lead β is defined as follows

$$\dot{W}_\beta^{(1)} = \dot{\varepsilon}_d N_\beta^{(0)} = \frac{1}{2\pi}(\dot{\varepsilon}_d) \int A_{dd}(\varepsilon) \frac{\Gamma_\beta}{\Gamma} f_\beta(\varepsilon) d\varepsilon \tag{78}$$

and the corresponding second order correction is

$$\dot{W}_\beta^{(2)} = \dot{\varepsilon}_d N_\beta^{(1)} = -\frac{1}{4\pi}(\dot{\varepsilon}_d)^2 \hbar \int A_{dd}^2(\varepsilon) \frac{\Gamma_\beta}{\Gamma} \partial(f_\beta(\varepsilon)) d\varepsilon \tag{79}$$

Because $\partial(f_\beta(\varepsilon)) < 0$, this correction $\dot{W}_\beta^{(2)}$ is always positive in the wide band limit.

Finally, the rate of heat produced in the quasistatic limit and the corresponding second order non-adiabatic correction are obtained from the first law (see Ref.27 and SI, Section K)

$$\dot{Q}_\beta^{(1)} = \dot{\varepsilon}_d \left(\partial_{\varepsilon_d} E_\beta^{(0)} - \mu_\beta \partial_{\varepsilon_d} N_\beta^{(0)}\right) - \dot{W}_\beta^{(1)} = \dot{\varepsilon}_d \int (\varepsilon - \mu_\beta) D_\beta(\varepsilon) \partial(f_\beta(\varepsilon)) d\varepsilon \tag{80}$$

$$\dot{Q}_\beta^{(2)} = \dot{\varepsilon}_d \left(\partial_{\varepsilon_d} E_\beta^{(1)} - \mu_\beta \partial_{\varepsilon_d} N_\beta^{(1)}\right) - \dot{W}_\beta^{(2)} = -\frac{(\dot{\varepsilon}_d)^2}{4\pi} \hbar \int (\varepsilon - \mu_\beta) A_{dd}^2(\varepsilon) \frac{\Gamma_\beta}{\Gamma} (\partial^2 f_\beta(\varepsilon)) d\varepsilon \tag{81}$$

Using these results, we can recast the entropy change in different orders as related to the work and heat fluxes. In zero order (static dot level) the change in entropy represents the steady state heats produced in the different baths under the steady state electronic currents

$$\dot{S}^{(0)} = \sum_\beta \frac{\dot{Q}_\beta^{(0)}}{T_\beta} \tag{82}$$

where $\dot{Q}_\beta^{(0)}$ is a steady state current coming from lead β into the dot. It should be emphasized that Eqs. (70) and (82) are not compatible, and represent two different views of the final state of the scattering process: Eq. (70) is the mathematical consequence of the time independent nature of the scattering process, while Eq. (82) reflects our understanding that all fluxes entering a given bath eventually thermalize in that bath in processes that that are not part of the scattering dynamics.

To first order in the driving speed we find from (70) and (80) a similar expression (SI, Section K)

$$\dot{S}^{(1)} = \dot{\varepsilon}_d \partial_{\varepsilon_d} \left( S^{(0)} \right) = \dot{\varepsilon}_d k_B \sum_\beta \int D_\beta(\varepsilon) \frac{\partial \sigma(x)}{\partial x}\bigg|_{x=f_\beta(\varepsilon)} \partial \left( f_\beta(\varepsilon) \right) d\varepsilon = \sum_\beta \frac{\dot{Q}_\beta^{(1)}}{T_\beta} \tag{83}$$

which expresses the fact that also to this order the entropy change is associated with the heat fluxes into or out from the baths, and no additional entropy is produced. Going to second order we find using Eqs.(74), (79) and (81) (SI, Section K) :

$$\dot{S}^{(2)} = \sum_\beta \frac{\dot{Q}_\beta^{(2)}}{T_\beta} + \sum_\beta \frac{\dot{W}_\beta^{(2)}}{T_\beta} \tag{84}$$

This identifies the (always positive) term $\dot{S}^{(2)}_{produced} = \sum_\beta \dot{W}_\beta^{(2)}/T_\beta$ as the entropy produced in the system due to the excess work done while driving the dot. Expressions (83) and (84) are a natural generalization for multiple baths of the results obtained in Ref 27 for one bath. It should be noted however that the result for $\dot{S}^{(2)}_{produced}$ is a non-trivial generalization, which requires to write the excess work as a sum of contributions (79) from the different baths. It should be emphasized that the positivity of $\dot{W}_\beta^{(2)}/T_\beta$ and its interpretation as the entropy production rate hold only in the wide band limit considered here.

## VI. Conclusions

We have obtained a general expression for the non-equilibrium steady state density matrix of multiple infinite baths coupled through a general interaction. Using the Moller (wave) operator, the non-equilibrium steady state density operator is expressed as a product of equilibrium (Gibbs) density operators associated with the different baths, expressed in terms of the corresponding incoming field operators. The developed framework recovers standard results obtained from the Landauer-Buttiker S-matrix theory or the non-equilibrium Green function formalism, as well as recent results obtained for slowly driven systems.

The results previously derived in the wide band approximation and for a single bath have now been obtained for multiple baths without taking the wide-band limit (WBL). In particular, a general expression for the dissipated power for the driven non-interacting resonant level were derived for general, multiple non-WBL baths connected through a driven dot, where both the dot energy level and its couplings to the baths are driven. It is also shown that, in the wide-band limit, the effective symmetric splitting of the system-bath interaction, used to determine the effective system Hamiltonian for the case of one bath[12,27,30], is valid also for a system coupled to multiple baths[50]. Importantly, we have generalized the thermodynamics of the driven dot[27,28] to the multiple bath case.

Driving a system coupled to multiple colored (non WBL) baths under bias has been found to have non-trivial energetic consequences. In particular, the excess power, $\dot{W}^{(2)}$, calculated to second order in the driving speed, can become negative, implying that it is possible to extract work or pump charge for a colored bath by driving the dot level alone. This development will make it possible to consider full engine/refrigerator cycles based on this model for non-equilibrium quantum thermodynamics of strongly coupled systems. It also implies that the driving induced entropy production, expressed in the WBL to this order by $\dot{W}^{(2)}/T$ (or in the many bath case as a sum over baths, $\sum_\beta \dot{W}^{(2)}_\beta / T_\beta$, cf. Eq. (84), which is always positive in the WBL, will need to be modified in a non-trivial way in the case of colored baths. We defer this issue to future work. Another issue for further future studies is the implications of driving the phases of the system-baths interaction, for which we have provided a preliminary discussion in Section IV and in the SI, Sections H and I.

To end this discussion, a conceptual issue should be pointed out. The driven resonance level model was constructed to represent the physics of leads connected to a bridging system, where each lead is assumed to be in its own thermal equilibrium. The physics behind the latter assumption reflects the microscopic size of the dot and the contact region relative to the macroscopic leads. To create a corresponding mathematical construct, one may assume that the leads are coupled to some external 'superbaths' that determines their intensive properties – temperature and chemical potential[25,51] This procedure works well so long as the process under consideration is near steady state so the dynamics at the interfaces between the leads and the superbaths is inconsequential. However, when the system is strongly driven, the dynamics at the dot-lead interface may become decoupled from that at the boundary between the leads and the superbaths, making definition of 'heat' and 'entropy' ambiguous in the sense that the heat $Q$ exchanged with the external superbaths (and the associated entropy $Q/T$) does not reflect the instantaneous dynamics at the dot-lead interface. This in turn results in the observation that expansion in the driving speed (Sect. IV) fails to yield consistent thermodynamics beyond second order[19]. The manifestation of this issue within the scattering approach will be considered in another publication.

**Acknowledgement**

This research was supported by the U.S. National Science Foundation (Grant No. CHE1665291 to A.N.), the Israel-U.S. Binational Science Foundation, the German Research Foundation (Grant No.DFG TH 820/11-1), and the University of Pennsylvania

**Appendix A. Derivation of the non-equilibrium steady state density matrix**

Here we prove that Eq. (8) with the Hamiltonian (6) gives a steady state density operator for all times $t > 0$.

Consider the following operator:

$$\hat{\Omega}(t) = \exp(-i\hat{H}t/\hbar)\exp(i\hat{H}_0 t/\hbar) \tag{A1}$$

where (note the difference from (6)) $\hat{H} = \hat{H}_0 + \hat{V}$. From (A1) it follows that

$$\partial_t \hat{\Omega}(t) = -\frac{i}{\hbar}\exp(-i\hat{H}t/\hbar)\hat{V}\exp(i\hat{H}_0 t/\hbar)$$
$$= -\frac{i}{\hbar}\exp(-i\hat{H}t/\hbar)\exp(i\hat{H}_0 t/\hbar)\exp(-i\hat{H}_0 t/\hbar)\hat{V}\exp(i\hat{H}_0 t/\hbar) = -\frac{i}{\hbar}\exp(-i\hat{H}t/\hbar)\exp(i\hat{H}_0 t/\hbar)\hat{V}_I(-t)$$
$$\tag{A2}$$

where $\hat{V}_I$ denotes the interaction representation of the coupling, $V_I(t) = \exp(i\hat{H}_0 t/\hbar)V\exp(-i\hat{H}_0 t/\hbar)$. An integral form of (A2) is

$$\hat{\Omega}(T_2) - \hat{\Omega}(T_1) = -\frac{i}{\hbar}\int_{T_1}^{T_2} \exp(-i\hat{H}t/\hbar)\exp(i\hat{H}_0 t/\hbar)\hat{V}_I(-t)dt \tag{A3}$$

Assuming that $\int_{T_1}^{T_2}\left\|\exp(-i\hat{H}t/\hbar)\exp(i\hat{H}_0 t/\hbar)\hat{V}_I(-t)\right\|dt < \infty$ we can re-write (A3) as follows:

$$\hat{\Omega}(T_2) - \hat{\Omega}(T_1) = -\frac{i}{\hbar}\lim_{\eta \to +0}\int_{T_1}^{T_2} a_\eta(t)\exp(-i\hat{H}t/\hbar)\exp(i\hat{H}_0 t/\hbar)\hat{V}_I(-t)dt \tag{A4}$$

where $\lim_{\eta\to+0} a_\eta(t)=1$, $|a_\eta(t)|<\infty$ and $\lim_{|T_{1(2)}|\to\infty} a_\eta(t)=0$. $a_\eta(t)$ is introduced to insure uniform convergence of the integral in the limit $|T_{1(2)}|\to\infty$.

Choosing $t_1=0$ and $a_\eta(t)=\exp(-\eta|\tau|)$, Eq. (A4) becomes

$$\hat{\Omega}(t)=\hat{I}-\frac{i}{\hbar}\lim_{\eta\to+0}\int_0^t \exp(-\eta|\tau|)\exp(-i\hat{H}\tau/\hbar)\exp(i\hat{H}_0\tau/\hbar)\hat{V}_I(-\tau)d\tau$$

$$=\hat{I}-\frac{i}{\hbar}\lim_{\eta\to+0}\int_0^t \exp(-\eta|\tau|)\hat{\Omega}(\tau)\hat{V}_I(-\tau)d\tau$$

$$=\hat{I}-\frac{i}{\hbar}\lim_{\eta\to+0}\int_0^t \exp(-\eta|\tau|)\hat{V}_I(-\tau)d\tau+\left(-\frac{i}{\hbar}\right)^2\lim_{\eta\to+0}\int_0^t\int_0^\tau \exp(-\eta|\tau|)\exp(-\eta|\tau_1|)\hat{\Omega}(\tau_1)\hat{V}_I(-\tau_1)d\tau_1\hat{V}_I(-\tau)d\tau$$

(A5)

Changing $\tau\to-\tau$ the first integral in (A5) can be re-written as follows:

$$-\frac{i}{\hbar}\lim_{\eta\to+0}\int_0^t \exp(-\eta|\tau|)\hat{V}_I(-\tau)d\tau=-\frac{i}{\hbar}\int_{-t}^0 \hat{\tilde{V}}_I(\tau)d\tau \tag{A6}$$

where

$$\hat{\tilde{V}}_I(\tau)=\exp(-\eta|\tau|)\hat{V}_I(\tau) \tag{A7}$$

In the second integral, change of variables $\tau\to-\tau$ and $\tau_1\to-\tau_1$ and swapping $\tau\leftrightarrow\tau_1$ leads to

$$\left(\frac{i}{\hbar}\right)^2\lim_{\eta\to+0}\int_0^t\int_0^\tau \exp(-\eta|\tau|)\exp(-\eta|\tau_1|)\hat{\Omega}(\tau_1)\hat{V}_I(-\tau_1)d\tau_1\hat{V}_I(-\tau)d\tau$$

$$=\left(\frac{i}{\hbar}\right)^2\lim_{\eta\to+0}\int_{-t}^0\int_{-\tau}^0 \hat{\Omega}(\tau_1)\hat{\tilde{V}}_I(\tau_1)d\tau_1\hat{\tilde{V}}_I(\tau)d\tau$$

(A8)

By continuing the recursion process with respect to $\hat{\Omega}(\tau_1)$, we obtain the following expansion:

$$\hat{\Omega}_+\equiv\hat{\Omega}(\infty)=\hat{I}-\frac{i}{\hbar}\int_{-\infty}^0 \hat{\tilde{V}}_I(\tau)d\tau+\left(-\frac{i}{\hbar}\right)^2\int_{-\infty}^0 \hat{\tilde{V}}_I(\tau)\int_{-\infty}^\tau \hat{\tilde{V}}_I(\tau_1)d\tau_1 d\tau+\ldots \tag{A9}$$

which constitutes an expansion of the Moller operator. One thing should be emphasized here: expression (A9) makes sense only if the series (A9) converges and the limit $\hat{\Omega}_+ \equiv \hat{\Omega}(\infty) = \lim_{t\to\infty} \exp(-i\hat{H}t)\exp(i\hat{H}_0 t)$ exists.

Introducing the evolution operator $\hat{U}_I(t_2,t_1) = \exp\left\{T\int_{t_1}^{t_2}\hat{\tilde{V}}_I(t)dt\right\}$ where $T$ stands for the time ordering, the solution Eq.(7) can be written as follows:

$$\hat{\rho}(t=0) = \hat{\rho}_I(t=0) = \hat{U}(0,-\infty)\hat{\rho}(t=-\infty)\hat{U}^\dagger(0,-\infty)$$
$$= \hat{U}_I(0,-\infty)\hat{\rho}_I(t=-\infty)\hat{U}_I^\dagger(0,-\infty) \quad (A10)$$

where index $I$ stands for the interaction representation. The evolution operator $\hat{U}_I(t_2,t_1)$ satisfies the following equation:

$$\partial_{t_2}\hat{U}_I(t_2,t_1) = -\frac{i}{\hbar}\hat{\tilde{V}}_I(t_2)\hat{U}_I(t_2,t_1) \quad (A11)$$

Thus

$$\hat{U}_I(t_2,t_1) = \hat{I} - \frac{i}{\hbar}\int_{t_1}^{t_2}\hat{\tilde{V}}_I(t_2)\hat{U}_I(t,t_1)dt \quad (A12)$$

Using recursion procedure, we can obtain the Dyson series for the evolution operator (A12):

$$\hat{U}_I(t,-\infty) = 1 - \frac{i}{\hbar}\int_{-\infty}^t \hat{\tilde{V}}_I(\tau)d\tau + \frac{(-i)^2}{\hbar^2}\int_{-\infty}^t \hat{\tilde{V}}_I(\tau)\int_{-\infty}^\tau \hat{\tilde{V}}_I(\tau_1)d\tau_1 d\tau + \ldots \quad (A13)$$

From (A9) and (A13) we see that

$$\hat{\Omega}_+ = \hat{U}_I(0,-\infty) \quad (A14)$$

which implies

$$\hat{\rho}(t=0) = \hat{\Omega}_+ \hat{\rho}_I(t=-\infty)\hat{\Omega}_+^\dagger \quad (A15)$$

Now we re-write the Moller operator a bit differently

$$\hat{\Omega}_+ = \hat{I} - \frac{i}{\hbar} \lim_{\eta \to +0} \int_{-\infty}^{0} \exp(\eta\tau)\exp(i\hat{H}\tau/\hbar)\exp(-i\hat{H}_0\tau/\hbar)\hat{V}_I(\tau)d\tau =$$

$$= \lim_{\eta \to +0} \eta \int_{-\infty}^{0} \exp(\eta\tau)\exp(i\hat{H}\tau/\hbar)\exp(-i\hat{H}_0\tau/\hbar)d\tau \quad (A16)$$

which is obtained from (A5) where the time was reversed $\tau \to -\tau$. In deriving (A16) we have integrated by parts using the equalities $\exp(i\hat{H}\tau/\hbar)\exp(-i\hat{H}_0\tau/\hbar)\hat{V}_I(\tau)$

$$= \frac{\hbar}{i}\frac{d}{d\tau}\left(\exp(i\hat{H}\tau/\hbar)\exp(-i\hat{H}_0\tau/\hbar)\right) \text{ and } \left[\exp(\eta\tau)\exp(i\hat{H}\tau/\hbar)\exp(-i\hat{H}_0\tau/\hbar)\right]_{-\infty}^{0} = \hat{I} \text{ . Thus}$$

$$\hat{H}\hat{\Omega}_+ = \lim_{\eta \to +0} \eta \int_{-\infty}^{0} \exp(\eta\tau)\exp(i\hat{H}\tau/\hbar)\hat{H}\exp(-i\hat{H}_0\tau/\hbar)d\tau$$

$$= \lim_{\eta \to +0} \eta \int_{-\infty}^{0} \exp(\eta\tau)\exp(i\hat{H}\tau/\hbar)(\hat{H}_0 + \hat{V})\exp(-i\hat{H}_0\tau/\hbar)d\tau$$

$$= \lim_{\eta \to +0} \eta \int_{-\infty}^{0} \exp(\eta\tau)\exp(i\hat{H}\tau/\hbar)\hat{H}_0 \exp(-i\hat{H}_0\tau/\hbar)d\tau \quad (A17)$$

$$+ \lim_{\eta \to +0} \eta \int_{-\infty}^{0} \exp(\eta\tau)\exp(i\hat{H}\tau/\hbar)\exp(-i\hat{H}_0\tau/\hbar)\hat{V}_I(\tau)d\tau$$

$$= \hat{\Omega}_+\hat{H}_0 + \frac{\hbar}{i}\lim_{\eta \to +0}\eta(\hat{I} - \hat{\Omega}_+) = \hat{\Omega}_+\hat{H}_0$$

which immediately leads to the well-known intertwining relation:

$$\hat{H}\hat{\Omega}_+ = \hat{\Omega}_+\hat{H}_0 \quad (A18)$$

Using Eq.(A16) the density matrix derivative at $t=0$ is evaluated as follows:

$$\partial_t \hat{\rho}(t=0) = -\frac{i}{\hbar}[\hat{H}(0), \hat{\rho}(0)] = \frac{i}{\hbar}\left(\hat{\Omega}_+\hat{\rho}_I(t=-\infty)\hat{\Omega}_+^\dagger \hat{H} - \hat{H}\hat{\Omega}_+\hat{\rho}_I(t=-\infty)\hat{\Omega}_+^\dagger\right)$$

$$= \frac{i}{\hbar}\left(\hat{\Omega}_+\hat{\rho}_I(t=-\infty)\hat{H}_0\hat{\Omega}_+^\dagger - \hat{\Omega}_+\hat{H}_0\hat{\rho}_I(t=-\infty)\hat{\Omega}_+^\dagger\right)$$

$$= \frac{i}{\hbar}\hat{\Omega}_+\left([\hat{\rho}_0, \hat{H}_0]\right)_I \hat{\Omega}_+^\dagger = 0$$

(A19)

where the last equality is obtained by assuming that $\left[\hat{\rho}_0, \hat{H}_0\right] = 0$. Since Eq. (7) is a first order differential equation, by recalling the existence and uniqueness theorem it follows from Eq. (A19) that $\hat{\rho}(t \geq 0) = \hat{\rho}(t = 0)$. Thus, the solution of Eq.(7) at $t > 0$ indeed yields a steady state given by Eq. (8).

**Appendix B. Equivalence of McLennan-Zubarev and Hershfield approaches to the present scattering method**

Here we show that the present scattering-theory based method is equivalent to the McLennan-Zubarev and Hershfield approaches for calculating the non-equilibrium steady-state density matrix[52,53].

In Appendix A it was shown that the solution of Eq.(7) under the adiabatic switching (6) of the inter-bath coupling yields a steady state at positive times. Alternatively, we can also write the time evolution of Eq. (7) in the interaction representation

$$\partial_t \hat{\rho}_I(t) = -\frac{i}{\hbar} \left[ \hat{\tilde{V}}_I(t), \hat{\rho}_I(t) \right] \tag{B1}$$

where $\tilde{V}$ is given by Eq. (A7) and includes a convergence factor. Integrating (B1) we have

$$\hat{\rho}_I(t) = \hat{\rho}_I(-\infty) - \int_{-\infty}^{t} \frac{i}{\hbar} \left[ \hat{\tilde{V}}_I(\tau), \hat{\rho}_I(\tau) \right] d\tau \tag{B2}$$

And continuing by recursion, we get a Dyson-like expression for the density matrix:

$$\hat{\rho}_I(t) = \hat{\rho}_I(-\infty) - \int_{-\infty}^{t} \frac{i}{\hbar} \left[ \hat{\tilde{V}}_I(\tau), \hat{\rho}_I(-\infty) \right] d\tau + \int_{-\infty}^{t} \frac{i}{\hbar} \left[ \hat{\tilde{V}}_I(\tau), \int_{-\infty}^{\tau} \frac{i}{\hbar} \left[ \hat{\tilde{V}}_I(\tau_1), \hat{\rho}_I(-\infty) \right] d\tau_1 \right] d\tau + \ldots \tag{B3}$$

Based on Appendix A, setting $t = 0$ in (B3), gives a steady state solution for $t > 0$. On the other hand, Eq. (B3) is exactly the series used by Hershfield for non-equilibrium steady state matrix[38]. This indicates the equivalence of our results and Heshfield's ones.

Next, we show the equivalence of our approach to that of McLennan and Zubarev[36,37]. To this end, we start from $\hat{H} = \hat{H}_0 + \hat{V}$ and consider the following exponential operator:

$$\hat{\tilde{U}}(T_2, T_1) = \exp\left(\frac{-i}{\hbar}\hat{H}(T_2 - T_1)/\hbar\right) \tag{B4}$$

which can be expanded into the following series:

$$\hat{\tilde{U}}(T_2, T_1) = \hat{I} + \left(\frac{-i}{\hbar}\right)\int_{T_1}^{T_2} \hat{H} dt + \left(\frac{-i}{\hbar}\right)^2 \int_{T_1}^{T_2} \hat{H} \int_{T_1}^{t} \hat{H} d\tau dt + \ldots \tag{B5}$$

We proceed by introducing the exponential factor $\exp(-\eta|\tau|)$ in each integral as we did in Appendix A, where the limit $\eta \to +0$ should be taken at the end of any calculation[54].

$$\hat{\tilde{U}}(T_2, T_1) = \lim_{\eta \to +0}\left(\hat{I} + \left(\frac{-i}{\hbar}\right)\int_{T_1}^{T_2} \hat{H}\exp(-\eta|t|)dt + \left(\frac{-i}{\hbar}\right)^2 \int_{T_1}^{T_2} \exp(-\eta|t|)\hat{H}\int_{T_1}^{t}\exp(-\eta|\tau|)\hat{H}d\tau dt + \ldots\right) \tag{B6}$$

The same expansion can be written for $\hat{H}_0$:

$$\hat{\tilde{U}}_0(T_2, T_1) = \exp\left(-\frac{i}{\hbar}\hat{H}_0(T_2 - T_1)\right)$$

$$= \lim_{\eta \to +0}\left(\hat{I} + \left(\frac{-i}{\hbar}\right)\int_{T_1}^{T_2} \hat{H}_0\exp(-\eta|t|)dt + \left(\frac{-i}{\hbar}\right)^2 \int_{T_1}^{T_2} \exp(-\eta|t|)\hat{H}_0\int_{T_1}^{t}\exp(-\eta|\tau|)\hat{H}_0 d\tau dt + \ldots\right) \tag{B7}$$

Next, consider the operator $\hat{\tilde{\Omega}}(T_1) = \hat{\tilde{U}}(0, T_1)\hat{\tilde{U}}_0(T_1, 0)$. Its time derivative is given by

$$\partial_{T_1}\left(\hat{\tilde{U}}(0, T_1)\hat{\tilde{U}}_0(T_1, 0)\right) = \frac{i}{\hbar}\hat{\tilde{U}}(0, T_1)\exp(-\eta|T_1|)\hat{V}\hat{\tilde{U}}_0(T_1, 0)$$

$$= \lim_{\eta \to +0}\frac{i}{\hbar}\hat{\tilde{U}}(0, T_1)\hat{\tilde{U}}_0(T_1, 0)\hat{\tilde{U}}_0(0, T_1)\exp(-\eta|T_1|)\hat{V}\hat{\tilde{U}}_0(T_1, 0) \tag{B8}$$

Using $\hat{\tilde{V}}_I(T_1) = \hat{\tilde{U}}_0(0, T_1)\exp(-\eta|T_1|)\hat{V}\hat{\tilde{U}}_0(T_1, 0)$ (see Eq.(A7)), Eq. (B8) leads to

$$\partial_{T_1}\hat{\tilde{\Omega}}(T_1) = \frac{i}{\hbar}\hat{\tilde{\Omega}}(T_1)\hat{\tilde{V}}_I(T_1) \tag{B9}$$

which can be expanded in the Dyson-like series:

$$\hat{\tilde{\Omega}}(T_1) = \lim_{\eta \to +0}\left\{\hat{I} - \frac{i}{\hbar}\int_{T_1}^{0}\hat{\tilde{V}}_I(\tau)d\tau + \left(\frac{-i}{\hbar}\right)^2\int_{T_1}^{0}\hat{\tilde{V}}_I(\tau)\int_{T_1}^{\tau}\hat{\tilde{V}}_I(\tau_1)d\tau_1 d\tau + \ldots\right\} \tag{B10}$$

Eq. (B10) is similar to the interaction representation evolution operator given by (A13). In particular, from Eqs. (A9) and (A13) we see that $\hat{\tilde{\Omega}}(-\infty) = \hat{\tilde{U}}(0,-\infty)\hat{\tilde{U}}_0(-\infty,0) = \hat{\Omega}_+$. This implies that Eq. (A15) is equivalent (since $\hat{\tilde{U}}_0$ commutes with $\hat{\rho}_0$) to

$$\hat{\rho} = \hat{\tilde{U}}(0,-\infty)\hat{\rho}_0\hat{\tilde{U}}^\dagger(0,-\infty) \tag{B11}$$

which is the "standard" solution of Eq. (7). We have thus shown that the derivation along the steps taken here reproduces the results of Appendix A. Note that to show this equivalence we need to demand that $\hat{\rho}_0$ commutes with $\hat{H}_0$, although this is not a condition for (B11) to be valid. [55]

To show the equivalence to the McLennan Zubarev formalism consider the operator:

$$\hat{\tilde{\rho}}(x) = \hat{\tilde{U}}(0,x)\hat{\rho}_0\hat{\tilde{U}}^\dagger(0,x) \tag{B12}$$

Its derivative with respect to $x$ is

$$\begin{aligned}\partial_x\hat{\tilde{\rho}}(x) &= \left(\partial_x\hat{\tilde{U}}(0,x)\right)\hat{\rho}_0\hat{\tilde{U}}^\dagger(0,x) + \hat{\tilde{\Omega}}(0,x)\hat{\rho}_0\left(\partial_x\hat{\tilde{U}}^\dagger(0,x)\right) \\ &= \frac{i}{\hbar}\hat{H}\exp(-\eta|x|)\hat{\tilde{U}}(0,x)\hat{\rho}_0\hat{\tilde{U}}^\dagger(0,x) - \frac{i}{\hbar}\exp(-\eta|x|)\hat{\tilde{U}}(0,x)\hat{\rho}_0\hat{\tilde{U}}^\dagger(0,x)\hat{H} \\ &= \frac{i}{\hbar}\left[\hat{H}\exp(-\eta|x|), \hat{\tilde{U}}(0,x)\hat{\rho}_0\hat{\tilde{U}}^\dagger(0,x)\right]\end{aligned} \tag{B13}$$

Again, in (B13), the limit $\eta\to +0$ is assumed. An integral form of (B13) is

$$\hat{\tilde{\rho}}(x) = \hat{\rho}_0 - \int_x^0 \frac{i}{\hbar}\exp(-|\eta||t|)\left[\hat{H}, \exp(\frac{i}{\hbar}\hat{H}t)\hat{\rho}_0\exp(-\frac{i}{\hbar}\hat{H}t)\right]dt \tag{B14}$$

which, in the limit $x\to -\infty$, becomes

$$\hat{\rho} = \hat{\rho}_0 - \lim_{\eta\to +0}\int_{-\infty}^0 \frac{i}{\hbar}\exp(\eta t)\left[\hat{H}, \exp(\frac{i}{\hbar}\hat{H}t)\hat{\rho}_0\exp(-\frac{i}{\hbar}\hat{H}t)\right]dt \tag{B15}$$

Eq. (B15) can be generalized: from Eq.(B11) it follows that

$$f(\hat{\rho}) = f\left(\hat{\tilde{U}}(0,-\infty)\hat{\rho}_0\hat{\tilde{U}}^\dagger(0,-\infty)\right) = \hat{\tilde{U}}(0,-\infty)f(\hat{\rho}_0)\hat{\tilde{U}}^\dagger(0,-\infty) =$$
$$f(\hat{\rho}_0) - \lim_{\eta \to +0}\int_{-\infty}^{0}\frac{i}{\hbar}\exp(\eta t)\left[\hat{H},\exp\left(\frac{i}{\hbar}\hat{H}t\right)f(\hat{\rho}_0)\exp\left(-\frac{i}{\hbar}\hat{H}t\right)\right]dt \quad (B16)$$

for any analytic $f$. In deriving (B16) we assumed that $\hat{\tilde{U}}(0,-\infty)$ is a unitary operator. By putting $f(\hat{\rho}) = \ln(\hat{\rho})$ in (B16) we get the following expression for the NESS density matrix

$$\hat{\rho} = \lim_{\eta \to +0}\prod_{\alpha}\frac{1}{Z_\alpha}\exp\left\{-\beta_\alpha\left(\hat{H}_0^\alpha - \mu_\alpha\hat{N}_0^\alpha - \int_{-\infty}^{0}\frac{i}{\hbar}\exp(\eta t)\left[\hat{H},\exp\left(\frac{i}{\hbar}\hat{H}t\right)(\hat{H}_0^\alpha - \mu_\alpha\hat{N}_0^\alpha)\exp\left(-\frac{i}{\hbar}\hat{H}t\right)\right]dt\right)\right\}$$
(B17)

Eq. (B17) is the McLennan-Zubarev non-equilibrium steady state density matrix. We note that throughout the derivation we assumed that the series (B6) converges.

**Appendix C. The Lippmann-Schwinger equation and creation/annihilation operators in the scattering states representation of the resonant level model**

Here we derive Eqs. (20)-(21) of the main text. We start by showing that

$$\left|\psi_{k\beta}\right\rangle = \hat{\Omega}_+\left|c_{k\beta}\right\rangle \quad (C1)$$

Indeed,

$$\hat{\Omega}_+\hat{H}_0\left|c_{k\beta}\right\rangle = \varepsilon_{k\beta}\hat{\Omega}_+\left|c_{k\beta}\right\rangle \quad (C2)$$

because $\left|c_{k\beta}\right\rangle$ is an eigenstate of $\hat{H}_0$ with the eigenenergy $\varepsilon_{k\beta}$.

Using the intertwining relation:

$$\hat{\Omega}_+\hat{H}_0\left|c_{k\beta}\right\rangle = \hat{H}\hat{\Omega}_+\left|c_{k\beta}\right\rangle = \varepsilon_{k\beta}\hat{\Omega}_+\left|c_{k\beta}\right\rangle \quad (C3)$$

gives (C1). Note that the relationship (C1) holds more generally: for any (scattering) many-body eigenstate of $\hat{H}_0$ operating with $\hat{\Omega}_+$ yields a corresponding eigenstate of $\hat{H}$ with the same eigenenergy.

The expression for the Moller operator (A16) is written in the time domain. It can be rewritten in the energy domain (assumed $\eta \to +0$, $2\eta$ is used instead of $\eta$ and $\hbar=1$):

$$\hat{\Omega}_+ = \frac{2\eta}{2\pi} \int\limits_{-\infty}^{\infty} \int\limits_{-\infty}^{0} \int\limits_{-\infty}^{0} \exp(i\hat{H}t)\exp(\eta t)\exp(\eta t')\exp(-i\hat{H}_0 t')\exp(i\varepsilon\{t-t'\}) dt dt' d\varepsilon =$$

$$= \frac{2\eta}{2\pi} \int\limits_{-\infty}^{\infty} \frac{1}{\varepsilon - \hat{H} + i\eta} \frac{1}{\varepsilon - \hat{H}_0 - i\eta} d\varepsilon$$

(C4)

Recalling the Dyson equation

$$\frac{1}{\varepsilon - \hat{H} \pm i\eta} = \frac{1}{\varepsilon - \hat{H}_0 \pm i\eta} + \frac{1}{\varepsilon - \hat{H} \pm i\eta} \hat{V} \frac{1}{\varepsilon - \hat{H}_0 \pm i\eta}$$

(C5)

we have for (C4):

$$\hat{\Omega}_+ = \frac{2\eta}{2\pi} \int\limits_{-\infty}^{\infty} \left( \frac{1}{\varepsilon - \hat{H}_0 + i\eta} + \frac{1}{\varepsilon - \hat{H} + i\eta} \hat{V} \frac{1}{\varepsilon - \hat{H}_0 + i\eta} \right) \frac{1}{\varepsilon - \hat{H}_0 - i\eta} d\varepsilon$$

$$= \frac{1}{2\pi} \int\limits_{-\infty}^{\infty} \frac{2\eta}{\{\varepsilon - \hat{H}_0\}^2 + \eta^2} d\varepsilon + \frac{1}{2\pi} \int\limits_{-\infty}^{\infty} \frac{1}{\varepsilon - \hat{H} + i\eta} \hat{V} \frac{2\eta}{\{\varepsilon - \hat{H}_0\}^2 + \eta^2} d\varepsilon$$

(C6)

$$= \int\limits_{-\infty}^{\infty} \delta(\varepsilon - \hat{H}_0) d\varepsilon + \int\limits_{-\infty}^{\infty} \hat{G}^r(\varepsilon) \hat{V} \delta(\varepsilon - \hat{H}_0) d\varepsilon = \hat{I} + \int\limits_{-\infty}^{\infty} \hat{G}^r(\varepsilon) \hat{V} \delta(\varepsilon - \hat{H}_0) d\varepsilon$$

Substituting the expression above into (C1) we have:

$$\left| \psi_{k\beta} \right\rangle = \left( \hat{I} + \int\limits_{-\infty}^{\infty} \hat{G}^r(\varepsilon) \hat{V} \delta(\varepsilon - \hat{H}_0) d\varepsilon \right) \left| c_{k\beta} \right\rangle = \left| c_{k\beta} \right\rangle + \int\limits_{-\infty}^{\infty} \hat{G}^r(\varepsilon) \hat{V} \delta(\varepsilon - \hat{H}_0) \left| c_{k\beta} \right\rangle d\varepsilon$$

$$= \left| c_{k\beta} \right\rangle + \hat{G}^r(\varepsilon_{k\beta}) \hat{V} \left| c_{k\beta} \right\rangle$$

(C7)

Eq. (C7) is also correct for an arbitrary many-body eigenstate[56] (assuming it belongs to the continuous spectrum of $\hat{H}_0$ i.e. it is a scattering state).

There is an alternative route of deriving (C7): from the Schrodinger equation it follows:

$$\hat{H} \left| \psi_{k\beta} \right\rangle - \hat{H}_0 \left| c_{k\beta} \right\rangle = \varepsilon_{k\beta} \left( \left| \psi_{k\beta} \right\rangle - \left| c_{k\beta} \right\rangle \right)$$

(C8)

or

$$\left| \psi_{k\beta} \right\rangle = \left| c_{k\beta} \right\rangle + \left( \varepsilon_{k\beta} - \hat{H}_0 \right)^{-1} \hat{V} \left| \psi_{k\beta} \right\rangle$$

(C9)

To avoid singularity, $i\eta$ must be added to the denominator which leads to the textbook version of Lippmann-Schwinger equation:

$$|\psi_{k\beta}\rangle = |c_{k\beta}\rangle + \hat{G}_0^{r/a}\hat{V}|\psi_{k\beta}\rangle \tag{C10}$$

where $\hat{G}_0^{r/a} = \lim_{\eta \to +0}(\varepsilon - \hat{H}_0 \pm i\eta)^{-1}$. Substituting iteratively $|\psi_{k\beta}\rangle$ into (C10) one can obtain an infinite (Born) series for (C10):

$$\begin{aligned}|\psi_{k\beta}\rangle &= \left(\hat{I} + \hat{G}_0^{r/a}\hat{V} + \hat{G}_0^{r/a}\hat{V}\hat{G}_0^{r/a}\hat{V} + ...\right)|c_{k\beta}\rangle \\ &= \left(\hat{I} + \hat{G}^{r/a}\hat{V}\right)|c_{k\beta}\rangle\end{aligned} \tag{C11}$$

which coincides with (C7). Note that this textbook derivation has an ambiguity with regard of the sign of $i\eta$, i.e. whether the solution we seek is incoming or outgoing.

For the non-interacting resonant level model Eq. (C7) can be solved analytically:

$$\begin{aligned}\hat{G}^r(\varepsilon_{k\beta})\hat{V}|c_{k\beta}\rangle &= \hat{G}^r(\varepsilon_{k\beta})|d\rangle V_{k\beta}^* \\ &= G_{dd}^r(\varepsilon_{k\beta})|d\rangle V_{k\beta}^* + V_{k\beta}^*\sum_{n\alpha} G_{n\alpha d}^r(\varepsilon_{k\beta})|c_{n\alpha}\rangle \\ &= G_{dd}^r(\varepsilon_{k\beta})|d\rangle V_{k\beta}^* + V_{k\beta}^* G_{dd}^r(\varepsilon_{k\beta})\sum_{n\alpha} G_{0,n\alpha n\alpha}^r(\varepsilon_{k\beta})V_{n\alpha}|c_{n\alpha}\rangle \\ &= G_{dd}^r(\varepsilon_{k\beta})|d\rangle V_{k\beta}^* + V_{k\beta}^* G_{dd}^r(\varepsilon_{k\beta})\sum_{n\alpha} \frac{1}{\varepsilon_{k\beta} - \varepsilon_{n\alpha} + i\eta}V_{n\alpha}|c_{n\alpha}\rangle\end{aligned} \tag{C12}$$

Thus,

$$|\psi_{k\beta}\rangle = G_{dd}^r(\varepsilon_{k\beta})V_{k\beta}^*|d\rangle + \sum_{n\alpha}\left\{\delta_{k\beta n\alpha} + G_{dd}^r(\varepsilon_{k\beta})V_{n\alpha}V_{k\beta}^*\frac{1}{\varepsilon_{k\beta} - \varepsilon_{n\alpha} + i\eta}\right\}|c_{n\alpha}\rangle \tag{C13}$$

or

$$\hat{\psi}_{k\beta}^\dagger = V_{k\beta}^* G_{dd}^r(\varepsilon_{k\beta})\hat{d}^\dagger + \sum_{n\alpha}\left\{\delta_{k\beta n\alpha} + V_{n\alpha}\frac{V_{k\beta}^* G_{dd}^r(\varepsilon_{k\beta})}{\varepsilon_{k\beta} - \varepsilon_{n\alpha} + i\eta}\right\}\hat{c}_{n\alpha}^\dagger$$

(C14)

which gives Eq. (20)

[35] Note that though the limit $t \to \infty$ is taken in Eq. (9), it is shown here that the form (8) represents the steady state density matrix reached at $t \geq 0$.

[41] In principal, a central region (bound states) can be included into the uncoupled Hamiltonian $\hat{H}_0 = \sum_\alpha \hat{H}_0^\alpha + \hat{H}_C$. In this case a projection of the Moller operator onto the

central region is no longer unitary and $\hat{\Omega}_+ \hat{\rho}_C \hat{\Omega}_+^\dagger = \hat{\mathbf{I}}$.

[42] Note that at equilibrium, representing the environment as comprising many baths only serves to keep our notation uniform

[43] A rigorous derivation of (39) can be obtained writing the energy as sum of contributions from the different leads (as implied by (13)), and calculating the $\varepsilon_d$-dependent part of the energy associated with each lead from the Grand canonical ensemble as in Ref. 27

[44] Note that this first order solution corresponds to the approximation obtained by assuming that the Hamiltonian in Eq. (45) is constant.

[45] The expansion in powers of the driving speed $\dot{\varepsilon}_d$ is valid when $\frac{\Gamma_\beta(\varepsilon)}{\partial f_\beta(\varepsilon)} \gg \hbar\dot{\varepsilon}_d$ for every lead $\beta$ and in the range of energies $\varepsilon_{min} < \varepsilon < \varepsilon_{max}$ where $\varepsilon_{max}$ and $\varepsilon_{min}$ are the maximum and the minimum values of $\varepsilon_d$ respectively. Near the Fermi surface the inequality above can be re-written as $k_B T_\beta \Gamma_\beta(\varepsilon) \gg \hbar\dot{\varepsilon}_d$. Assuming an ac-driving $\varepsilon_d = E_0 \sin(2\pi \frac{t}{\tau_c})$ and taking typical values as $k_B T_\beta = 0.026 eV$, $\tau_c = 1 ps$, $E_0 = 0.1 eV$ one obtains $\Gamma \gg 10^{-5} eV$ which holds, for such driving speeds, for most weakly ($\Gamma \ll k_B T_\beta$) and strongly coupled systems.

[46] P. Haughian, M. Esposito, and T.L. Schmidt, Phys. Rev. B **97**, 085435 (2018).

[47] N. Bode, S.V. Kusminskiy, R. Egger, and F. Von Oppen, Phys. Rev. Lett. **107**, 036804 (2011).

[48] The forms (64) are obtained by augmenting state population, normalization and speed into the standard expression for a 1-dimensional flux

[49] T. Kita: Prog. Theor. Phys. 123, **581** (2010).

[50] A symmetric splitting of energy current between system and baths was pointed out in Ref. 30

[51] A. Oz, O. Hod, and A. Nitzan, J. Chem. Theory and Comp. **16**, 1232 (2020).

[52] H. Ness, Phys. Rev. E **90**, (2014).

[53] H. Ness, Phys. Rev. E **88**, (2013).

[54] Including this factor is, as in Appendix A, to ensure that the series (B6) and (B7) converge uniformly, which allows us to evaluate derivatives of these operators as well as taking the limits $T_{1,2} \to \pm\infty$

[55] It proves that by introducing the interaction suddenly also leads to the same non-equilibrium steady-state density matrix.

[56] While in the present work only the non-interacting case is considered, Eq. (C7) is also valid for interacting many-body systems.

[57] S. Weinberg, *The Quantum Theory of Fields, Vol. 1: Foundations*, 1st ed. (Cambridge University Press, 1995).

[58] (F10) is the formal definition of the S operator written in term of the field operator and is equivalent to (F14) which expresses it in terms of overlap between scattering states. Both give S through the relationships $\hat{\psi}^\dagger_{n,+} = \sum_k S_{nk} \hat{\psi}^\dagger_{k,-}$ and $|\psi_{n,+}\rangle = \sum_k S_{nk} |\psi_{k,-}\rangle$

**Supplementary Information for**

"Transport and thermodynamics in quantum junctions: A scattering approach"
by
Alexander Semenov[†] and Abraham Nitzan[†‡1]

[†]*Department of Chemistry, University of Pennsylvania, Philadelphia, Pennsylvania 19104, USA*

[‡]*School of Chemistry, The Sackler Faculty of Science, Tel Aviv University, Tel Aviv 69978, Israel*

**Section A. Proof of Eqs.(27)**

Here we establish the relation (27) that connects between the Hamiltonian in the local (free) and scattering states representations. The calculation procedure is most easily demonstrated by starting from the sum

$$\sum_{n\alpha} \hat{c}_{n\alpha}^\dagger \hat{c}_{n\alpha} = \sum_{n\alpha} \sum_{m\gamma} \left\{ \delta_{m\gamma n\alpha} + V_{n\alpha} \frac{V_{m\gamma}^* G_{dd}^r(\varepsilon_{m\gamma})}{\varepsilon_{m\gamma} - \varepsilon_{n\alpha} + i\eta} \right\} \sum_{k\beta} \left\{ \delta_{k\beta n\alpha} + V_{n\alpha}^* \frac{V_{k\beta} G_{dd}^a(\varepsilon_{k\beta})}{\varepsilon_{k\beta} - \varepsilon_{n\alpha} - i\eta} \right\} \hat{\psi}_{k\beta}^\dagger \hat{\psi}_{m\gamma} \quad \text{(S 1)}$$

To proceed further it is useful to employ the Sokhotski–Plemelj theorem. Its integral form:

$$\lim_{\eta \to +0} \int_a^b \frac{F(x)}{x \pm i\eta} dx = \mp i\pi F(0) + \text{PP} \int_a^b \frac{F(x)}{x} dx \quad \text{(S 2)}$$

and the equivalent functional from:

$$\lim_{\eta \to +0} \frac{1}{x \pm i\eta} = \mp i\pi \delta(0) + \text{PP} \frac{1}{x} \quad \text{(S 3)}$$

where PP stands for the principal value, $F(x)$ is an analytical function and $\mp i\pi F(0)$ is a half of a residue with respect to variable $x$ and limit $\eta \to +0$.

Consider the individual terms:

$$\hat{N}_1 = \sum_{n\alpha} \sum_{m\gamma} \delta_{m\gamma n\alpha} \sum_{k\beta} \delta_{k\beta n\alpha} \hat{\psi}_{k\beta}^\dagger \hat{\psi}_{m\gamma} = \sum_{n\alpha} \hat{\psi}_{n\alpha}^\dagger \hat{\psi}_{n\alpha} \quad \text{(S 4)a}$$

$$\hat{N}_2 = \sum_{n\alpha} \sum_{m\gamma} \delta_{m\gamma n\alpha} \sum_{k\beta} \left\{ V_{n\alpha}^* \frac{V_{k\beta} G_{dd}^a(\varepsilon_{k\beta})}{\varepsilon_{k\beta} - \varepsilon_{n\alpha} - i\eta} \right\} \hat{\psi}_{k\beta}^\dagger \hat{\psi}_{m\gamma} \quad \text{(S 4)b}$$

---

[1] Author to whom correspondence should be addressed: anitzan@sas.upenn.edu

$$\hat{N}_3 = \sum_{n\alpha}\sum_{m\gamma}\left\{V_{n\alpha}\frac{V_{m\gamma}^* G_{dd}^r(\varepsilon_{m\gamma})}{\varepsilon_{m\gamma} - \varepsilon_{n\alpha} + i\eta}\right\}\sum_{k\beta}\delta_{k\beta n\alpha}\hat{\psi}_{k\beta}^\dagger\hat{\psi}_{m\gamma} \tag{S 4)c}$$

$$\hat{N}_4 = \sum_{n\alpha}\sum_{m\gamma}\left\{V_{n\alpha}\frac{V_{m\gamma}^* G_{dd}^r(\varepsilon_{m\gamma})}{\varepsilon_{m\gamma} - \varepsilon_{n\alpha} + i\eta}\right\}\sum_{k\beta}\left\{V_{n\alpha}^*\frac{V_{k\beta}G_{dd}^a(\varepsilon_{k\beta})}{\varepsilon_{k\beta} - \varepsilon_{n\alpha} - i\eta}\right\}\hat{\psi}_{k\beta}^\dagger\hat{\psi}_{m\gamma} \tag{S 4)d}$$

For the last term we have

$$\hat{N}_4 = \sum_{k\beta}\sum_{m\gamma}\sum_{n\alpha}\left\{V_{n\alpha}\frac{V_{m\gamma}^* G_{dd}^r(\varepsilon_{m\gamma})}{\varepsilon_{m\gamma} - \varepsilon_{n\alpha} + i\eta}\right\}\left\{V_{n\alpha}^*\frac{V_{k\beta}G_{dd}^a(\varepsilon_{k\beta})}{\varepsilon_{k\beta} - \varepsilon_{n\alpha} - i\eta}\right\}\hat{\psi}_{k\beta}^\dagger\hat{\psi}_{m\gamma}$$

$$= \sum_{k\beta}\sum_{m\gamma}V_{m\gamma}^* G_{dd}^r(\varepsilon_{m\gamma})V_{k\beta}G_{dd}^a(\varepsilon_{k\beta})\sum_{n\alpha}\left(\frac{|V_{n\alpha}|^2}{\varepsilon_{m\gamma} - \varepsilon_{n\alpha} + i\eta}\frac{1}{\varepsilon_{k\beta} - \varepsilon_{n\alpha} - i\eta}\right)\hat{\psi}_{k\beta}^\dagger\hat{\psi}_{m\gamma}$$

$$= \sum_{k\beta}\sum_{m\gamma}V_{m\gamma}^* G_{dd}^r(\varepsilon_{m\gamma})V_{k\beta}G_{dd}^a(\varepsilon_{k\beta})\frac{1}{\varepsilon_{k\beta} - \varepsilon_{m\gamma} - 2i\eta}\sum_{n\alpha}|V_{n\alpha}|^2\left(\frac{1}{\varepsilon_{m\gamma} - \varepsilon_{n\alpha} + i\eta} - \frac{1}{\varepsilon_{k\beta} - \varepsilon_{n\alpha} - i\eta}\right)\hat{\psi}_{k\beta}^\dagger\hat{\psi}_{m\gamma}$$

$$= \sum_{k\beta}\sum_{m\gamma}V_{m\gamma}^* G_{dd}^r(\varepsilon_{m\gamma})V_{k\beta}G_{dd}^a(\varepsilon_{k\beta})\left\{\mathrm{PP}\frac{1}{\varepsilon_{k\beta} - \varepsilon_{m\gamma}} + i\pi\delta(\varepsilon_{k\beta} - \varepsilon_{m\gamma})\right\}\left(\Sigma_{dd}^r(\varepsilon_{m\gamma}) - \Sigma_{dd}^a(\varepsilon_{k\beta})\right)\hat{\psi}_{k\beta}^\dagger\hat{\psi}_{m\gamma}$$

$$\tag{S 5}$$

where the self-energy functions are defined by Eq. (24) and Eq. (S 3) was used.

The second term can be cast as

$$\hat{N}_2 = \sum_{n\alpha}\sum_{m\gamma}\delta_{m\gamma n\alpha}\sum_{k\beta}\left\{V_{n\alpha}^*\frac{V_{k\beta}G_{dd}^a(\varepsilon_{k\beta})}{\varepsilon_{k\beta} - \varepsilon_{n\alpha} - i\eta}\right\}\hat{\psi}_{k\beta}^\dagger\hat{\psi}_{m\gamma}$$

$$= \sum_{m\gamma}\sum_{k\beta}\left\{V_{m\gamma}^*\frac{V_{k\beta}G_{dd}^r(\varepsilon_{m\gamma})G_{dd}^a(\varepsilon_{k\beta})}{\varepsilon_{k\beta} - \varepsilon_{m\gamma} - i\eta}\right\}\hat{\psi}_{k\beta}^\dagger\hat{\psi}_{m\gamma}\left\{G_{dd}^r(\varepsilon_{m\gamma})\right\}^{-1} \tag{S 6}$$

$$= \sum_{m\gamma}\sum_{k\beta}\left\{V_{m\gamma}^*\frac{V_{k\beta}G_{dd}^r(\varepsilon_{m\gamma})G_{dd}^a(\varepsilon_{k\beta})}{\varepsilon_{k\beta} - \varepsilon_{m\gamma} - i\eta}\right\}\hat{\psi}_{k\beta}^\dagger\hat{\psi}_{m\gamma}\left\{\varepsilon_{m\gamma} - \varepsilon_d - \Sigma_{dd}^r(\varepsilon_{m\gamma})\right\}$$

and the third term becomes

$$\hat{N}_3 = \sum_{n\alpha}\sum_{m\gamma}\left\{V_{n\alpha}\frac{V_{m\gamma}^* G_{dd}^r(\varepsilon_{m\gamma})}{\varepsilon_{m\gamma} - \varepsilon_{n\alpha} + i\eta}\right\}\sum_{k\beta}\delta_{k\beta n\alpha}\hat{\psi}_{k\beta}^\dagger\hat{\psi}_{m\gamma}$$

$$= \sum_{m\gamma}\sum_{k\beta}\left\{V_{k\beta}\frac{V_{m\gamma}^* G_{dd}^r(\varepsilon_{m\gamma})G_{dd}^a(\varepsilon_{k\beta})}{\varepsilon_{m\gamma} - \varepsilon_{k\beta} + i\eta}\right\}\hat{\psi}_{k\beta}^\dagger\hat{\psi}_{m\gamma}\left\{\varepsilon_{k\beta} - \varepsilon_d - \Sigma_{dd}^a(\varepsilon_{k\beta})\right\} \tag{S 7}$$

Using Eqs. (S 4)-(S 7) and (S 3) in (SA1) one obtains

$$\sum_{n\alpha} \hat{c}^{\dagger}_{n\alpha} \hat{c}_{n\alpha} = \sum_{n\alpha} \hat{\psi}^{\dagger}_{n\alpha} \hat{\psi}_{n\alpha} - \sum_{m\gamma} \sum_{k\beta} \{V^*_{m\gamma} V_{k\beta} G^r_{dd}(\varepsilon_{m\gamma}) G^a_{dd}(\varepsilon_{k\beta})\} \hat{\psi}^{\dagger}_{k\beta} \hat{\psi}_{m\gamma} \left\{ \text{PP} \frac{1}{\varepsilon_{k\beta} - \varepsilon_{m\gamma}} + \pi i \delta(\varepsilon_{k\beta} - \varepsilon_{m\gamma}) \right\}$$

$$\times \left\{ (\varepsilon_{k\beta} - \varepsilon_{m\gamma}) + \Sigma^r_{dd}(\varepsilon_{m\gamma}) - \Sigma^r_{dd}(\varepsilon_{m\gamma}) + \Sigma^a_{dd}(\varepsilon_{k\beta}) - \Sigma^a_{dd}(\varepsilon_{k\beta}) + \varepsilon_d - \varepsilon_d \right\}$$

$$= \sum_{n\alpha} \hat{\psi}^{\dagger}_{n\alpha} \hat{\psi}_{n\alpha} - \sum_{m\gamma} \sum_{k\beta} \{V^*_{m\gamma} V_{k\beta} G^r_{dd}(\varepsilon_{m\gamma}) G^a_{dd}(\varepsilon_{k\beta})\} \hat{\psi}^{\dagger}_{k\beta} \hat{\psi}_{m\gamma} (\varepsilon_{k\beta} - \varepsilon_{m\gamma}) \text{PP} \frac{1}{\varepsilon_{k\beta} - \varepsilon_{m\gamma}}$$

$$= \sum_{n\alpha} \hat{\psi}^{\dagger}_{n\alpha} \hat{\psi}_{n\alpha} - \sum_{m\gamma} \sum_{k\beta} \{V^*_{m\gamma} V_{k\beta} G^r_{dd}(\varepsilon_{m\gamma}) G^a_{dd}(\varepsilon_{k\beta})\} \hat{\psi}^{\dagger}_{k\beta} \hat{\psi}_{m\gamma}$$

(S 8)

and

$$\hat{d}^{\dagger} \hat{d} = \sum_{k\beta} G^a_{dd}(\varepsilon_{k\beta}) V_{k\beta} \hat{\psi}^{\dagger}_{k\beta} \sum_{m\gamma} G^r_{dd}(\varepsilon_{m\gamma}) V^*_{m\gamma} \hat{\psi}_{m\gamma} \tag{S 9}$$

which leads to

$$\sum_{n\alpha} \hat{c}^{\dagger}_{n\alpha} \hat{c}_{n\alpha} = \sum_{n\alpha} \hat{\psi}^{\dagger}_{n\alpha} \hat{\psi}_{n\alpha} - \hat{d}^{\dagger} \hat{d} \tag{S 10}$$

We can employ the same procedure to evaluate $\hat{H}_0 = \sum_{k\alpha} \varepsilon_{k\alpha} \hat{c}^{\dagger}_{k\alpha} \hat{c}_{k\alpha}$ which is written as

$$\hat{H}_0 = \hat{H}_{01} + \hat{H}_{02} + \hat{H}_{03} + \hat{H}_{04} \quad \text{where}$$

$$\hat{H}_{01} = \sum_{n\alpha} \sum_{m\gamma} \delta_{m\gamma n\alpha} \sum_{k\beta} \delta_{k\beta n\alpha} \hat{\psi}^{\dagger}_{k\beta} \hat{\psi}_{m\gamma} = \sum_{n\alpha} \varepsilon_{n\alpha} \hat{\psi}^{\dagger}_{n\alpha} \hat{\psi}_{n\alpha} \tag{S 11)a}$$

$$\hat{H}_{02} = \sum_{n\alpha} \sum_{m\gamma} \varepsilon_{n\alpha} \delta_{m\gamma n\alpha} \sum_{k\beta} V^*_{n\alpha} \frac{V_{k\beta} G^a_{dd}(\varepsilon_{k\beta})}{\varepsilon_{k\beta} - \varepsilon_{n\alpha} - i\eta} \hat{\psi}^{\dagger}_{k\beta} \hat{\psi}_{m\gamma} \tag{S 11)b}$$

$$\hat{H}_{03} = \sum_{n\alpha} \sum_{m\gamma} \varepsilon_{n\alpha} V_{n\alpha} \frac{V^*_{m\gamma} G^r_{dd}(\varepsilon_{m\gamma})}{\varepsilon_{m\gamma} - \varepsilon_{n\alpha} + i\eta} \sum_{k\beta} \delta_{k\beta n\alpha} \hat{\psi}^{\dagger}_{k\beta} \hat{\psi}_{m\gamma} \tag{S 11)c}$$

$$\hat{H}_{04} = \sum_{n\alpha} \sum_{m\gamma} \varepsilon_{n\alpha} V_{n\alpha} \frac{V^*_{m\gamma} G^r_{dd}(\varepsilon_{m\gamma})}{\varepsilon_{m\gamma} - \varepsilon_{n\alpha} + i\eta} \sum_{k\beta} V^*_{n\alpha} \frac{V_{k\beta} G^a_{dd}(\varepsilon_{k\beta})}{\varepsilon_{k\beta} - \varepsilon_{n\alpha} - i\eta} \hat{\psi}^{\dagger}_{k\beta} \hat{\psi}_{m\gamma} \tag{S 11)d}$$

The last term can be cast in the form

$$\hat{H}_{04} = \sum_{n\alpha} |V_{n\alpha}|^2 \varepsilon_{n\alpha} \sum_{k\beta} \sum_{m\gamma} V^*_{m\gamma} G^r_{dd}(\varepsilon_{m\gamma}) V_{k\beta} G^a_{dd}(\varepsilon_{k\beta}) \frac{1}{\varepsilon_{k\beta} - \varepsilon_{m\gamma} - 2i\eta} \left( \frac{1}{\varepsilon_{m\gamma} - \varepsilon_{n\alpha} + i\eta} - \frac{1}{\varepsilon_{k\beta} - \varepsilon_{n\alpha} - i\eta} \right)$$

$$= \sum_{k\beta} \sum_{m\gamma} \hat{\psi}^\dagger_{k\beta} \hat{\psi}_{m\gamma} V^*_{m\gamma} G^r_{dd}(\varepsilon_{m\gamma}) V_{k\beta} G^a_{dd}(\varepsilon_{k\beta}) \frac{1}{\varepsilon_{k\beta} - \varepsilon_{m\gamma} - 2i\eta} \sum_{n\alpha} |V_{n\alpha}|^2 \left( \frac{\varepsilon_{n\alpha}}{\varepsilon_{m\gamma} - \varepsilon_{n\alpha} + i\eta} - \frac{\varepsilon_{n\alpha}}{\varepsilon_{k\beta} - \varepsilon_{n\alpha} - i\eta} \right)$$

$$= \sum_{k\beta} \sum_{m\gamma} \hat{\psi}^\dagger_{k\beta} \hat{\psi}_{m\gamma} V^*_{m\gamma} G^r_{dd}(\varepsilon_{m\gamma}) V_{k\beta} G^a_{dd}(\varepsilon_{k\beta}) \frac{1}{\varepsilon_{k\beta} - \varepsilon_{m\gamma} - 2i\eta} \sum_{n\alpha} |V_{n\alpha}|^2$$

$$\times \left( \frac{\varepsilon_{n\alpha} - \varepsilon_{m\gamma} - i\eta + \varepsilon_{m\gamma} + i\eta}{\varepsilon_{m\gamma} - \varepsilon_{n\alpha} + i\eta} - \frac{\varepsilon_{n\alpha} - \varepsilon_{k\beta} + i\eta + \varepsilon_{k\beta} - i\eta}{\varepsilon_{k\beta} - \varepsilon_{n\alpha} - i\eta} \right)$$

$$= \sum_{k\beta} \sum_{m\gamma} \hat{\psi}^\dagger_{k\beta} \hat{\psi}_{m\gamma} V^*_{m\gamma} G^r_{dd}(\varepsilon_{m\gamma}) V_{k\beta} G^a_{dd}(\varepsilon_{k\beta}) \frac{1}{\varepsilon_{k\beta} - \varepsilon_{m\gamma} - 2i\eta} \sum_{n\alpha} |V_{n\alpha}|^2 \left( \frac{\varepsilon_{m\gamma} + i\eta}{\varepsilon_{m\gamma} - \varepsilon_{n\alpha} + i\eta} - \frac{\varepsilon_{k\beta} - i\eta}{\varepsilon_{k\beta} - \varepsilon_{n\alpha} - i\eta} \right)$$

$$= \sum_{k\beta} \sum_{m\gamma} \hat{\psi}^\dagger_{k\beta} \hat{\psi}_{m\gamma} V^*_{m\gamma} G^r_{dd}(\varepsilon_{m\gamma}) V_{k\beta} G^a_{dd}(\varepsilon_{k\beta}) \left\{ PP \frac{1}{\varepsilon_{k\beta} - \varepsilon_{m\gamma}} + i\pi \delta(\varepsilon_{k\beta} - \varepsilon_{m\gamma}) \right\} \left( \Sigma^r_{dd}(\varepsilon_{m\gamma}) \varepsilon_{m\gamma} - \Sigma^a_{dd}(\varepsilon_{k\beta}) \varepsilon_{k\beta} \right)$$

(S 12)

For the 2nd and 3rd terms the summation over $n\alpha$ yields:

$$\hat{H}_{02} = \sum_{m\gamma} \sum_{k\beta} \left\{ V^*_{m\gamma} \frac{V_{k\beta} G^r_{dd}(\varepsilon_{m\gamma}) G^a_{dd}(\varepsilon_{k\beta})}{\varepsilon_{k\beta} - \varepsilon_{m\gamma} - i\eta} \right\} \hat{\psi}^\dagger_{k\beta} \hat{\psi}_{m\gamma} \varepsilon_{m\gamma} \left\{ \varepsilon_{m\gamma} - \varepsilon_d - \Sigma^r_{dd}(\varepsilon_{m\gamma}) \right\} \qquad \text{(S 13)}$$

$$\hat{H}_{03} = \sum_{m\gamma} \sum_{k\beta} \left\{ V_{m\gamma} \frac{V^*_{m\gamma} G^r_{dd}(\varepsilon_{m\gamma}) G^a_{dd}(\varepsilon_{k\beta})}{\varepsilon_{m\gamma} - \varepsilon_{k\beta} + i\eta} \right\} \hat{\psi}^\dagger_{k\beta} \hat{\psi}_{m\gamma} \varepsilon_{k\beta} \left\{ \varepsilon_{k\beta} - \varepsilon_d - \Sigma^a_{dd}(\varepsilon_{k\beta}) \right\} \qquad \text{(S 14)}$$

Using again (S 3) one gets

$$\sum_{n\alpha} \varepsilon_{n\alpha} \hat{c}^\dagger_{n\alpha} \hat{c}_{n\alpha} = \sum_{n\alpha} \varepsilon_{n\alpha} \hat{\psi}^\dagger_{n\alpha} \hat{\psi}_{n\alpha} - \sum_{m\gamma} \sum_{k\beta} \left\{ V^*_{m\gamma} V_{k\beta} G^r_{dd}(\varepsilon_{m\gamma}) G^a_{dd}(\varepsilon_{k\beta}) \right\} \hat{\psi}^\dagger_{k\beta} \hat{\psi}_{m\gamma} \left\{ PP \frac{1}{\varepsilon_{k\beta} - \varepsilon_{m\gamma}} + \pi i \delta(\varepsilon_{k\beta} - \varepsilon_{m\gamma}) \right\}$$

$$\times \left\{ (\varepsilon^2_{k\beta} - \varepsilon^2_{m\gamma}) + \varepsilon_{m\gamma} \Sigma^r_{dd}(\varepsilon_{m\gamma}) - \varepsilon_{m\gamma} \Sigma^r_{dd}(\varepsilon_{m\gamma}) + \varepsilon_{k\beta} \Sigma^a_{dd}(\varepsilon_{k\beta}) - \varepsilon_{k\beta} \Sigma^a_{dd}(\varepsilon_{k\beta}) - \varepsilon_d(\varepsilon_{k\beta} - \varepsilon_{m\gamma}) \right\}$$

$$= \sum_{n\alpha} \varepsilon_{n\alpha} \hat{\psi}^\dagger_{n\alpha} \hat{\psi}_{n\alpha} - \sum_{m\gamma} \sum_{k\beta} \left\{ V^*_{m\gamma} V_{k\beta} G^r_{dd}(\varepsilon_{m\gamma}) G^a_{dd}(\varepsilon_{k\beta}) \right\} \hat{\psi}^\dagger_{k\beta} \hat{\psi}_{m\gamma} \left\{ \varepsilon_{k\beta} + \varepsilon_{m\gamma} - \varepsilon_d \right\}$$

$$= \sum_n \varepsilon_n \hat{\psi}^\dagger_n \hat{\psi}_n - \sum_m \sum_k \left\{ V^*_m V_k G^r_{dd}(\varepsilon_m) G^a_{dd}(\varepsilon_k) \right\} \hat{\psi}^\dagger_k \hat{\psi}_m \left\{ \varepsilon_k + \varepsilon_m - \varepsilon_d \right\}$$

(S 15)

In the last line of (S 15) we just shortened the notation by omitting the leads indexes.

Finally, consider the term:

$$\bar{V} \equiv \hat{V} - \varepsilon_d \hat{d}^\dagger \hat{d} = \sum_{k\alpha} \left( V_{k\alpha} \hat{c}^\dagger_{k\alpha} \hat{d} + V^*_{k\alpha} \hat{d}^\dagger \hat{c}_{k\alpha} \right) \qquad \text{(S 16)}$$

Using Eqs. (21)-(22) it becomes

$$\bar{V} = \sum_m G_{dd}^r(\varepsilon_m) V_m^* \sum_n V_n \sum_k \hat{\psi}_k^\dagger \hat{\psi}_m \left\{ \delta_{kn} + V_n^* \frac{V_k G_{dd}^a(\varepsilon_k)}{\varepsilon_k - \varepsilon_n - i\eta} \right\}$$

$$+ \sum_m G_{dd}^a(\varepsilon_k) V_k \sum_n V_n^* \sum_m \hat{\psi}_k^\dagger \hat{\psi}_m \left\{ \delta_{nm} + V_n \frac{V_m^* G_{dd}^r(\varepsilon_m)}{\varepsilon_m - \varepsilon_n + i\eta} \right\}$$

$$= \sum_k \sum_m G_{dd}^r(\varepsilon_m) V_k V_m^* \hat{\psi}_k^\dagger \hat{\psi}_m \left\{ 1 + G_{dd}^a(\varepsilon_k) \sum_n V_n V_n^* \frac{1}{\varepsilon_k - \varepsilon_n - i\eta} \right\}$$

$$+ \sum_k \sum_m G_{dd}^a(\varepsilon_k) V_m^* V_k \hat{\psi}_k^\dagger \hat{\psi}_m \left\{ 1 + G_{dd}^r(\varepsilon_m) \sum_n V_n V_n^* \frac{1}{\varepsilon_m - \varepsilon_n + i\eta} \right\} \quad \text{(S 17)}$$

$$= \sum_k \sum_m G_{dd}^a(\varepsilon_k) G_{dd}^r(\varepsilon_m) V_k V_m^* \hat{\psi}_k^\dagger \hat{\psi}_m \left\{ \left( G_{dd}^a(\varepsilon_k) \right)^{-1} + \Sigma_{dd}^a(\varepsilon_k) \right\}$$

$$+ \sum_k \sum_m G_{dd}^a(\varepsilon_k) G_{dd}^r(\varepsilon_m) V_k V_m^* \hat{\psi}_k^\dagger \hat{\psi}_m \left\{ \left( G_{dd}^r(\varepsilon_m) \right)^{-1} + \Sigma_{dd}^r(\varepsilon_m) \right\}$$

$$= \sum_k \sum_m G_{dd}^a(\varepsilon_k) G_{dd}^r(\varepsilon_m) V_k V_m^* \hat{\psi}_k^\dagger \hat{\psi}_m \left\{ \varepsilon_k + \varepsilon_m - 2\varepsilon_d \right\}$$

Using (S 17) together with Eqs. (S 9) we get

$$\sum_{n\alpha} \varepsilon_{n\alpha} \hat{c}_{n\alpha}^\dagger \hat{c}_{n\alpha} + \bar{V} + \varepsilon_d \hat{d}^\dagger \hat{d}$$

$$= \sum_n \varepsilon_n \hat{\psi}_n^\dagger \hat{\psi}_n - \sum_m \sum_k \left\{ V_m^* V_k G_{dd}^r(\varepsilon_m) G_{dd}^a(\varepsilon_k) \right\} \hat{\psi}_k^\dagger \hat{\psi}_m \left\{ \varepsilon_k + \varepsilon_m - \varepsilon_d \right\}$$

$$+ \sum_k \sum_m G_{dd}^a(\varepsilon_k) G_{dd}^r(\varepsilon_m) V_k V_m^* \hat{\psi}_k^\dagger \hat{\psi}_m \left\{ \varepsilon_k + \varepsilon_m - 2\varepsilon_d \right\} + \sum_m \sum_k \left\{ V_m^* V_k G_{dd}^r(\varepsilon_m) G_{dd}^a(\varepsilon_k) \right\} \hat{\psi}_k^\dagger \hat{\psi}_m \varepsilon_d$$

$$= \sum_n \varepsilon_n \hat{\psi}_n^\dagger \hat{\psi}_n$$

(S 18)

Taken together, Eqs. (S 10) and (S 18) prove Eq.(27).

**Section B. Calculation of the particle current**

The current into $\alpha$ lead can be expressed as follows:

$$J_\alpha = i\text{Tr}\left\{\hat{\rho}[\hat{V}_\alpha, \hat{N}_\alpha]\right\} = i\sum_n \left\{\text{Tr}\{\hat{\rho}V_{n\alpha}\hat{c}^\dagger_{n\alpha}\hat{d} - \hat{\rho}V^*_{n\alpha}\hat{d}^\dagger\hat{c}_{n\alpha}\}\right\}$$

$$= i\sum_n \text{Tr}\left[\hat{\rho}V_{n\alpha}\sum_{k\beta}\left\{\delta_{k\beta n\alpha} + V^*_{n\alpha}\frac{V_{k\beta}G^a_{dd}(\varepsilon_{k\beta})}{\varepsilon_{k\beta} - \varepsilon_{n\alpha} - i\eta}\right\}\hat{\psi}^\dagger_{k\beta}\sum_{m\gamma}G^r_{dd}(\varepsilon_{m\gamma})V^*_{m\gamma}\hat{\psi}_{m\gamma}\right]$$

$$-i\sum_n \text{Tr}\left[\hat{\rho}V^*_{n\alpha}\sum_{m\gamma}G^a_{dd}(\varepsilon_{m\gamma})V_{m\gamma}\hat{\psi}^\dagger_{m\gamma}\sum_{k\beta}\left\{\delta_{k\beta n\alpha} + V_{n\alpha}\frac{V^*_{k\beta}G^r_{dd}(\varepsilon_{k\beta})}{\varepsilon_{k\beta} - \varepsilon_{n\alpha} + i\eta}\right\}\hat{\psi}_{k\beta}\right]$$

$$= J^{(1)}_\alpha + J^{(2)}_\alpha$$

(SA1)

Where

$$J^{(1)}_\alpha = \sum_n V_{n\alpha}\sum_{k\beta}\sum_{m\gamma}G^r_{dd}(\varepsilon_{m\gamma})V^*_{m\gamma}\delta_{k\beta n\alpha}\text{Tr}\{\hat{\rho}\hat{\psi}^\dagger_{k\beta}\hat{\psi}_{m\gamma}\} - \sum_n V^*_{n\alpha}\sum_{k\beta}\sum_{m\gamma}G^a_{dd}(\varepsilon_{m\gamma})V_{m\gamma}\delta_{k\beta n\alpha}\text{Tr}\{\hat{\rho}\hat{\psi}^\dagger_{m\gamma}\hat{\psi}_{k\beta}\}$$

(SA2)

and

$$J^{(2)}_\alpha = \sum_n \text{Tr}\left\{\hat{\rho}V_{n\alpha}\sum_{k\beta}V^*_{n\alpha}\frac{V_{k\beta}G^a_{dd}(\varepsilon_{k\beta})}{\varepsilon_{k\beta} - \varepsilon_{n\alpha} - i\eta}\hat{\psi}^\dagger_{k\beta}\sum_{m\gamma}G^r_{dd}(\varepsilon_{m\gamma})V^*_{m\gamma}\hat{\psi}_{m\gamma}\right\}$$

$$-\sum_n \text{Tr}\left\{\hat{\rho}V^*_{n\alpha}\sum_{m\gamma}G^a_{dd}(\varepsilon_{m\gamma})V_{m\gamma}\hat{\psi}^\dagger_{m\gamma}\sum_{k\beta}V_{n\alpha}\frac{V^*_{k\beta}G^r_{dd}(\varepsilon_{k\beta})}{\varepsilon_{k\beta} - \varepsilon_{n\alpha} + i\eta}\hat{\psi}_{k\beta}\right\}$$

(SA3)

For $J^{(1)}_\alpha$ we have

$$J^{(1)}_\alpha = \sum_n V_{n\alpha}\sum_{k\beta}\sum_{m\gamma}G^r_{dd}(\varepsilon_{m\gamma})V^*_{k\gamma}\delta_{k\beta n\alpha}\delta_{k\beta m\gamma}f_\gamma(\varepsilon_{m\gamma}) - \sum_n V^*_{n\alpha}\sum_{k\beta}\sum_{m\gamma}G^a_{dd}(\varepsilon_{m\gamma})V_{k\gamma}\delta_{k\beta n\alpha}\delta_{k\beta m\gamma}f_\gamma(\varepsilon_{m\gamma})$$

$$= \sum_n |V_{n\alpha}|^2 f_\alpha(\varepsilon_{n\alpha})\{G^r_{dd}(\varepsilon_{n\alpha}) - G^a_{dd}(\varepsilon_{n\alpha})\} = \frac{1}{2\pi}\int_{-\infty}^{\infty}\sum_n 2\pi|V_{n\alpha}|^2\delta(\varepsilon - \varepsilon_{n\alpha})f_\alpha(\varepsilon)\{G^r_{dd}(\varepsilon) - G^a_{dd}(\varepsilon)\}d\varepsilon =$$

$$= \frac{1}{2\pi}\int_{-\infty}^{\infty}\Gamma_\alpha(\varepsilon)f_\alpha(\varepsilon)\{G^r_{dd}(\varepsilon) - G^a_{dd}(\varepsilon)\}d\varepsilon$$

(SA4)

while $J^{(2)}_\alpha$ takes the form:

$$J^{(2)}_\alpha = \sum_n V_{n\alpha}\sum_{k\beta}V^*_{n\alpha}\frac{V_{k\beta}G^a_{dd}(\varepsilon_{k\beta})}{\varepsilon_{k\beta} - \varepsilon_{n\alpha} - i\eta}\sum_{m\gamma}G^r_{dd}(\varepsilon_{m\gamma})V^*_{m\gamma}\text{Tr}\{\hat{\rho}\hat{\psi}^\dagger_{k\beta}\hat{\psi}_{m\gamma}\}$$

$$-\sum_n V^*_{n\alpha}\sum_{m\gamma}G^a_{dd}(\varepsilon_{m\gamma})V_{m\gamma}\sum_{k\beta}V_{n\alpha}\frac{V^*_{k\beta}G^r_{dd}(\varepsilon_{k\beta})}{\varepsilon_{k\beta} - \varepsilon_{n\alpha} + i\eta}\text{Tr}\{\hat{\rho}\hat{\psi}^\dagger_{m\gamma}\hat{\psi}_{k\beta}\}$$

$$= \sum_n |V_{n\alpha}|^2 \sum_{k\beta} \sum_{m\gamma} G^r_{dd}(\varepsilon_{m\gamma})V^*_{m\gamma}\delta_{k\beta m\gamma}f_\gamma(\varepsilon_{m\gamma})\frac{V_{k\beta}G^a_{dd}(\varepsilon_{k\beta})}{\varepsilon_{k\beta}-\varepsilon_{n\alpha}-i\eta}$$

$$-\sum_n |V_{n\alpha}|^2 \sum_{k\beta}\sum_{m\gamma} G^a_{dd}(\varepsilon_{m\gamma})V_{m\gamma}V_{n\alpha}\frac{V^*_{k\beta}G^r_{dd}(\varepsilon_{k\beta})}{\varepsilon_{k\beta}-\varepsilon_{n\alpha}+i\eta}\delta_{k\beta m\gamma}f_\gamma(\varepsilon_{m\gamma})$$

$$=\sum_n |V_{n\alpha}|^2 \sum_{k\beta} G^r_{dd}(\varepsilon_{k\beta})V^*_{k\beta}f_\beta(\varepsilon_{k\beta})\frac{V_{k\beta}G^a_{dd}(\varepsilon_{k\beta})}{\varepsilon_{k\beta}-\varepsilon_{n\alpha}-i\eta}-\sum_n |V_{n\alpha}|^2 \sum_{k\beta} G^a_{dd}(\varepsilon_{k\beta})V_{k\beta}f_\beta(\varepsilon_{k\beta})\frac{V^*_{k\beta}G^r_{dd}(\varepsilon_{k\beta})}{\varepsilon_{k\beta}-\varepsilon_{n\alpha}+i\eta}$$

$$=\sum_n |V_{n\alpha}|^2 \sum_{k\beta} G^r_{dd}(\varepsilon_{k\beta})G^a_{dd}(\varepsilon_{k\beta})f_\beta(\varepsilon_{k\beta})|V_{k\beta}|^2 \left\{\frac{1}{\varepsilon_{k\beta}-\varepsilon_{n\alpha}-i\eta}-\frac{1}{\varepsilon_{k\beta}-\varepsilon_{n\alpha}+i\eta}\right\}$$

$$=\sum_{k\beta} iG^r_{dd}(\varepsilon_{k\beta})G^a_{dd}(\varepsilon_{k\beta})f_\beta(\varepsilon_{k\beta})|V_{k\beta}|^2 \sum_n 2\pi |V_{n\alpha}|^2 \delta(\varepsilon_{k\beta}-\varepsilon_{n\alpha})$$

$$=\sum_{k\beta} iG^r_{dd}(\varepsilon_{k\beta})G^a_{dd}(\varepsilon_{k\beta})f_\beta(\varepsilon_{k\beta})|V_{k\beta}|^2 \Gamma_\alpha(\varepsilon_{k\beta})$$

$$=\frac{1}{2\pi}\int_{-\infty}^{\infty} G^r_{dd}(\varepsilon)G^a_{dd}(\varepsilon)i\sum_{k\beta}2\pi f_\beta(\varepsilon)|V_{k\beta}|^2 \delta(\varepsilon_{k\beta}-\varepsilon)\Gamma_\alpha(\varepsilon)d\varepsilon$$

$$=\frac{1}{2\pi}\int_{-\infty}^{\infty} G^r_{dd}(\varepsilon)G^a_{dd}(\varepsilon)\Sigma^<_{dd}(\varepsilon)\Gamma_\alpha(\varepsilon)d\varepsilon=\frac{1}{2\pi}\int_{-\infty}^{\infty} G^<_{dd}(\varepsilon)\Gamma_\alpha(\varepsilon)d\varepsilon$$

(SA5)

Combining (SA4) and (SA5) we have:

$$J_\alpha = \frac{i}{2\pi}\int_{-\infty}^{\infty}\left\{G^<_{dd}(\varepsilon)\Gamma_\alpha(\varepsilon)+\Gamma_\alpha(\varepsilon)f_\alpha(\varepsilon)\left(G^r_{dd}(\varepsilon)-G^a_{dd}(\varepsilon)\right)\right\}d\varepsilon \tag{SA6}$$

which holds for any number of thermal baths. In the case of a two terminals junction (L,R), using $G^<_{dd}(\varepsilon)=iG^r_{dd}(\varepsilon)\left(f_L(\varepsilon)\Gamma_L(\varepsilon)+f_R(\varepsilon)\Gamma_R(\varepsilon)\right)G^a_{dd}(\varepsilon)$ and $G^r_{dd}(\varepsilon)-G^a_{dd}(\varepsilon)=-i(\Gamma_L+\Gamma_R)G^r_{dd}(\varepsilon)G^a_{dd}(\varepsilon)$, one gets

$$J_L = \frac{i}{2\pi}\int_{-\infty}^{\infty}\left\{G^<_{dd}(\varepsilon)\Gamma_L(\varepsilon)+\Gamma_L(\varepsilon)f_L(\varepsilon)\left(G^r_{dd}(\varepsilon)-G^a_{dd}(\varepsilon)\right)\right\}d\varepsilon$$

$$=\frac{1}{2\pi}\int_{-\infty}^{\infty}\Gamma_R(\varepsilon)\Gamma_L(\varepsilon)G^r_{dd}(\varepsilon)G^a_{dd}(\varepsilon)\left(f_L(\varepsilon)-f_R(\varepsilon)\right)d\varepsilon$$

(SA7)

where Eq. (23) was used to get the final symmetric form. The result is the Landauer expression for the current

$$J_L = \frac{1}{2\pi}\int T(\varepsilon)\left(f_L(\varepsilon)-f_R(\varepsilon)\right)d\varepsilon \tag{SA8}$$

With the transmission coefficient $T(\varepsilon)$ given by

$$T(\varepsilon) = \Gamma_R(\varepsilon)\Gamma_L(\varepsilon)G_{dd}^r(\varepsilon)G_{dd}^a(\varepsilon) \tag{SA9}$$

**Section C. Equivalence of Landauer-Buttiker formalism to the present method**

The original Landauer-Buttiker scattering theory approach to junction transport has been formulated in terms of the *S*-matrix. Here we demonstrate the equivalence of the two formalisms. We start by introducing the incoming and outgoing scattering solutions:

the incoming one

$$\hat{\psi}_{k\beta,+}^\dagger = V_{k\beta}^* G_{dd}^r(\varepsilon_{k\beta})\hat{d}^\dagger + \sum_{n\alpha}\left\{\delta_{k\beta n\alpha} + V_{n\alpha}\frac{V_{k\beta}^* G_{dd}^r(\varepsilon_{k\beta})}{\varepsilon_{k\beta} - \varepsilon_{n\alpha} + i\eta}\right\}\hat{c}_{n\alpha}^\dagger \tag{SB1}$$

$$\hat{\psi}_{k\beta,+} = V_{k\beta} G_{dd}^a(\varepsilon_{k\beta})\hat{d} + \sum_{n\alpha}\left\{\delta_{k\beta n\alpha} + V_{n\alpha}^*\frac{V_{k\beta} G_{dd}^a(\varepsilon_{k\beta})}{\varepsilon_{k\beta} - \varepsilon_{n\alpha} - i\eta}\right\}\hat{c}_{n\alpha} \tag{SB2}$$

and the outgoing one

$$\hat{\psi}_{k\beta,-}^\dagger = V_{k\beta}^* G_{dd}^a(\varepsilon_{k\beta})\hat{d}^\dagger + \sum_{n\alpha}\left\{\delta_{k\beta n\alpha} + V_{n\alpha}\frac{V_{k\beta}^* G_{dd}^a(\varepsilon_{k\beta})}{\varepsilon_{k\beta} - \varepsilon_{n\alpha} - i\eta}\right\}\hat{c}_{n\alpha}^\dagger \tag{SB3}$$

$$\hat{\psi}_{k\beta,-} = V_{k\beta} G_{dd}^r(\varepsilon_{k\beta})\hat{d} + \sum_{n\alpha}\left\{\delta_{k\beta n\alpha} + V_{n\alpha}^*\frac{V_{k\beta} G_{dd}^r(\varepsilon_{k\beta})}{\varepsilon_{k\beta} - \varepsilon_{n\alpha} + i\eta}\right\}\hat{c}_{n\alpha} \tag{SB4}$$

Note that the operators that appear in Eqs. (20) correspond to the incoming states, where for simplification of presentation, the incoming "+" labels in $\hat{\psi}_{k\beta,+}^\dagger$ and $\hat{\psi}_{k\beta,+}$ were omitted. The outgoing solutions correspond to time reversed solutions, where the baths are uncoupled from each other and in their own equilibrium in the future, and they become coupled as time propagates backwards

$$\hat{\psi}_{k\beta,-}^\dagger = \hat{\Omega}_- \hat{c}_{k\beta}^\dagger \hat{\Omega}_-^\dagger \tag{SB5}$$

with the corresponding Moller operator

$$\hat{\Omega}_- = \lim_{t \to -\infty} \exp(-i\hat{H}t)\exp(i\hat{H}_0 t) \tag{SB6}$$

Thus, Eqs. (SB3)-(SB4) are obtained from (20) by replacing the retarded Green's function with the advanced one (and vice versa) and change the sign of $\eta$ in (20).

Next, we introduce the energy renormalized operators:

$$\hat{\chi}^{\dagger}_{\varepsilon_k\beta,\pm} = \hat{\psi}^{\dagger}_{k\beta,\pm}\sqrt{2\pi D_{\varepsilon_k\beta}} \tag{SB7}$$

where $D_{\varepsilon_k\beta} = D_\beta(\varepsilon_k)$ is the density of energy states in lead $\beta$. It is easy to verify that

$$\left[\hat{\chi}^{\dagger}_{\varepsilon\beta,\pm},\hat{\chi}_{\varepsilon'\alpha,\pm}\right]_{+} = 2\pi\delta_{\alpha\beta}\delta(\varepsilon-\varepsilon') \tag{SB8}$$

where for definiteness, here and below we specify to fermions, and

$$\mathrm{Tr}\left\{\hat{\rho}_{ss}\hat{\chi}^{\dagger}_{\varepsilon\beta,+}\hat{\chi}_{\varepsilon'\alpha,+}\right\} = 2\pi\delta_{\alpha\beta}\delta(\varepsilon-\varepsilon')f_\alpha(\varepsilon) \tag{SB9}$$

The scattering matrix can be defined as follows[54]:

$$S_{k\beta n\alpha} = \left[\hat{\psi}^{\dagger}_{k\beta,+},\hat{\psi}_{n\alpha,-}\right]_{+} \tag{SB10}$$

and can be evaluated using (SB1)-(SB4). An easier way is to employ Lippmann-Schwinger equations[55]. For the incoming eigenfunction associated with lead $\beta$ we have

$$\left|\psi_{k\beta,+}\right\rangle = \left|c_{k\beta}\right\rangle + \hat{G}^r(\varepsilon_k)\hat{V}\left|c_{k\beta}\right\rangle \tag{SB11}$$

and the corresponding outgoing wavefunction is

$$\left|\psi_{k\beta,-}\right\rangle = \left|c_{k\beta}\right\rangle + \hat{G}^a(\varepsilon_k)\hat{V}\left|c_{k\beta}\right\rangle \tag{SB12}$$

Subtracting (SB12) from (SB11) we have:

$$\left|\psi_{k\beta,+}\right\rangle = \left|\psi_{k\beta,-}\right\rangle + \left\{\hat{G}^r(\varepsilon_k)-\hat{G}^a(\varepsilon_k)\right\}\hat{V}\left|c_{k\beta}\right\rangle \tag{SB13}$$

Thus,

$$\begin{aligned}S_{k\beta n\alpha} &= \frac{\left\langle\psi_{n\alpha,-}|\psi_{k\beta,+}\right\rangle}{\left\langle\psi_{n\alpha,-}|\psi_{n\alpha,-}\right\rangle} = \frac{\left\langle\psi_{n\alpha,-}|\psi_{k\beta,-}\right\rangle + \left\langle\psi_{n\alpha,-}\left|\left\{\hat{G}^r(\varepsilon_k)-\hat{G}^a(\varepsilon_k)\right\}\hat{V}\right|c_{k\beta}\right\rangle}{\left\langle\psi_{n\alpha,-}|\psi_{n\alpha,-}\right\rangle} \\ &= \delta_{n\alpha k\beta} + \left\langle\psi_{n\alpha,-}\left|\left\{\hat{G}^r(\varepsilon_k)-\hat{G}^a(\varepsilon_k)\right\}\right|\psi_{n\alpha,-}\right\rangle\left\langle\psi_{n\alpha,-}|d\right\rangle V_{k\beta} \\ &= \delta_{n\alpha k\beta} - 2\pi i\delta(\varepsilon_k-\varepsilon_n)G^r_{dd}(\varepsilon_k)V_{k\beta}V^*_{n\alpha}\end{aligned} \tag{SB14}$$

In the energy representation

$$S_{\beta\alpha}(\varepsilon) = \delta_{\alpha\beta} - 2\pi i D_\varepsilon G^r_{dd}(\varepsilon)V_{\varepsilon\beta}V^*_{\varepsilon\alpha} = \delta_{\alpha\beta} - iG^r_{dd}(\varepsilon)\sqrt{\Gamma_\alpha(\varepsilon)\Gamma_\beta(\varepsilon)}\exp\left\{i\Phi_{\alpha\beta}(\varepsilon)\right\} \tag{SB15}$$

where $\Phi_{\alpha\beta}(\varepsilon) = \arg(V_{\varepsilon\beta}V^*_{\varepsilon\alpha})$

This is the Mahaux-Weidenmueller formula used by von Oppen and co-workers[28,34,46].

The particle current out of lead $\alpha$ is

$$J_\alpha(t) = \left(\frac{1}{2\pi}\right)^2 \int_{-\infty}^{\infty}\int_{-\infty}^{\infty} \text{Tr}\left\{\hat{\rho}_{ss}\left(\hat{\chi}^\dagger_{\varepsilon\alpha,-}\hat{\chi}_{\varepsilon'\alpha,-} - \hat{\chi}^\dagger_{\varepsilon\alpha,+}\hat{\chi}_{\varepsilon'\alpha,+}\right)\right\}\exp(i(\varepsilon'-\varepsilon)t)d\varepsilon d\varepsilon' \quad \text{(SB16)}$$

At steady state it coincides with the Landauer-Buttiker expression for the current. Indeed, using the notation of Ref. [17] (slightly renormalized),

$$J^{LB}_\alpha(t) = \left(\frac{1}{2\pi}\right)^2 \int_{-\infty}^{\infty}\int_{-\infty}^{\infty} \text{Tr}\left\{\hat{\rho}_0\left(\hat{b}^\dagger_{\varepsilon\alpha}\hat{b}_{\varepsilon'\alpha} - \hat{a}^\dagger_{\varepsilon\alpha}\hat{a}_{\varepsilon'\alpha}\right)\right\}\exp(i(\varepsilon'-\varepsilon)t)d\varepsilon d\varepsilon' \quad \text{(SB17)}$$

where

$$\left[\hat{a}^\dagger_{\varepsilon\alpha}\hat{a}_{\varepsilon'\beta}\right]_+ = 2\pi\delta_{\alpha\beta}\delta(\varepsilon-\varepsilon') \quad \text{(SB18)}$$

$$\text{Tr}\{\hat{\rho}_0\hat{a}^\dagger_{\varepsilon\alpha}\hat{a}_{\varepsilon'\beta}\} = 2\pi\delta_{\alpha\beta}\delta(\varepsilon-\varepsilon')f_\alpha(\varepsilon) \quad \text{(SB19)}$$

$$\hat{b}^\dagger_{\varepsilon\beta} = \sum_\beta S_{\alpha\beta}(\varepsilon)\hat{a}^\dagger_{\varepsilon\alpha} \quad \text{(SB20)}$$

Substitution (SB20) into (SB17) gives:

$$J^{LB}_\alpha(t) = \left(\frac{1}{2\pi}\right)^2 \int_{-\infty}^{\infty}\int_{-\infty}^{\infty} \sum_{\beta\beta'}\left(S_{\beta\alpha}(\varepsilon)S^\dagger_{\beta'\alpha}(\varepsilon') - \delta_{\alpha\beta}\delta_{\alpha\beta'}\right)\text{Tr}\left\{\hat{\rho}_0\hat{a}^\dagger_{\varepsilon\beta}\hat{a}_{\varepsilon'\beta'}\right\}\exp(i(\varepsilon'-\varepsilon)t)d\varepsilon d\varepsilon'$$

$$= \left(\frac{1}{2\pi}\right)\int_{-\infty}^{\infty}\sum_\beta \left(S_{\beta\alpha}(\varepsilon)S^\dagger_{\beta\alpha}(\varepsilon) - \delta_{\alpha\beta}\right)f_\beta(\varepsilon)d\varepsilon \quad \text{(SB21)}$$

On the other hand, for (SB16) with (SB10) one gets:

$$J_\alpha(t) = \left(\frac{1}{2\pi}\right)^2 \int_{-\infty}^{\infty}\int_{-\infty}^{\infty}\sum_{\beta\beta'}\left(S_{\beta\alpha}(\varepsilon)S^\dagger_{\beta'\alpha}(\varepsilon') - \delta_{\alpha\beta}\delta_{\alpha\beta'}\right)\text{Tr}\left\{\hat{\rho}_{ss}\hat{\chi}^\dagger_{\varepsilon\beta,+}\hat{\chi}_{\varepsilon'\beta',+}\right\}\exp(i(\varepsilon'-\varepsilon)t)d\varepsilon d\varepsilon'$$

$$= \left(\frac{1}{2\pi}\right)\int_{-\infty}^{\infty}\sum_\beta\left(S_{\beta\alpha}(\varepsilon)S^\dagger_{\beta\alpha}(\varepsilon) - \delta_{\alpha\beta}\right)f_\beta(\varepsilon)d\varepsilon \quad \text{(SB22)}$$

which coincides with (SB21).

For completeness, we also introduce, following Ref. [28], the outgoing and the incoming distribution matrixes:

$$\phi_{\alpha\beta,out}(\varepsilon,t) = \left(\frac{1}{2\pi}\right)\int_{-\infty}^{\infty}\text{Tr}\left\{\hat{\rho}_{ss}\hat{\chi}^{\dagger}_{(\varepsilon-\omega/2)\alpha,-}\hat{\chi}_{(\varepsilon+\omega/2)\beta,-}\right\}\exp(i\omega t)d\omega \quad \text{(SB23)}$$

$$\phi_{\alpha\beta,inc}(\varepsilon,t) = \left(\frac{1}{2\pi}\right)\int_{-\infty}^{\infty}\text{Tr}\left\{\hat{\rho}_{ss}\hat{\chi}^{\dagger}_{(\varepsilon-\omega/2)\alpha,+}\hat{\chi}_{(\varepsilon+\omega/2)\beta,+}\right\}\exp(i\omega t)d\omega \quad \text{(SB24)}$$

In steady state both (SB23) and (SB24) are time-independent.

**Section D. Evaluation of $D_\beta(\varepsilon)$ and $\partial_{\varepsilon_d}D_\beta(\varepsilon)$ for a given bath β.**

The density of states associated with lead β is given by $D_\beta(\varepsilon) = \frac{1}{\pi}\text{Tr}_\beta\left\{\text{Im}\{\hat{G}^r(\varepsilon)\}\right\}$, where the partial trace is taken over the scattering states of β lead. Consequently

$$\partial_{\varepsilon_d}D_\beta(\varepsilon) = \frac{1}{\pi}\partial_{\varepsilon_d}\text{Tr}_\beta\left\{\text{Im}(\hat{G}^r)\right\} = \frac{1}{\pi}\sum_k\left\{\langle\psi_{k\beta}|\partial_{\varepsilon_d}\text{Im}(\hat{G}^r)|\psi_{k\beta}\rangle + \langle\partial_{\varepsilon_d}\psi_{k\beta}|\text{Im}(\hat{G}^r)|\psi_{k\beta}\rangle\right.$$

$$\left.+\langle\psi_{k\beta}|\text{Im}(\hat{G}^r)|\partial_{\varepsilon_d}\psi_{k\beta}\rangle\right\} = \frac{1}{\pi}\text{Im}\sum_k\left\{\langle\psi_{k\beta}|\partial_{\varepsilon_d}\hat{G}^r|\psi_{k\beta}\rangle + \right. \quad \text{(SC1)}$$

$$\left.\langle\partial_{\varepsilon_d}\psi_{k\beta}|\psi_{k\beta}\rangle\langle\psi_{k\beta}|\hat{G}^r|\psi_{k\beta}\rangle + \langle\psi_{k\beta}|\hat{G}^r|\psi_{k\beta}\rangle\langle\psi_{k\beta}|\partial_{\varepsilon_d}\psi_{k\beta}\rangle\right\} = \frac{1}{\pi}\text{Im}\sum_k\langle\psi_{k\beta}|\partial_{\varepsilon_d}\hat{G}^r|\psi_{k\beta}\rangle$$

where the identities $\partial_{\varepsilon_d}\langle\psi_{k\beta}|\psi_{k\beta}\rangle = \langle\partial_{\varepsilon_d}\psi_{k\beta}|\psi_{k\beta}\rangle + \langle\psi_{k\beta}|\partial_{\varepsilon_d}\psi_{k\beta}\rangle = 0$ and

$\langle\psi_{k\beta}|\hat{G}^r|\psi_{n\alpha}\rangle = \langle\psi_{k\beta}|\hat{G}^r|\psi_{k\beta}\rangle\delta_{k\beta n\alpha}$ have been used. Using Eq.

**Error! Reference source not found.** and the identity, for an arbitrary operator, $\partial\{\hat{B}^{-1}\} = \hat{B}^{-1}(\partial\hat{B})\hat{B}^{-1}$ we have:

$$\sum_k\langle\psi_{k\beta}|\partial_{\varepsilon_d}\hat{G}^r|\psi_{k\beta}\rangle = \sum_k\langle\psi_{k\beta}|\hat{G}^r|d\rangle\langle d|\hat{G}^r|\psi_{k\beta}\rangle = \sum_k\langle\psi_{k\beta}|\hat{G}^r|\psi_{k\beta}\rangle\langle\psi_{k\beta}|d\rangle\langle d|\psi_{k\beta}\rangle\langle\psi_{k\beta}|\hat{G}^r|\psi_{k\beta}\rangle$$

$$=\sum_k\langle\psi_{k\beta}|\hat{G}^r|\psi_{k\beta}\rangle|V_{k\beta}|^2 G^r_{dd}(\varepsilon_k)G^a_{dd}(\varepsilon_k)\langle\psi_{k\beta}|\hat{G}^r|\psi_{k\beta}\rangle = \sum_k\langle\psi_{k\beta}|\hat{G}^r\hat{G}^r|\psi_{k\beta}\rangle|V_{k\beta}|^2 G^r_{dd}(\varepsilon_k)G^a_{dd}(\varepsilon_k)$$

$$=-\sum_k\partial_\varepsilon\langle\psi_{k\beta}|\hat{G}^r|\psi_{k\beta}\rangle|V_{k\beta}|^2 G^r_{dd}(\varepsilon_k)G^a_{dd}(\varepsilon_k)$$

$$=-\partial_\varepsilon\frac{1}{2\pi}\int_{-\infty}^{\infty}\frac{1}{\varepsilon-\varepsilon'+i\eta}\Gamma_\beta G^r_{dd}(\varepsilon')G^a_{dd}(\varepsilon')d\varepsilon' = -\frac{\Gamma_\beta}{2\pi\Gamma}\partial_\varepsilon\int_{-\infty}^{\infty}\frac{1}{\varepsilon-\varepsilon'+i\eta}A_{dd}(\varepsilon')d\varepsilon' = \frac{\Gamma_\beta}{2\pi\Gamma}\partial_{\varepsilon_d}\int_{-\infty}^{\infty}\frac{1}{\varepsilon-\varepsilon'+i\eta}A_{dd}(\varepsilon')d\varepsilon'$$

(SC2)

In the last line above we switched from summation to integration and evaluated the integral by parts.

On the other hand,

$$\frac{1}{2\pi}\int_{-\infty}^{\infty}\frac{1}{\varepsilon-\varepsilon'+i\eta}A_{dd}(\varepsilon')d\varepsilon' = \sum_{k\beta}\langle\psi_{k\beta}|\hat{G}^r(\varepsilon)|\psi_{k\beta}\rangle|V_{k\beta}|^2 G^r_{dd}(\varepsilon_k)G^a_{dd}(\varepsilon_k)$$
$$= \sum_{k\beta}\langle\psi_{k\beta}|\hat{G}^r|\psi_{k\beta}\rangle\langle\psi_{k\beta}|d\rangle\langle d|\psi_{k\beta}\rangle = \langle d|\hat{G}^r|d\rangle = G^r_{dd}(\varepsilon) \quad \text{(SC3)}$$

Substituting (SC3) into (SC1) leads to

$$\partial_{\varepsilon_d}D_\beta(\varepsilon) = \frac{\Gamma_\beta}{\pi\Gamma}\operatorname{Im}\{\partial_{\varepsilon_d}G^r_{dd}(\varepsilon)\} \quad \text{(SC4)}$$

Thus, the $\varepsilon_d$-dependent part of the total density is

$$D_\beta(\varepsilon)_{\varepsilon_d} = \frac{\Gamma_\beta}{\pi\Gamma}\operatorname{Im}\{G^r_{dd}(\varepsilon)\} = \frac{\Gamma_\beta}{2\pi\Gamma}A_{dd}(\varepsilon) \quad \text{(SC5)}$$

For one lead it yields the well-known result – the spectral density of the dot[1].

**Section E. Evaluation of the non-adiabatic correction to an expectation value.**

Her we evaluate the lowest order non-adiabatic correction to the expectation value of a single-particle operator of the general form (49). This correction is given by (here we set $\hbar=1$):

$$\langle\hat{A}\rangle^{(1)} = \operatorname{Tr}\{\hat{A}\hat{\rho}^{(1)}_{ss}\} = \sum_{k\beta n\alpha}\gamma_{k\beta n\alpha}\operatorname{Tr}\{\hat{\psi}^\dagger_{k\beta}\hat{\psi}_{n\alpha}\hat{\rho}^{(1)}_{ss}\}$$
$$= -\lim_{\eta\to+0}\sum_\nu \dot{R}^\nu \sum_{k\beta n\alpha}\gamma_{k\beta n\alpha}\int_{-\infty}^0 \exp(\eta\tau)\operatorname{Tr}\{\hat{\psi}^\dagger_{k\beta}\hat{\psi}_{n\alpha}\exp(i\hat{H}\tau)(\partial_{R^\nu}\hat{\rho}_{ss})\exp(-i\hat{H}\tau)\}d\tau \quad \text{(SD1)}$$

using $\hat{H}=\sum_{k\alpha}\varepsilon_{k\alpha}\hat{\psi}^\dagger_{k\alpha}\hat{\psi}_{k\alpha}$ and $e^{-it\varepsilon_{k\alpha}\hat{\psi}^\dagger_{k\alpha}\hat{\psi}_{k\alpha}}\hat{\psi}^\dagger_{k\alpha}e^{it\varepsilon_{k\alpha}\hat{\psi}^\dagger_{k\alpha}\hat{\psi}_{k\alpha}} = e^{-it\varepsilon_{k\alpha}}\hat{\psi}^\dagger_{k\alpha}$ as well as its Hermitian conjugate one gets, for both fermions and bosons,

$$\operatorname{Tr}\{\hat{\psi}^\dagger_{k\beta}\hat{\psi}_{n\alpha}\exp(i\hat{H}\tau)(\partial_{R^\nu}\hat{\rho}_{ss})\exp(-i\hat{H}\tau)\} = \operatorname{Tr}\{\exp(-i\hat{H}\tau)\hat{\psi}^\dagger_{k\beta}\hat{\psi}_{n\alpha}\exp(i\hat{H}\tau)(\partial_{R^\nu}\hat{\rho}_{ss})\}$$
$$= \exp\{-i\tau(\varepsilon_{k\beta}-\varepsilon_{n\alpha})\}\operatorname{Tr}\{\hat{\psi}^\dagger_{k\beta}\hat{\psi}_{n\alpha}(\partial_{R^\nu}\hat{\rho}_{ss})\} \quad \text{(SD2)}$$

Next use $\partial_{R^\nu}\operatorname{Tr}\{\hat{\psi}^\dagger_{k\beta}\hat{\psi}_{n\alpha}\hat{\rho}_{ss}\} = \partial_{R^\nu}(f(\varepsilon_{k\beta}))\delta_{k\beta n\alpha} = 0$ to transform the last term in (SD2)

$$\mathrm{Tr}\left\{\hat{\psi}^{\dagger}_{k\beta}\hat{\psi}_{n\alpha}\left(\partial_{R^{\nu}}\hat{\rho}_{ss}\right)\right\}=-\mathrm{Tr}\left\{\partial_{R^{\nu}}\left(\hat{\psi}^{\dagger}_{k\beta}\hat{\psi}_{n\alpha}\right)\hat{\rho}_{ss}\right\} \tag{SD3}$$

Using Eqs. (SD2)-(SD3) for the integral in (SD1) we get:

$$\int_{-\infty}^{0}\exp(\eta\tau)\,\mathrm{Tr}\left\{\hat{\psi}^{\dagger}_{k\beta}\hat{\psi}_{n\alpha}\exp\left(i\hat{H}\tau\right)\left(\partial_{R^{\nu}}\hat{\rho}_{ss}\right)\exp\left(-i\hat{H}\tau\right)\right\}d\tau$$

$$=-\int_{-\infty}^{0}\exp\left\{-i\tau\left(\varepsilon_{k\beta}-\varepsilon_{n\alpha}\right)+\eta\tau\right\}d\tau\,\mathrm{Tr}\left\{\partial_{R^{\nu}}\left(\hat{\psi}^{\dagger}_{k\beta}\hat{\psi}_{n\alpha}\right)\hat{\rho}_{ss}\right\}=-\frac{1}{\eta-i\left(\varepsilon_{k\beta}-\varepsilon_{n\alpha}\right)}\mathrm{Tr}\left\{\partial_{R^{\nu}}\left(\hat{\psi}^{\dagger}_{k\beta}\hat{\psi}_{n\alpha}\right)\hat{\rho}_{ss}\right\}$$

$$=-\left\{\frac{\eta}{\left(\varepsilon_{k\beta}-\varepsilon_{n\alpha}\right)^{2}+\eta^{2}}+i\frac{\varepsilon_{k\beta}-\varepsilon_{n\alpha}}{\left(\varepsilon_{k\beta}-\varepsilon_{n\alpha}\right)^{2}+\eta^{2}}\right\}\mathrm{Tr}\left\{\partial_{R^{\nu}}\left(\hat{\psi}^{\dagger}_{k\beta}\hat{\psi}_{n\alpha}\right)\hat{\rho}_{ss}\right\}$$

(SD4)

This finally leads to

$$\left\langle\hat{A}\right\rangle^{(1)}=\lim_{\eta\to+0}\sum_{\nu}\dot{R}^{\nu}\sum_{k\beta n\alpha}\left\{\frac{\eta}{\left(\varepsilon_{k\beta}-\varepsilon_{n\alpha}\right)^{2}+\eta^{2}}+i\frac{\varepsilon_{k\beta}-\varepsilon_{n\alpha}}{\left(\varepsilon_{k\beta}-\varepsilon_{n\alpha}\right)^{2}+\eta^{2}}\right\}\gamma_{k\beta n\alpha}\,\mathrm{Tr}\left\{\partial_{R^{\nu}}\left(\hat{\psi}^{\dagger}_{k\beta}\hat{\psi}_{n\alpha}\right)\hat{\rho}_{ss}\right\}$$

$$=\lim_{\eta\to+0}\sum_{\nu}\dot{R}^{\nu}\,\mathrm{Tr}\left\{\hat{\rho}_{ss}\sum_{k\beta n\alpha}\left(\frac{\eta}{\left(\varepsilon_{k\beta}-\varepsilon_{n\alpha}\right)^{2}+\eta^{2}}+i\frac{\varepsilon_{k\beta}-\varepsilon_{n\alpha}}{\left(\varepsilon_{k\beta}-\varepsilon_{n\alpha}\right)^{2}+\eta^{2}}\right)\gamma_{k\beta n\alpha}\partial_{R^{\nu}}\left(\hat{\psi}^{\dagger}_{k\beta}\hat{\psi}_{n\alpha}\right)\right\}$$

(SD5)

One can carry a similar procedure in the Heisenberg picture where the time evolution of the scattering field operators needs to be considered:

$$\left\langle\hat{A}\right\rangle=\mathrm{Tr}\left\{\hat{\rho}(t)\hat{A}\right\}=\sum_{k\beta n\alpha}\mathrm{Tr}\left(\hat{\rho}(t)\gamma_{k\beta n\alpha}\hat{\psi}^{\dagger}_{k\beta}\hat{\psi}_{n\alpha}\right)=\sum_{k\beta n\alpha}\mathrm{Tr}\left(\hat{U}(t,T)\hat{\rho}(T)\hat{U}(T,t)\gamma_{k\beta n\alpha}\hat{\psi}^{\dagger}_{k\beta}\hat{\psi}_{n\alpha}\right)$$

$$=\sum_{k\beta n\alpha}\mathrm{Tr}\left(\hat{\rho}(T)\gamma_{k\beta n\alpha}\hat{U}(T,t)\hat{\psi}^{\dagger}_{k\beta}\hat{\psi}_{n\alpha}\hat{U}(t,T)\right)=\sum_{k\beta n\alpha}\mathrm{Tr}\left(\hat{\rho}(T)\gamma_{k\beta n\alpha}\hat{\psi}^{\dagger}_{k\beta}(t)\hat{\psi}_{n\alpha}(t)\right)$$

(SD6)

or

$$\partial_{t}\left\{\hat{\psi}^{\dagger}_{k\beta}(t)\hat{\psi}_{n\alpha}(t)\right\}=i\left[\hat{H}\left(R^{\nu}(t)\right),\hat{\psi}^{\dagger}_{k\beta}(t)\hat{\psi}_{n\alpha}(t)\right] \tag{SD7}$$

where $\hat{\psi}^{\dagger}_{k\beta}(t)=\hat{U}(T,t)\hat{\psi}^{\dagger}_{k\beta}$.

Introducing the ansatz

$$\hat{\psi}^{\dagger}_{k\beta}(t)\hat{\psi}_{n\alpha}(t)=\exp\left\{i\hat{H}\left(R^{\nu}(t)\right)t\right\}\left(\hat{\psi}^{\dagger}_{k\beta}\left(R^{\nu}(t)\right)\hat{\psi}_{n\alpha}\left(R^{\nu}(t)\right)+\Delta\left(\hat{\psi}^{\dagger}_{k\beta}\hat{\psi}_{n\alpha}\right)(t)\right)\exp\left\{-i\hat{H}\left(R^{\nu}(t)\right)t\right\}$$

(SD8)

and inserting the expression above into (SD7) one gets, in analogy to (45):

$$\partial_t \left( \exp\{i\hat{H}(R^\nu(t))t\} \Delta(\hat{\psi}^\dagger_{k\beta}\hat{\psi}_{n\alpha})(t) \exp\{-i\hat{H}(R^\nu(t))t\} \right)$$
$$= i \exp\{i\hat{H}(R^\nu(t))t\} \left[ \hat{H}(R^\nu(t)), \Delta(\hat{\psi}^\dagger_{k\beta}\hat{\psi}_{n\alpha})(t) \right] \exp\{-i\hat{H}(R^\nu(t))t\} \quad \text{(SD9)}$$
$$- \sum_\nu \dot{R}^\nu \partial_{R^\nu} \left( \exp\{i\hat{H}(R^\nu(t))t\} \hat{\psi}^\dagger_{k\beta}(R^\nu(t)) \hat{\psi}_{n\alpha}(R^\nu(t)) \exp\{-i\hat{H}(R^\nu(t))t\} \right)$$

Integrating of (SD9) leads to:

$$\exp\{i\hat{H}(R^\nu(t))t\} \Delta(\hat{\psi}^\dagger_{k\beta}\hat{\psi}_{n\alpha})(t) \exp\{-i\hat{H}(R^\nu(t))t\}$$
$$= \exp\{i\hat{H}(R^\nu(T))T\} \Delta(\hat{\psi}^\dagger_{k\beta}\hat{\psi}_{n\alpha})(T) \exp\{-i\hat{H}(R^\nu(T))T\}$$
$$- \sum_\nu \dot{R}^\nu \int_T^t U^\dagger(t,\tau) \partial_{R^\nu} \left\{ \exp(i\hat{H}(R^\nu(\tau))\tau) \hat{\psi}^\dagger_{k\beta}(R^\nu(\tau)) \hat{\psi}_{n\alpha}(R^\nu(\tau)) \exp(-i\hat{H}(R^\nu(\tau))\tau) \right\} U^\dagger(\tau,t) d\tau$$
$$= \exp\{i\hat{H}(R^\nu(T))T\} \Delta(\hat{\psi}^\dagger_{k\beta}\hat{\psi}_{n\alpha})(T) \exp\{-i\hat{H}(R^\nu(T))T\}$$
$$- \sum_\nu \dot{R}^\nu \int_T^t U(\tau,t) \partial_{R^\nu} \left\{ \hat{\psi}^\dagger_{k\beta}(R^\nu(\tau)) \hat{\psi}_{n\alpha}(R^\nu(\tau)) \right\} U(t,\tau) \exp\{i(\varepsilon_{k\beta}-\varepsilon_{n\alpha})\tau\} d\tau$$

(SD10)

or

$$\exp\{i\hat{H}(R^\nu(T))T\} \Delta(\hat{\psi}^\dagger_{k\beta}\hat{\psi}_{n\alpha})(T) \exp\{-i\hat{H}(R^\nu(T))T\}$$
$$= \exp\{i\hat{H}(R^\nu(t))t\} \Delta(\hat{\psi}^\dagger_{k\beta}\hat{\psi}_{n\alpha})(t) \exp\{-i\hat{H}(R^\nu(t))t\} \quad \text{(SD11)}$$
$$+ \sum_\nu \dot{R}^\nu \int_T^t \hat{U}(\tau,t) \partial_{R^\nu} \left\{ \hat{\psi}^\dagger_{k\beta}(R^\nu(\tau)) \hat{\psi}_{n\alpha}(R^\nu(\tau)) \right\} \hat{U}(t,\tau) \exp\{i(\varepsilon_{k\beta}-\varepsilon_{n\alpha})\tau\} d\tau$$

Eq. (SD11) is exact. Setting the boundary conditions $\Delta(\hat{\psi}^\dagger_{k\beta}\hat{\psi}_{n\alpha})(t)|_{t=\infty}=0$ and introducing the adiabatic approximation, analogues to the one described below Eq. (45), the adiabatic correction for the operator takes the following form

$$\hat{A}^{(1)}(T) = \lim_{\eta \to +0} \sum_\nu \dot{R}^\nu$$
$$\times \sum_{k\beta n\alpha} \int_0^\infty \exp\{(-\eta + i(\varepsilon_{k\beta}-\varepsilon_{n\alpha}))\tau\} \quad \text{(SD12)}$$
$$\times \exp(i\hat{H}(R^\nu(T))(\tau-T)) \partial_{R^\nu} \left\{ \hat{\psi}^\dagger_{k\beta}(R^\nu(T)) \hat{\psi}_{n\alpha}(R^\nu(T)) \right\} \exp(-i\hat{H}(R^\nu(T))(\tau-T)) d\tau$$

Thus,

$$\langle \hat{A} \rangle^{(1)} = \text{Tr}\{\hat{A}^{(1)}(T)\hat{\rho}_{ss}(T)\}$$

$$= \lim_{\eta \to +0} \sum_{\nu} \dot{R}^{\nu} \sum_{k\beta n\alpha} \int_0^{\infty} \exp\{(-\eta + i(\varepsilon_{k\beta} - \varepsilon_{n\alpha}))\tau\} \text{Tr}\{\hat{\rho}_{ss}(R^{\nu}(T))\partial_{R^{\nu}}\{\hat{\psi}^{\dagger}_{k\beta}(R^{\nu}(T))\hat{\psi}_{n\alpha}(R^{\nu}(T))\}\}d\tau$$

$$= \lim_{\eta \to +0} \sum_{\nu} \dot{R}^{\nu} \text{Tr}\left\{\hat{\rho}_{ss} \sum_{k\beta n\alpha} \left(\frac{\eta}{(\varepsilon_{k\beta} - \varepsilon_{n\alpha})^2 + \eta^2} + i\frac{\varepsilon_{k\beta} - \varepsilon_{n\alpha}}{(\varepsilon_{k\beta} - \varepsilon_{n\alpha})^2 + \eta^2}\right)\gamma_{k\beta n\alpha}\partial_{R^{\nu}}\left(\hat{\psi}^{\dagger}_{k\beta}\hat{\psi}_{n\alpha}\right)\right\}$$

(SD13)

yielding again the result (SD5)

**Section F. Evaluation of** $\partial_{\varepsilon_d}(\hat{\psi}^{\dagger}_{k\beta}\hat{\psi}_{n\alpha})$ **(Eq. (51))**

From Eq.**Error! Reference source not found.** we have:

$$\partial_{\varepsilon_d}\hat{\psi}^{\dagger}_{k\beta} = \partial_{\varepsilon_d}G^r_{dd}(\varepsilon_{k\beta})\left\{V^*_{k\beta}\hat{d}^{\dagger} + V^*_{k\beta}\sum_{n\alpha}V_{n\alpha}\frac{\hat{c}^{\dagger}_{n\alpha}}{\varepsilon_{k\beta} - \varepsilon_{n\alpha} + i\eta}\right\}$$

$$= G^r_{dd}(\varepsilon_{k\beta})G^r_{dd}(\varepsilon_{k\beta})\left\{V^*_{k\beta}\hat{d}^{\dagger} + V^*_{k\beta}\sum_{n\alpha}V_{n\alpha}\frac{\hat{c}^{\dagger}_{n\alpha}}{\varepsilon_{k\beta} - \varepsilon_{n\alpha} + i\eta}\right\}$$

(SE1)

From Eq. (20)a it follows that

$$G^r_{dd}(\varepsilon_{k\beta})\left\{V^*_{k\beta}\hat{d}^{\dagger} + V^*_{k\beta}\sum_{n\alpha}V_{n\alpha}\frac{\hat{c}^{\dagger}_{n\alpha}}{\varepsilon_{k\beta} - \varepsilon_{n\alpha} + i\eta}\right\} = \hat{\psi}^{\dagger}_{k\beta} - \hat{c}^{\dagger}_{k\beta}$$

(SE2)

This is the last term in (SE2). Thus

$$\partial_{\varepsilon_d}\hat{\psi}^{\dagger}_{k\beta} = G^r_{dd}(\varepsilon_{k\beta})\{\hat{\psi}^{\dagger}_{k\beta} - \hat{c}^{\dagger}_{k\beta}\}$$

(SE3)

From Eq.**Error! Reference source not found.**a of the main text it follows that

$$\hat{c}^{\dagger}_{k\beta} - \hat{\psi}^{\dagger}_{k\beta} = \sum_{m\gamma}V^*_{k\beta}\frac{V_{m\gamma}G^a_{dd}(\varepsilon_{m\gamma})}{\varepsilon_{m\gamma} - \varepsilon_{k\beta} - i\eta}\hat{\psi}^{\dagger}_{m\gamma}$$

(SE4)

Hence, combining (SE4) and (SE3)

$$\partial_{\varepsilon_d}\hat{\psi}^{\dagger}_{k\beta} = -\sum_{m\gamma}V^*_{k\beta}\frac{V_{m\gamma}G^r_{dd}(\varepsilon_{k\beta})G^a_{dd}(\varepsilon_{m\gamma})}{\varepsilon_{m\gamma} - \varepsilon_{k\beta} - i\eta}\hat{\psi}^{\dagger}_{m\gamma}$$

(SE5)

Consider now the following expression:

$$-\text{Tr}\left\{\partial_{\varepsilon_d}\left(\hat{\psi}^{\dagger}_{k\beta}\hat{\psi}_{n\alpha}\right)\hat{\rho}\right\} = -\text{Tr}\left\{\partial_{\varepsilon_d}\left(\hat{\psi}^{\dagger}_{k\beta}\right)\hat{\psi}_{n\alpha}\hat{\rho}\right\} - \text{Tr}\left\{\hat{\psi}^{\dagger}_{k\beta}\partial_{\varepsilon_d}\left(\hat{\psi}_{n\alpha}\right)\hat{\rho}\right\} \quad \text{(SE6)}$$

Substituting (SE5) into the first term on the right of (SE6) leads to

$$-\text{Tr}\left\{\partial_{\varepsilon_d}\left(\hat{\psi}^{\dagger}_{k\beta}\right)\hat{\psi}_{n\alpha}\hat{\rho}\right\} = \sum_{m\gamma} V^*_{k\beta}\frac{V_{m\gamma}G^r_{dd}(\varepsilon_{k\beta})G^a_{dd}(\varepsilon_{m\gamma})}{\varepsilon_{m\gamma} - \varepsilon_{k\beta} - i\eta}\text{Tr}\left\{\hat{\psi}^{\dagger}_{m\gamma}\hat{\psi}_{n\alpha}\hat{\rho}\right\} =$$

$$= \sum_{m\gamma} V^*_{k\beta}\frac{V_{m\gamma}G^r_{dd}(\varepsilon_{k\beta})G^a_{dd}(\varepsilon_{m\gamma})}{\varepsilon_{m\gamma} - \varepsilon_{k\beta} - i\eta}f_{\gamma}(\varepsilon_{m\gamma})\delta_{m\gamma n\alpha} \quad \text{(SE7)}$$

$$= V^*_{k\beta}\frac{V_{n\alpha}G^r_{dd}(\varepsilon_{k\beta})G^a_{dd}(\varepsilon_{n\alpha})}{\varepsilon_{n\alpha} - \varepsilon_{k\beta} - i\eta}f_{\alpha}(\varepsilon_{n\alpha})$$

By analogy, the second term in (SE6) is

$$-\text{Tr}\left\{\hat{\psi}^{\dagger}_{k\beta}\partial_{\varepsilon_d}\left(\hat{\psi}_{n\alpha}\right)\hat{\rho}\right\} = V_{k\beta}\frac{V^*_{n\alpha}G^r_{dd}(\varepsilon_{k\beta})G^a_{dd}(\varepsilon_{n\alpha})}{\varepsilon_{k\beta} - \varepsilon_{n\alpha} + i\eta}f_{\beta}(\varepsilon_{k\beta})$$

$$= -V_{k\beta}\frac{V^*_{n\alpha}G^r_{dd}(\varepsilon_{k\beta})G^a_{dd}(\varepsilon_{n\alpha})}{\varepsilon_{n\alpha} - \varepsilon_{k\beta} - i\eta}f_{\beta}(\varepsilon_{k\beta}) \quad \text{(SE8)}$$

Using (SE7) and (SE8) in (SE6) leads to

$$-\text{Tr}\left\{\partial_v\left(\hat{\psi}^{\dagger}_{k\beta}\hat{\psi}_{n\alpha}\right)\hat{\rho}\right\} = V^*_{k\beta}\frac{V_{n\alpha}G^r_{dd}(\varepsilon_{k\beta})G^a_{dd}(\varepsilon_{n\alpha})}{\varepsilon_{n\alpha} - \varepsilon_{k\beta} - i\eta}f_{\alpha}(\varepsilon_{n\alpha}) - V_{k\beta}\frac{V^*_{n\alpha}G^r_{dd}(\varepsilon_{k\beta})G^a_{dd}(\varepsilon_{n\alpha})}{\varepsilon_{n\alpha} - \varepsilon_{k\beta} - i\eta}f_{\beta}(\varepsilon_{k\beta})$$

$$= V^*_{k\beta}\frac{V_{n\alpha}G^r_{dd}(\varepsilon_{k\beta})G^a_{dd}(\varepsilon_{n\alpha})}{\varepsilon_{n\alpha} - \varepsilon_{k\beta} - i\eta}\left\{f_{\alpha}(\varepsilon_{n\alpha}) - f_{\beta}(\varepsilon_{k\beta})\right\}$$

(SE9)

**Section G. Evaluation of non-adiabatic corrections due to finite driving speed: the lowest order correction to the particle number and the 2nd order correction to the dissipated power.**

Using Eqs.(21) and (50) with $\hat{A} = \hat{d}^{\dagger}\hat{d}$, then converting summations to integrations in the standard way, the first order correction to the particle number in the driven dot takes the form

$$N^{(1)}$$

$$= -\dot{\varepsilon}_d \sum_{k\alpha} \sum_{n\beta} |V_{k\alpha}|^2 |V_{n\beta}|^2 G^r_{dd}(\varepsilon_k) G^a_{dd}(\varepsilon_k) G^r_{dd}(\varepsilon_n) G^a_{dd}(\varepsilon_n) \frac{f_\beta(\varepsilon_n) - f_\alpha(\varepsilon_k)}{\varepsilon_n - \varepsilon_k - i\eta_1} \left\{ \frac{\eta_2}{(\varepsilon_n - \varepsilon_k)^2 + \eta_2^2} + i \frac{\varepsilon_n - \varepsilon_k}{(\varepsilon_n - \varepsilon_k)^2 + \eta_2^2} \right\}$$

$$= -\dot{\varepsilon}_d \left(\frac{1}{2\pi}\right)^2 \int d\varepsilon \int d\varepsilon' \sum_{k\alpha} \sum_{n\beta} |V_{k\alpha}|^2 |V_{n\beta}|^2 \, 2\pi\delta(\varepsilon - \varepsilon_n) 2\pi\delta(\varepsilon' - \varepsilon_k)$$

$$\times G^r_{dd}(\varepsilon') G^a_{dd}(\varepsilon') G^r_{dd}(\varepsilon) G^a_{dd}(\varepsilon) \frac{f_\beta(\varepsilon) - f_\alpha(\varepsilon')}{\varepsilon - \varepsilon' - i\eta_1} \left\{ \frac{\eta_2}{(\varepsilon - \varepsilon')^2 + \eta_2^2} + i \frac{\varepsilon - \varepsilon'}{(\varepsilon - \varepsilon')^2 + \eta_2^2} \right\}$$

$$= -\dot{\varepsilon}_d \left(\frac{1}{2\pi}\right)^2 \int d\varepsilon \int d\varepsilon' \frac{A_{dd}(\varepsilon) A_{dd}(\varepsilon')}{\Gamma(\varepsilon) \Gamma(\varepsilon')} \sum_\alpha \sum_\beta \Gamma_\beta(\varepsilon) \Gamma_\alpha(\varepsilon') \frac{f_\beta(\varepsilon) - f_\alpha(\varepsilon')}{\varepsilon - \varepsilon' - i\eta_1} \left\{ \frac{\eta_2}{(\varepsilon - \varepsilon')^2 + \eta_2^2} + i \frac{\varepsilon - \varepsilon'}{(\varepsilon - \varepsilon')^2 + \eta_2^2} \right\}$$

(SF1)

The energy integrals may be taken over the complete real energy axis, $-\infty < \varepsilon < \infty$. This does not imply making the wide band approximation, which is determined by the energy dependence of the couplings and state densities as expressed by the energy dependence of the functions $G^{r/a}_{dd}$, $A_{dd}$ and $\Gamma$ in Eq. (SF1).

Swapping the indexes $\alpha$ and $\beta$ in the double sum in (SF1) we get:

$$N^{(1)}$$

$$= -\frac{\dot{\varepsilon}_d}{2} \left(\frac{1}{2\pi}\right)^2 \int d\varepsilon \int d\varepsilon' \frac{A_{dd}(\varepsilon) A_{dd}(\varepsilon')}{\Gamma(\varepsilon) \Gamma(\varepsilon')}$$

$$\times \sum_\alpha \sum_\beta \left\{ \Gamma_\beta(\varepsilon) \Gamma_\alpha(\varepsilon') \frac{f_\beta(\varepsilon) - f_\alpha(\varepsilon')}{\varepsilon - \varepsilon' - i\eta_1} + \Gamma_\alpha(\varepsilon) \Gamma_\beta(\varepsilon') \frac{f_\alpha(\varepsilon) - f_\beta(\varepsilon')}{\varepsilon - \varepsilon' - i\eta_1} \right\} \times \left\{ \frac{\eta_2}{(\varepsilon - \varepsilon')^2 + \eta_2^2} + i \frac{\varepsilon - \varepsilon'}{(\varepsilon - \varepsilon')^2 + \eta_2^2} \right\}$$

$$= -\frac{\dot{\varepsilon}_d}{2} \left(\frac{1}{2\pi}\right)^2 \int d\varepsilon \int d\varepsilon' \frac{A_{dd}(\varepsilon) A_{dd}(\varepsilon')}{\Gamma(\varepsilon) \Gamma(\varepsilon')}$$

$$\times \sum_\alpha \sum_\beta \frac{1}{\varepsilon - \varepsilon' - i\eta_1} \left\{ \Gamma_\beta(\varepsilon) \Gamma_\alpha(\varepsilon') f_\beta(\varepsilon) - \Gamma_\alpha(\varepsilon) \Gamma_\beta(\varepsilon') f_\beta(\varepsilon') + \Gamma_\alpha(\varepsilon) \Gamma_\beta(\varepsilon') f_\alpha(\varepsilon) - \Gamma_\beta(\varepsilon) \Gamma_\alpha(\varepsilon') f_\alpha(\varepsilon') \right\}$$

$$\times \left\{ \frac{\eta_2}{(\varepsilon - \varepsilon')^2 + \eta_2^2} + i \frac{\varepsilon - \varepsilon'}{(\varepsilon - \varepsilon')^2 + \eta_2^2} \right\}$$

(SF2)

With (S 2) and taking the (first) limit $\eta_1 \to +0$, Eq. (SF2) becomes

$$N^{(1)} = -\frac{\dot{\varepsilon}_d}{2}\left(\frac{1}{2\pi}\right)^2 \int d\varepsilon \left[ i\pi \left\{ \frac{\eta_2}{(\varepsilon-\varepsilon')^2+\eta_2^2} + i\frac{\varepsilon-\varepsilon'}{(\varepsilon-\varepsilon')^2+\eta_2^2} \right\} \right.$$

$$+ \frac{A_{dd}(\varepsilon)A_{dd}(\varepsilon')}{\Gamma(\varepsilon)\Gamma(\varepsilon')} \sum_\alpha \sum_\beta \{\Gamma_\beta(\varepsilon)\Gamma_\alpha(\varepsilon')f_\beta(\varepsilon) - \Gamma_\alpha(\varepsilon)\Gamma_\beta(\varepsilon')f_\beta(\varepsilon') + \Gamma_\alpha(\varepsilon)\Gamma_\beta(\varepsilon')f_\alpha(\varepsilon) - \Gamma_\beta(\varepsilon)\Gamma_\alpha(\varepsilon')f_\alpha(\varepsilon')\} \Big|_{\varepsilon=\varepsilon'}$$

$$+ \int d\varepsilon' \frac{A_{dd}(\varepsilon)A_{dd}(\varepsilon')}{\Gamma(\varepsilon)\Gamma(\varepsilon')}$$

$$\times \sum_\alpha \sum_\beta \left\{ \frac{\Gamma_\beta(\varepsilon)\Gamma_\alpha(\varepsilon')f_\beta(\varepsilon)-\Gamma_\alpha(\varepsilon)\Gamma_\beta(\varepsilon')f_\beta(\varepsilon')}{\varepsilon-\varepsilon'} + \frac{\Gamma_\alpha(\varepsilon)\Gamma_\beta(\varepsilon')f_\alpha(\varepsilon)-\Gamma_\beta(\varepsilon)\Gamma_\alpha(\varepsilon')f_\alpha(\varepsilon')}{\varepsilon-\varepsilon'} \right\}$$

$$\times \left\{ \frac{\eta_2}{(\varepsilon-\varepsilon')^2+\eta_2^2} + i\frac{\varepsilon-\varepsilon'}{(\varepsilon-\varepsilon')^2+\eta_2^2} \right\} \Bigg]$$

$$= -\frac{\dot{\varepsilon}_d}{2}\left(\frac{1}{2\pi}\right)^2 \int d\varepsilon \int d\varepsilon' \frac{A_{dd}(\varepsilon)A_{dd}(\varepsilon')}{\Gamma(\varepsilon)\Gamma(\varepsilon')}$$

$$\times \sum_\alpha \sum_\beta \left\{ \frac{\Gamma_\beta(\varepsilon)\Gamma_\alpha(\varepsilon')f_\beta(\varepsilon)-\Gamma_\alpha(\varepsilon)\Gamma_\beta(\varepsilon')f_\beta(\varepsilon')}{\varepsilon-\varepsilon'} + \frac{\Gamma_\alpha(\varepsilon)\Gamma_\beta(\varepsilon')f_\alpha(\varepsilon)-\Gamma_\beta(\varepsilon)\Gamma_\alpha(\varepsilon')f_\alpha(\varepsilon')}{\varepsilon-\varepsilon'} \right\}$$

$$\times \left\{ \frac{\eta_2}{(\varepsilon-\varepsilon')^2+\eta_2^2} + i\frac{\varepsilon-\varepsilon'}{(\varepsilon-\varepsilon')^2+\eta_2^2} \right\}$$

(SF3)

In what follows will also use the identity

$$\frac{F(\varepsilon)-F(\varepsilon')}{\varepsilon-\varepsilon'} = \partial_\varepsilon F(\varepsilon') + (\varepsilon-\varepsilon')\frac{\partial^2 F(\tilde{\varepsilon})}{2\partial\varepsilon^2} \tag{SF4}$$

for some $\tilde{\varepsilon} \in (\varepsilon',\varepsilon)$. This leads to

$$\frac{\Gamma_\beta(\varepsilon)\Gamma_\alpha(\varepsilon')f_\beta(\varepsilon)-\Gamma_\alpha(\varepsilon)\Gamma_\beta(\varepsilon')f_\beta(\varepsilon')}{\varepsilon-\varepsilon'}$$

$$= \Gamma_\alpha(\varepsilon')\partial_\varepsilon\{f_\beta(\varepsilon')\Gamma_\beta(\varepsilon')\} - f_\beta(\varepsilon')\Gamma_\beta(\varepsilon')\partial_\varepsilon\Gamma_\alpha(\varepsilon') + (\varepsilon-\varepsilon')\theta_{\beta\alpha}(\varepsilon,\varepsilon') \tag{SF5}$$

and the equivalent expression obtained from interchanging $\alpha$ and $\beta$, where $\theta_{\beta\alpha}(\varepsilon,\varepsilon')$ stands for the sum of second derivatives obtained from the second term in (SF4).

We use these relationships to evaluate (SF3). Consider first the contribution associated with the term $\lim_{\eta_2 \to +0} \frac{\eta_2}{(\varepsilon-\varepsilon')^2 + \eta_2^2} = \pi\delta(\varepsilon-\varepsilon')$ in the last brackets of (SF3). With (SF5) it leads to

$$-\frac{\dot{\varepsilon}_d}{2}\left(\frac{1}{2\pi}\right)^2 \int d\varepsilon \int d\varepsilon' \frac{A_{dd}(\varepsilon)A_{dd}(\varepsilon')}{\Gamma(\varepsilon)\Gamma(\varepsilon')}$$

$$\times \sum_\alpha \sum_\beta \left\{ \frac{\Gamma_\beta(\varepsilon)\Gamma_\alpha(\varepsilon')f_\beta(\varepsilon) - \Gamma_\alpha(\varepsilon)\Gamma_\beta(\varepsilon')f_\beta(\varepsilon')}{\varepsilon-\varepsilon'} + \frac{\Gamma_\alpha(\varepsilon)\Gamma_\beta(\varepsilon')f_\alpha(\varepsilon) - \Gamma_\beta(\varepsilon)\Gamma_\alpha(\varepsilon')f_\alpha(\varepsilon')}{\varepsilon-\varepsilon'} \right\}$$

$$\times \pi\delta(\varepsilon-\varepsilon') = -\frac{\dot{\varepsilon}_d}{4}\left(\frac{1}{2\pi}\right)^2 \int d\varepsilon \int d\varepsilon' \frac{A_{dd}(\varepsilon)A_{dd}(\varepsilon')}{\Gamma(\varepsilon)\Gamma(\varepsilon')}$$

$$\times \pi\delta(\varepsilon-\varepsilon') \sum_\alpha \sum_\beta \Big( \Gamma_\alpha(\varepsilon')\partial_\varepsilon\{f_\beta(\varepsilon')\Gamma_\beta(\varepsilon')\} - f_\beta(\varepsilon')\Gamma_\beta(\varepsilon')\big(\partial_\varepsilon\Gamma_\alpha(\varepsilon')\big) + (\varepsilon-\varepsilon')\theta_{\beta\alpha}(\varepsilon,\varepsilon')$$

$$+\Gamma_\beta(\varepsilon')\partial_\varepsilon\{f_\alpha(\varepsilon')\Gamma_\alpha(\varepsilon')\} - f_\alpha(\varepsilon')\Gamma_\alpha(\varepsilon')\big(\partial_\varepsilon\Gamma_\beta(\varepsilon')\big) + (\varepsilon-\varepsilon')\theta_{\alpha\beta}(\varepsilon,\varepsilon') \Big)$$

$$= \frac{-\dot{\varepsilon}_d}{8\pi} \int d\varepsilon \frac{A_{dd}^2(\varepsilon)}{\Gamma^2(\varepsilon)} \sum_\alpha \sum_\beta \Big( \Gamma_\alpha(\varepsilon)\partial_\varepsilon\{f_\beta(\varepsilon)\Gamma_\beta(\varepsilon)\} + \Gamma_\beta(\varepsilon)\partial_\varepsilon\{f_\alpha(\varepsilon)\Gamma_\alpha(\varepsilon)\}$$

$$- f_\beta(\varepsilon)\Gamma_\beta(\varepsilon)\partial_\varepsilon\Gamma_\alpha(\varepsilon) - f_\alpha(\varepsilon)\Gamma_\alpha(\varepsilon)\partial_\varepsilon\Gamma_\beta(\varepsilon) \Big)$$

(SF6)

In obtaining (I6) the contribution from $(\varepsilon-\varepsilon')\theta_{\beta\alpha}(\varepsilon,\varepsilon')$ disappeared since

$$\lim_{\eta_2 \to +0} \frac{\eta_2}{(\varepsilon-\varepsilon')^2 + \eta_2^2}(\varepsilon-\varepsilon')\theta_{\beta\alpha}(\varepsilon,\varepsilon') = \pi\delta(\varepsilon-\varepsilon')(\varepsilon-\varepsilon)\theta_{\beta\alpha}(\varepsilon,\varepsilon) = 0$$

Next, consider the contribution arising from the term $i\frac{\varepsilon-\varepsilon'}{(\varepsilon-\varepsilon')^2 + \eta_2^2}$ in the last bracket of (SF3). We can swap $\varepsilon$ and $\varepsilon'$ to cast this contribution in the form

$$-\frac{\dot{\varepsilon}_d}{2}\left(\frac{1}{2\pi}\right)^2 \int d\varepsilon \int d\varepsilon' \frac{A_{dd}(\varepsilon)A_{dd}(\varepsilon')}{\Gamma(\varepsilon)\Gamma(\varepsilon')}$$

$$\times \sum_\alpha \sum_\beta \left\{\frac{\Gamma_\beta(\varepsilon)\Gamma_\alpha(\varepsilon')f_\beta(\varepsilon)-\Gamma_\alpha(\varepsilon)\Gamma_\beta(\varepsilon')f_\beta(\varepsilon')+\Gamma_\alpha(\varepsilon)\Gamma_\beta(\varepsilon')f_\alpha(\varepsilon)-\Gamma_\beta(\varepsilon)\Gamma_\alpha(\varepsilon')f_\alpha(\varepsilon')}{(\varepsilon-\varepsilon')^2 + \eta_2^2}\right\}$$

$$=-\frac{\dot{\varepsilon}_d}{2}\left(\frac{1}{2\pi}\right)^2 \int d\varepsilon \int d\varepsilon' \frac{A_{dd}(\varepsilon)A_{dd}(\varepsilon')}{\Gamma(\varepsilon)\Gamma(\varepsilon')}$$

$$\times \sum_\alpha \sum_\beta \left\{\frac{\Gamma_\beta(\varepsilon')\Gamma_\alpha(\varepsilon)f_\beta(\varepsilon')-\Gamma_\alpha(\varepsilon')\Gamma_\beta(\varepsilon)f_\beta(\varepsilon)+\Gamma_\alpha(\varepsilon')\Gamma_\beta(\varepsilon)f_\alpha(\varepsilon')-\Gamma_\beta(\varepsilon')\Gamma_\alpha(\varepsilon)f_\alpha(\varepsilon)}{(\varepsilon'-\varepsilon)^2 + \eta_2^2}\right\}$$

(SF7)

It is easily seen that the integrand in (SF7) is antisymmetric under the interchange $\varepsilon \leftrightarrow \varepsilon'$, hence the double integral over $\varepsilon$ and $\varepsilon'$, and therefore this contribution to (SF3) vanishes. The correction to the particle number is therefore determined by the term (SF6):

$$N^{(1)} = \frac{-\dot{\varepsilon}_d}{8\pi}\int d\varepsilon \frac{A_{dd}^2(\varepsilon)}{\Gamma^2(\varepsilon)}\sum_\alpha \sum_\beta \left[\Gamma_\alpha(\varepsilon)\partial_\varepsilon\left(f_\beta(\varepsilon)\Gamma_\beta(\varepsilon)\right)+\Gamma_\beta(\varepsilon)\partial_\varepsilon\left(f_\alpha(\varepsilon)\Gamma_\alpha(\varepsilon)\right)\right.$$

$$\left.-f_\beta(\varepsilon)\Gamma_\beta(\varepsilon)\partial_\varepsilon\Gamma_\alpha(\varepsilon)-f_\alpha(\varepsilon)\Gamma_\alpha(\varepsilon)\partial_\varepsilon\Gamma_\beta(\varepsilon)\right]$$

$$=\frac{-\dot{\varepsilon}_d}{8\pi}\int d\varepsilon \frac{A_{dd}^2(\varepsilon)}{\Gamma^2(\varepsilon)}\sum_\alpha \sum_\beta \left[\Gamma_\alpha(\varepsilon)\Gamma_\beta(\varepsilon)\left(\partial_\varepsilon f_\beta(\varepsilon)+\partial_\varepsilon f_\alpha(\varepsilon)\right)+\right.$$

$$\left.+\left(f_\beta(\varepsilon)-f_\alpha(\varepsilon)\right)\left(\Gamma_\alpha(\varepsilon)\partial_\varepsilon\Gamma_\beta(\varepsilon)-\Gamma_\beta(\varepsilon)\partial_\varepsilon\Gamma_\alpha(\varepsilon)\right)\right]$$

(SF8)

which may be further simplified as follows

$$N^{(1)} = \frac{-\dot{\varepsilon}_d}{8\pi}\int d\varepsilon \frac{A_{dd}^2(\varepsilon)}{\Gamma^2(\varepsilon)}$$

$$\times \left[\sum_\beta \Gamma(\varepsilon)\Gamma_\beta(\varepsilon)\partial_\varepsilon f_\beta(\varepsilon)+\sum_\alpha \Gamma(\varepsilon)\Gamma_\alpha(\varepsilon)\partial_\varepsilon f_\alpha(\varepsilon)\right.$$

$$+\sum_\beta f_\beta(\varepsilon)\left(\Gamma(\varepsilon)\partial_\varepsilon\Gamma_\beta(\varepsilon)-\Gamma_\beta(\varepsilon)\partial_\varepsilon\Gamma(\varepsilon)\right)$$

$$\left.+\sum_\alpha f_\alpha(\varepsilon)\left(\Gamma(\varepsilon)\partial_\varepsilon\{\Gamma_\alpha(\varepsilon)\}-\Gamma_\alpha(\varepsilon)\partial_\varepsilon\{\Gamma(\varepsilon)\}\right)\right]$$

$$=\frac{-\dot{\varepsilon}_d}{4\pi}\int d\varepsilon \frac{A_{dd}^2(\varepsilon)}{\Gamma^2(\varepsilon)}\sum_\alpha \left\{\Gamma(\varepsilon)\Gamma_\alpha(\varepsilon)\partial_\varepsilon f_\alpha(\varepsilon)+f_\alpha(\varepsilon)\left(\Gamma(\varepsilon)\partial_\varepsilon\Gamma_\alpha(\varepsilon)-\Gamma_\alpha(\varepsilon)\partial_\varepsilon\Gamma(\varepsilon)\right)\right\}$$

$$=\frac{-\dot{\varepsilon}_d}{4\pi}\int d\varepsilon A_{dd}^2(\varepsilon)\partial_\varepsilon \sum_\alpha \frac{\Gamma_\alpha(\varepsilon)f_\alpha(\varepsilon)}{\Gamma(\varepsilon)} = \frac{-\dot{\varepsilon}_d}{4\pi}\int d\varepsilon A_{dd}^2(\varepsilon)\partial_\varepsilon \tilde{f}(\varepsilon)$$

(SF9)

where a weighted distribution was introduced $\tilde{f}(\varepsilon) = \sum_\alpha \dfrac{\Gamma_\alpha(\varepsilon) f_\alpha(\varepsilon)}{\Gamma(\varepsilon)}$

Finally, the generated power, which is second-order in driving rate, is obtained from (SF9) in the form

$$\dot{W}^{(2)} = \dot{\varepsilon}_d N^{(1)} = \frac{-(\dot{\varepsilon}_d)^2}{4\pi} \int d\varepsilon\, A_{dd}^2(\varepsilon)\, \partial_\varepsilon \tilde{f}(\varepsilon) \tag{SF10}$$

**Section H. Driving the interaction term**

Here we consider driving the system by a single time dependent parameter $R$ and assume that both $\varepsilon_d$ and the coupling elements $V_{k\alpha}$ depend on this parameter. From the Lippman – Schwinger equation, Eq. (18), it follows that

$$\begin{aligned}
\partial_R |\psi_{k\beta}\rangle &= \partial_R \left( |c_{k\beta}\rangle + \hat{G}^r(\varepsilon_k)\hat{V}|c_{k\beta}\rangle \right) = \left( \partial_R |c_{k\beta}\rangle \right) + \partial_R \left( \hat{G}^r(\varepsilon_k)\hat{V} \right) |c_{k\beta}\rangle + \hat{G}^r(\varepsilon_k)\hat{V}\left( \partial_R |c_{k\beta}\rangle \right) \\
&= \partial_R \left( \hat{G}^r(\varepsilon_k)\hat{V} \right) |c_{k\beta}\rangle = \left( \left( \partial_R \hat{G}^r(\varepsilon_k) \right) \hat{V} + \hat{G}^r(\varepsilon_k)\left( \partial_R \hat{V} \right) \right) |c_{k\beta}\rangle = \hat{G}^r(\varepsilon_k)\left( \partial_R \hat{V} \right) \hat{G}^r(\varepsilon_k) \hat{V} |c_{k\beta}\rangle + \hat{G}^r(\varepsilon_k)\left( \partial_R \hat{V} \right) |c_{k\beta}\rangle \\
&= \hat{G}^r(\varepsilon_k)\left( \partial_R \hat{V} \right)\left( |c_{k\beta}\rangle + \hat{G}^r(\varepsilon_k)\hat{V}|c_{k\beta}\rangle \right) = \hat{G}^r(\varepsilon_k)\left( \partial_R \hat{V} \right)|\psi_{k\beta}\rangle = \sum_{n\alpha} G^r_{n\alpha n\alpha}(\varepsilon_k) |\psi_{n\alpha}\rangle \langle \psi_{n\alpha}|\partial_R \hat{V}|\psi_{k\beta}\rangle
\end{aligned}$$

(SG1)

which implies that

$$\partial_R \hat{\psi}^\dagger_{k\beta} = \sum_{n\alpha} G^r_{n\alpha n\alpha}(\varepsilon_k) \langle \psi_{n\alpha}|\partial_R \hat{V}|\psi_{k\beta}\rangle \hat{\psi}^\dagger_{n\alpha} \tag{SG2}$$

Thus

$$\begin{aligned}
\mathrm{Tr}\left\{ \partial_R \left( \hat{\psi}^\dagger_{k\beta} \right) \hat{\psi}_{n\alpha} \hat{\rho}_{ss} \right\} &= \sum_{m\gamma} G^r_{m\gamma m\gamma}(\varepsilon_k) \langle \psi_{m\gamma}|\partial_R \hat{V}|\psi_{k\beta}\rangle \mathrm{Tr}\left\{ \hat{\psi}^\dagger_{m\gamma} \hat{\psi}_{n\alpha} \hat{\rho}_{ss} \right\} \\
&= \langle \psi_{n\alpha}|\partial_R \hat{V}|\psi_{k\beta}\rangle \frac{f_\alpha(\varepsilon_n)}{\varepsilon_k - \varepsilon_n + i\eta}
\end{aligned}$$

(SG3)

and

$$\mathrm{Tr}\left\{ \hat{\psi}^\dagger_{k\beta} \left( \partial_R \hat{\psi}_{n\alpha} \right) \hat{\rho}_{ss} \right\} = \langle \psi_{n\alpha}|\partial_R \hat{V}|\psi_{k\beta}\rangle \frac{f_\beta(\varepsilon_k)}{\varepsilon_n - \varepsilon_k - i\eta} \tag{SG4}$$

Combining together (SG3)-(SG4)

$$\text{Tr}\left\{\partial_R(\hat{\psi}_{k\beta}^\dagger\hat{\psi}_{n\alpha})\hat{\rho}_{ss}\right\} = -\left\langle\psi_{n\alpha}\left|\partial_R\hat{V}\right|\psi_{k\beta}\right\rangle\frac{f_\alpha(\varepsilon_n) - f_\beta(\varepsilon_k)}{\varepsilon_n - \varepsilon_k - i\eta} \tag{SG5}$$

Thus, from Eq.**Error! Reference source not found.** it follows that:

$$\dot{W}^{(2)} =$$

$$-\sum_{k\beta n\alpha}\left(\dot{R}\right)^2\left(\frac{\eta_2}{(\varepsilon_k - \varepsilon_n)^2 + \eta_2^2} + i\frac{\varepsilon_k - \varepsilon_n}{(\varepsilon_k - \varepsilon_n)^2 + \eta_2^2}\right) \tag{SG6}$$

$$\times\left\langle\psi_{k\beta}\left|\partial_R\hat{V}\right|\psi_{n\alpha}\right\rangle\left\langle\psi_{n\alpha}\left|\partial_R\hat{V}\right|\psi_{k\beta}\right\rangle\left\{\frac{f_\alpha(\varepsilon_n) - f_\beta(\varepsilon_k)}{\varepsilon_n - \varepsilon_k - i\eta_1}\right\}$$

Next, repeating a series of steps similar to the procedure outlined in Section G: swapping the indexes $\alpha$ and $\beta$, taking the limits $\eta_{1,2} \to +0$ and using the identities

$$(\varepsilon_n - \varepsilon_k)^{-1}\left(F_1(\varepsilon_n)F_2(\varepsilon_k) - F_1(\varepsilon_k)F_2(\varepsilon_n)\right)$$

$$= F_2(\varepsilon_k)\partial_\varepsilon F_1(\varepsilon_k) + \frac{1}{2}(\varepsilon_n - \varepsilon_k)F_2(\varepsilon_k)\partial^2 F_1(\tilde{\varepsilon}_1)/\partial\varepsilon^2 \tag{SG7}$$

$$-\partial_\varepsilon F_2(\varepsilon_k)F_1(\varepsilon_k) - \frac{1}{2}(\varepsilon_n - \varepsilon_k)F_1(\varepsilon_k)\partial^2 F_2(\tilde{\varepsilon}_2)/\partial\varepsilon^2$$

and

$$\delta(\varepsilon_n - \varepsilon_k)F(\varepsilon_n - \varepsilon_k) = F(0) \tag{SG8}$$

one gets:

$$\dot{W}^{(2)} = -\frac{1}{2} \sum_{k\beta n\alpha} (\dot{R})^2 \pi \delta(\varepsilon_n - \varepsilon_k)$$

$$\times \left( \langle \psi_{k\beta} | \partial_R \hat{V} | \psi_{n\alpha} \rangle \langle \psi_{n\alpha} | \partial_R \hat{V} | \psi_{k\beta} \rangle \left\{ \frac{f_\alpha(\varepsilon_n) - f_\beta(\varepsilon_k)}{\varepsilon_n - \varepsilon_k} \right\} \right.$$

$$\left. + \langle \psi_{k\alpha} | \partial_R \hat{V} | \psi_{n\beta} \rangle \langle \psi_{n\beta} | \partial_R \hat{V} | \psi_{k\alpha} \rangle \left\{ \frac{f_\beta(\varepsilon_n) - f_\alpha(\varepsilon_k)}{\varepsilon_n - \varepsilon_k} \right\} \right)$$

$$= -\frac{1}{2} \sum_{k\beta n\alpha} (\dot{R})^2 \pi \delta(\varepsilon_n - \varepsilon_k) \frac{1}{\varepsilon_n - \varepsilon_k}$$

$$\times \left( |\langle \psi_{k\beta} | \partial_R \hat{V} | \psi_{n\alpha} \rangle|^2 f_\alpha(\varepsilon_n) - |\langle \psi_{k\alpha} | \partial_R \hat{V} | \psi_{n\beta} \rangle|^2 f_\alpha(\varepsilon_k) + |\langle \psi_{k\alpha} | \partial_R \hat{V} | \psi_{n\beta} \rangle|^2 f_\beta(\varepsilon_n) - |\langle \psi_{k\beta} | \partial_R \hat{V} | \psi_{n\alpha} \rangle|^2 f_\beta(\varepsilon_k) \right)$$

$$= \dot{W}_I^{(2)} + \dot{W}_{II}^{(2)}$$

(SG9)

where

$$\dot{W}_I^{(2)} = -\frac{\pi (\dot{R})^2}{2} \sum_{n\alpha} \sum_{k\beta} \delta(\varepsilon_n - \varepsilon_k) \left\{ (\partial_\varepsilon f_\alpha(\varepsilon_n)) + (\partial_\varepsilon f_\beta(\varepsilon_n)) \right\} |\langle \psi_{k\beta} | \partial_R \hat{V} | \psi_{k\alpha} \rangle|^2$$

$$= -\pi (\dot{R})^2 \sum_{n\alpha} (\partial_\varepsilon f_\alpha(\varepsilon_n)) \sum_{k\beta} \delta(\varepsilon_n - \varepsilon_k) |\langle \psi_{k\beta} | \partial_R \hat{V} | \psi_{k\alpha} \rangle|^2$$

(SG10)

$$\dot{W}_{II}^{(2)} = -\frac{\pi (\dot{R})^2}{2} \sum_{n\alpha} \sum_{k\beta} (f_\alpha(\varepsilon_n) - f_\beta(\varepsilon_n)) \delta(\varepsilon_n - \varepsilon_k) \partial_{\varepsilon_n} \left( |\langle \psi_{k\beta} | \partial_R \hat{V} | \psi_{n\alpha} \rangle|^2 - |\langle \psi_{n\beta} | \partial_R \hat{V} | \psi_{k\alpha} \rangle|^2 \right)$$

(SG11)

Here $\partial_{\varepsilon_n}$ denotes a derivative with respect to the corresponding energy level, so in $\partial_{\varepsilon_n} \langle \psi_{k\beta} | \partial_R \hat{V} | \psi_{n\alpha} \rangle$ the derivative is taken with respect to the energy of state $\psi_{n\alpha}$. The first of these terms, $\dot{W}_I^{(2)}$ is always positive while the second, $\dot{W}_{II}^{(2)}$, can be negative, indicating the possibility to extract energy from the voltage bias[2]. In what follows we evaluate each of these terms separately.

First, one needs to obtain $\langle \psi_{k\beta} | \partial_R \hat{V} | \psi_{n\alpha} \rangle$. Recalling Eqs. (20):

$$|\psi_{n\alpha}\rangle = G_{dd}^r(\varepsilon_n) V_{n\alpha}^* |d\rangle + \sum_{m\gamma} \left\{ \delta_{n\alpha m\gamma} + G_{dd}^r(\varepsilon_n) V_{m\gamma} V_{n\alpha}^* \frac{1}{\varepsilon_n - \varepsilon_m + i\eta} \right\} |c_{m\gamma}\rangle \qquad \text{(SG12)a}$$

$$\langle \psi_{k\beta} | = G_{dd}^a(\varepsilon_k) V_{k\beta} \langle d | + \sum_{m\gamma} \left\{ \delta_{k\beta m\gamma} + G_{dd}^a(\varepsilon_k) V_{m\gamma}^* V_{k\beta} \frac{1}{\varepsilon_k - \varepsilon_m - i\eta} \right\} \langle c_{m\gamma} | \qquad \text{(SG12)b}$$

one gets (limit $\eta \to +0$ is assumed)

$$\langle \psi_{k\beta} | \partial_R \hat{V} | \psi_{n\alpha} \rangle = \langle \psi_{k\beta} | \left\{ (\partial_R \varepsilon_d) | d \rangle \langle d | + \sum_{m\gamma} (\partial_R V_{m\gamma}^*) | d \rangle \langle c_{m\gamma} | + (\partial_R V_{m\gamma}) | c_{m\gamma} \rangle \langle d | \right\} | \psi_{n\alpha} \rangle$$

$$= \sum_{m\gamma} \left[ (\partial_R V_{m\gamma}^*) \langle \psi_{k\beta} | d \rangle \langle c_{m\gamma} | \psi_{n\alpha} \rangle + (\partial_R V_{m\gamma}) \langle \psi_{k\beta} | c_{m\gamma} \rangle \langle d | \psi_{n\alpha} \rangle \right] + (\partial_R \varepsilon_d) \langle \psi_{k\beta} | d \rangle \langle d | \psi_{n\alpha} \rangle$$

$$= \sum_{m\gamma} \left[ (\partial_R V_{m\gamma}^*) G_{dd}^a(\varepsilon_k) V_{k\beta} \left\{ \delta_{m\gamma n\alpha} + G_{dd}^r(\varepsilon_n) V_{m\gamma} V_{n\alpha}^* \frac{1}{\varepsilon_n - \varepsilon_m + i\eta} \right\} \right.$$

$$+ (\partial_R V_{m\gamma}) \left\{ \delta_{k\beta m\gamma} + G_{dd}^a(\varepsilon_k) V_{m\gamma}^* V_{k\beta} \frac{1}{\varepsilon_k - \varepsilon_m - i\eta} \right\} G_{dd}^r(\varepsilon_n) V_{n\alpha}^* \right] + (\partial_R \varepsilon_d) \langle \psi_{k\beta} | d \rangle \langle d | \psi_{n\alpha} \rangle$$

$$= G_{dd}^a(\varepsilon_k) G_{dd}^r(\varepsilon_n) \sum_{m\gamma} \left[ (\partial_R V_{m\gamma}^*) V_{k\beta} \left\{ \delta_{m\gamma n\alpha} G_{dd}^r(\varepsilon_n)^{-1} + V_{m\gamma} V_{n\alpha}^* \frac{1}{\varepsilon_n - \varepsilon_m + i\eta} \right\} \right.$$

$$+ (\partial_R V_{m\gamma}) \left\{ \delta_{k\beta m\gamma} G_{dd}^a(\varepsilon_k)^{-1} + V_{m\gamma}^* V_{k\beta} \frac{1}{\varepsilon_k - \varepsilon_m - i\eta} \right\} V_{n\alpha}^* \right] + (\partial_R \varepsilon_d) V_{n\alpha}^* V_{k\beta} G_{dd}^a(\varepsilon_k) G_{dd}^r(\varepsilon_n)$$

$$= G_{dd}^a(\varepsilon_k) G_{dd}^r(\varepsilon_n) \left\{ V_{n\alpha}^* V_{k\beta} \left( \sum_{m\gamma} V_{m\gamma} (\partial_R V_{m\gamma}^*) \frac{1}{\varepsilon_n - \varepsilon_m + i\eta} + (\partial_R V_{m\gamma}) V_{m\gamma}^* \frac{1}{\varepsilon_k - \varepsilon_m - i\eta} + (\partial_R \varepsilon_d) \right) \right.$$ (SG13)

$$+ G_{dd}^r(\varepsilon_n)^{-1} (\partial_R V_{n\alpha}^*) V_{k\beta} + G_{dd}^a(\varepsilon_k)^{-1} V_{n\alpha}^* (\partial_R V_{k\beta}) \right\}$$

To proceed further, the phase $\Phi_{k\alpha}$ needs to be introduced:

$$V_{k\alpha} = |V_{k\alpha}| \exp(-i\Phi_{k\alpha}) \qquad \text{(SG14)}$$

Thus

$$\partial_R V_{k\alpha} = (\partial_R |V_{k\alpha}|) \exp(-i\Phi_{k\alpha}) - i|V_{k\alpha}| \exp(-i\Phi_{k\alpha})(\partial_R \Phi_{k\alpha}) \qquad \text{(SG15)}$$

Recalling that $\Gamma_\alpha(\varepsilon_k) = 2\pi |V_{k\alpha}|^2 D_\alpha(\varepsilon_k)$ where $D_\alpha(\varepsilon_k)$ is the density of states and $\partial_R D_\alpha(\varepsilon_k) / D_\alpha(\varepsilon_k) \to 0$ one also gets

$$\frac{\partial_R |V_{k\alpha}|}{|V_{k\alpha}|} = \frac{\partial_R \sqrt{\Gamma_\alpha(\varepsilon_k)}}{\sqrt{\Gamma_\alpha(\varepsilon_k)}} = \frac{\partial_R \Gamma_\alpha(\varepsilon_k)}{2\Gamma_\alpha(\varepsilon_k)} \qquad \text{(SG16)}$$

and

$$\frac{\partial_R V_{k\alpha}}{V_{k\alpha}} = \frac{\partial_R |V_{k\alpha}|}{|V_{k\alpha}|} - i\exp(-i\Phi_{k\alpha})(\partial_R \Phi_{k\alpha}) \tag{SG17}$$

With (SG14) - (SG17), Eq.(SG13) can be re-written in the following form:

$$\langle \psi_{k\beta} | \partial_R \hat{V} | \psi_{n\alpha} \rangle = G_{dd}^a(\varepsilon_k) G_{dd}^r(\varepsilon_n) V_{n\alpha}^* V_{k\beta} \left\{ \sum_{m\gamma} V_{m\gamma} (\partial_R V_{m\gamma}^*) \frac{1}{\varepsilon_n - \varepsilon_m + i\eta} + (\partial_R V_{m\gamma}) V_{m\gamma}^* \frac{1}{\varepsilon_k - \varepsilon_m - i\eta} + (\partial_R \varepsilon_d) \right.$$

$$\left. + G_{dd}^r(\varepsilon_n)^{-1} \frac{(\partial_R V_{n\alpha}^*)}{V_{n\alpha}^*} + G_{dd}^a(\varepsilon_k)^{-1} \frac{(\partial_R V_{k\beta})}{V_{k\beta}} \right\}$$

$$= G_{dd}^a(\varepsilon_k) G_{dd}^r(\varepsilon_n) V_{n\alpha}^* V_{k\beta} \left[ \frac{1}{2} \sum_{m\gamma} (\partial_R |V_{m\gamma}|^2) \left\{ \frac{1}{\varepsilon_n - \varepsilon_m + i\eta} + \frac{1}{\varepsilon_k - \varepsilon_m - i\eta} \right\} + (\partial_R \varepsilon_d) \right.$$

$$+ i \sum_{m\gamma} |V_{m\gamma}|^2 (\partial_R \Phi_{m\gamma}) \left\{ \frac{1}{\varepsilon_n - \varepsilon_m + i\eta} - \frac{1}{\varepsilon_k - \varepsilon_m - i\eta} \right\}$$

$$+ \left( \varepsilon_n - \varepsilon_d - \Lambda(\varepsilon_n) + i\frac{\Gamma(\varepsilon_n)}{2} \right) \left( i(\partial_R \Phi_{n\alpha}) + \frac{\partial_R \Gamma_\alpha(\varepsilon_n)}{2\Gamma_\alpha(\varepsilon_n)} \right) + \left( \varepsilon_k - \varepsilon_d - \Lambda(\varepsilon_k) - i\frac{\Gamma(\varepsilon_k)}{2} \right) \left( -i(\partial_R \Phi_{k\beta}) + \frac{\partial_R \Gamma_\beta(\varepsilon_k)}{2\Gamma_\beta(\varepsilon_k)} \right) \right]$$

(SG18)

In Eq. (SG18) the term in the bracket […] has the both imaginary and real parts. Its real part:

$$\text{Re}[...] = \text{Re}\left[\frac{1}{2}\sum_{m\gamma}\left(\partial_R |V_{m\gamma}|^2\right)\left\{\frac{1}{\varepsilon_n - \varepsilon_m + i\eta} + \frac{1}{\varepsilon_k - \varepsilon_m - i\eta}\right\} + \left(\partial_R \varepsilon_d\right)\right.$$

$$+ i\sum_{m\gamma}|V_{m\gamma}|^2 \left(\partial_R \Phi_{m\gamma}\right)\left\{\frac{1}{\varepsilon_n - \varepsilon_m + i\eta} - \frac{1}{\varepsilon_k - \varepsilon_m - i\eta}\right\}$$

$$+\left(\varepsilon_n - \varepsilon_d - \Lambda(\varepsilon_n) + i\frac{\Gamma(\varepsilon_n)}{2}\right)\left(i(\partial_R \Phi_{n\alpha}) + \frac{\partial_R \Gamma_\alpha(\varepsilon_n)}{2\Gamma_\alpha(\varepsilon_n)}\right) + \left(\varepsilon_k - \varepsilon_d - \Lambda(\varepsilon_k) - i\frac{\Gamma(\varepsilon_k)}{2}\right)\left(-i(\partial_R \Phi_{k\beta}) + \frac{\partial_R \Gamma_\beta(\varepsilon_k)}{2\Gamma_\beta(\varepsilon_k)}\right)\right]$$

$$= \partial_R\left(\frac{\Lambda(\varepsilon_n) + \Lambda(\varepsilon_k)}{2} + \varepsilon_d\right) + \frac{1}{2}\sum_\gamma \left\{\left(\partial_R \Phi_{n\gamma}\right)\Gamma_\gamma(\varepsilon_n) + \left(\partial_R \Phi_{k\gamma}\right)\Gamma_\gamma(\varepsilon_k)\right\}$$

$$+ \left(\varepsilon_n - \varepsilon_d - \Lambda(\varepsilon_n)\right)\frac{\partial_R \Gamma_\alpha(\varepsilon_n)}{2\Gamma_\alpha(\varepsilon_n)} + \left(\varepsilon_k - \varepsilon_d - \Lambda(\varepsilon_k)\right)\frac{\partial_R \Gamma_\beta(\varepsilon_k)}{2\Gamma_\beta(\varepsilon_k)}$$

$$- \left(\partial_R \Phi_{n\alpha}\right)\frac{\Gamma(\varepsilon_n)}{2} - \left(\partial_R \Phi_{k\beta}\right)\frac{\Gamma(\varepsilon_k)}{2}$$

$$= \partial_R\left(\frac{\Lambda(\varepsilon_n) + \Lambda(\varepsilon_k)}{2} + \varepsilon_d\right) + \frac{1}{2}\sum_\gamma \left\{\partial_R\left(\Phi_{n\gamma} - \Phi_{n\alpha}\right)\Gamma_\gamma(\varepsilon_n) + \partial_R\left(\Phi_{k\gamma} - \Phi_{k\beta}\right)\Gamma_\gamma(\varepsilon_k)\right\}$$

$$+ \left(\varepsilon_n - \varepsilon_d - \Lambda(\varepsilon_n)\right)\frac{\partial_R \Gamma_\alpha(\varepsilon_n)}{2\Gamma_\alpha(\varepsilon_n)} + \left(\varepsilon_k - \varepsilon_d - \Lambda(\varepsilon_k)\right)\frac{\partial_R \Gamma_\beta(\varepsilon_k)}{2\Gamma_\beta(\varepsilon_k)}$$

(SG19)

and the imaginary part:

$$\text{Im}[...]$$

$$= \frac{1}{4}\partial_R\left(\Gamma(\varepsilon_k) - \Gamma(\varepsilon_n)\right) + \sum_{m\gamma}|V_{m\gamma}|^2 \left(\partial_R \Phi_{m\gamma}\right)\text{PP}\left\{\frac{1}{\varepsilon_n - \varepsilon_m} - \frac{1}{\varepsilon_k - \varepsilon_m}\right\}$$

$$+ \left(\varepsilon_n - \varepsilon_d - \Lambda(\varepsilon_n)\right)\left(\partial_R \Phi_{n\alpha}\right) - \left(\varepsilon_k - \varepsilon_d - \Lambda(\varepsilon_k)\right)\left(\partial_R \Phi_{k\beta}\right)$$

$$+ \frac{\partial_R \Gamma_\alpha(\varepsilon_n)}{4\Gamma_\alpha(\varepsilon_n)}\Gamma(\varepsilon_n) - \frac{\partial_R \Gamma_\beta(\varepsilon_k)}{4\Gamma_\beta(\varepsilon_k)}\Gamma(\varepsilon_k)$$

(SG20)

Thus,

$$\left|\left\langle \psi_{k\beta} \left| \partial_R \hat{V} \right| \psi_{n\alpha} \right\rangle\right|^2 = \frac{A_{dd}(\varepsilon_n) A_{dd}(\varepsilon_k)}{\Gamma(\varepsilon_n) \Gamma(\varepsilon_k)} \left|V_{n\alpha} V_{k\beta}\right|^2 \left\{(\text{Re}[...])^2 + (\text{Im}[...])^2\right\}$$

$$= \frac{A_{dd}(\varepsilon_n) A_{dd}(\varepsilon_k)}{\Gamma(\varepsilon_n) \Gamma(\varepsilon_k)} \left|V_{n\alpha} V_{k\beta}\right|^2$$

$$\times \left[ \left\{ \partial_R \left( \frac{\Lambda(\varepsilon_n) + \Lambda(\varepsilon_k)}{2} + \varepsilon_d \right) + \frac{1}{2} \sum_\gamma \left\{ \partial_R \left( \Phi_{n\gamma} - \Phi_{n\alpha} \right) \Gamma_\gamma(\varepsilon_n) + \partial_R \left( \Phi_{k\gamma} - \Phi_{k\beta} \right) \Gamma_\gamma(\varepsilon_k) \right\} \right.$$

$$+ \left( \varepsilon_n - \varepsilon_d - \Lambda(\varepsilon_n) \right) \frac{\partial_R \Gamma_\alpha(\varepsilon_n)}{2\Gamma_\alpha(\varepsilon_n)} + \left( \varepsilon_k - \varepsilon_d - \Lambda(\varepsilon_k) \right) \frac{\partial_R \Gamma_\beta(\varepsilon_k)}{2\Gamma_\beta(\varepsilon_k)} \right\}^2$$

$$+ \left\{ \frac{1}{4} \partial_R \left( \Gamma(\varepsilon_k) - \Gamma(\varepsilon_n) \right) + \sum_{m\gamma} \left|V_{m\gamma}\right|^2 \left( \partial_R \Phi_{m\gamma} \right) \text{PP} \left\{ \frac{1}{\varepsilon_n - \varepsilon_m} - \frac{1}{\varepsilon_k - \varepsilon_m} \right\} \right.$$

$$+ \left( \varepsilon_n - \varepsilon_d - \Lambda(\varepsilon_n) \right) \left( \partial_R \Phi_{n\alpha} \right) - \left( \varepsilon_k - \varepsilon_d - \Lambda(\varepsilon_k) \right) \left( \partial_R \Phi_{k\beta} \right) + \frac{\partial_R \Gamma_\alpha(\varepsilon_n)}{4\Gamma_\alpha(\varepsilon_n)} \Gamma(\varepsilon_n) - \frac{\partial_R \Gamma_\beta(\varepsilon_k)}{4\Gamma_\beta(\varepsilon_k)} \Gamma(\varepsilon_k) \right\}^2 \right]$$

(SG21)

Using Eq. (SG21) the derivative $\partial_{\varepsilon_n} \left|\left\langle \psi_{k\beta} \left| \partial_R \hat{V} \right| \psi_{n\alpha} \right\rangle\right|^2$ can be evaluated as follows:

$$\partial_{\varepsilon_n}\left|\left\langle \psi_{k\beta}\left|\partial_R \hat{V}\right|\psi_{n\alpha}\right\rangle\right|^2$$

$$=\left(\partial_{\varepsilon_n}\left(\frac{A_{dd}(\varepsilon_n)}{\Gamma(\varepsilon_n)}\right)\frac{A_{dd}(\varepsilon_k)}{\Gamma(\varepsilon_k)}|V_{n\alpha}V_{k\beta}|^2+\frac{A_{dd}(\varepsilon_n)A_{dd}(\varepsilon_k)}{\Gamma(\varepsilon_n)\Gamma(\varepsilon_k)}\left(\partial_{\varepsilon_n}|V_{n\alpha}|^2\right)|V_{k\beta}|^2\right)$$

$$\times\Bigg[\Bigg\{\partial_R\left(\frac{\Lambda(\varepsilon_n)+\Lambda(\varepsilon_k)}{2}+\varepsilon_d\right)+\frac{1}{2}\sum_\gamma\Big\{\partial_R\left(\Phi_{n\gamma}-\Phi_{n\alpha}\right)\Gamma_\gamma(\varepsilon_n)+\partial_R\left(\Phi_{k\gamma}-\Phi_{k\beta}\right)\Gamma_\gamma(\varepsilon_k)\Big\}$$

$$+\left(\varepsilon_n-\varepsilon_d-\Lambda(\varepsilon_n)\right)\frac{\partial_R\Gamma_\alpha(\varepsilon_n)}{2\Gamma_\alpha(\varepsilon_n)}+\left(\varepsilon_k-\varepsilon_d-\Lambda(\varepsilon_k)\right)\frac{\partial_R\Gamma_\beta(\varepsilon_k)}{2\Gamma_\beta(\varepsilon_k)}\Bigg\}^2$$

$$+\Bigg\{\frac{1}{4}\partial_R\left(\Gamma(\varepsilon_k)-\Gamma(\varepsilon_n)\right)+\sum_{m\gamma}|V_{m\gamma}|^2\left(\partial_R\Phi_{m\gamma}\right)\mathrm{PP}\left\{\frac{1}{\varepsilon_n-\varepsilon_m}-\frac{1}{\varepsilon_k-\varepsilon_m}\right\}$$

$$+\left(\varepsilon_n-\varepsilon_d-\Lambda(\varepsilon_n)\right)\left(\partial_R\Phi_{n\alpha}\right)-\left(\varepsilon_k-\varepsilon_d-\Lambda(\varepsilon_k)\right)\left(\partial_R\Phi_{k\beta}\right)+\frac{\partial_R\Gamma_\alpha(\varepsilon_n)}{4\Gamma_\alpha(\varepsilon_n)}\Gamma(\varepsilon_n)-\frac{\partial_R\Gamma_\beta(\varepsilon_k)}{4\Gamma_\beta(\varepsilon_k)}\Gamma(\varepsilon_k)\Bigg\}^2\Bigg]$$

$$+\frac{A_{dd}(\varepsilon_n)A_{dd}(\varepsilon_k)}{\Gamma(\varepsilon_n)\Gamma(\varepsilon_k)}|V_{n\alpha}V_{k\beta}|^2$$

$$\times\Bigg[2\Bigg\{\partial_R\left(\frac{\Lambda(\varepsilon_n)+\Lambda(\varepsilon_k)}{2}+\varepsilon_d\right)+\frac{1}{2}\sum_\gamma\Big\{\partial_R\left(\Phi_{n\gamma}-\Phi_{n\alpha}\right)\Gamma_\gamma(\varepsilon_n)+\partial_R\left(\Phi_{k\gamma}-\Phi_{k\beta}\right)\Gamma_\gamma(\varepsilon_k)\Big\}$$

$$+\left(\varepsilon_n-\varepsilon_d-\Lambda(\varepsilon_n)\right)\frac{\partial_R\Gamma_\alpha(\varepsilon_n)}{2\Gamma_\alpha(\varepsilon_n)}+\left(\varepsilon_k-\varepsilon_d-\Lambda(\varepsilon_k)\right)\frac{\partial_R\Gamma_\beta(\varepsilon_k)}{2\Gamma_\beta(\varepsilon_k)}\Bigg\}$$

$$\times\Bigg\{\partial_R\partial_{\varepsilon_n}\left(\frac{\Lambda(\varepsilon_n)}{2}\right)+\frac{1}{2}\sum_\gamma\partial_{\varepsilon_n}\Big\{\partial_R\left(\Phi_{n\gamma}-\Phi_{n\alpha}\right)\Gamma_\gamma(\varepsilon_n)\Big\}$$

$$+\left(1-\partial_{\varepsilon_n}\Lambda(\varepsilon_n)\right)\frac{\partial_R\Gamma_\alpha(\varepsilon_n)}{2\Gamma_\alpha(\varepsilon_n)}+\left(\varepsilon_n-\varepsilon_d-\Lambda(\varepsilon_n)\right)\partial_{\varepsilon_n}\left(\frac{\partial_R\Gamma_\alpha(\varepsilon_n)}{2\Gamma_\alpha(\varepsilon_n)}\right)\Bigg\}$$

$$+2\Bigg\{\frac{1}{4}\partial_R\left(\Gamma(\varepsilon_k)-\Gamma(\varepsilon_n)\right)+\sum_{m\gamma}|V_{m\gamma}|^2\left(\partial_R\Phi_{m\gamma}\right)\mathrm{PP}\left\{\frac{1}{\varepsilon_n-\varepsilon_m}-\frac{1}{\varepsilon_k-\varepsilon_m}\right\}$$

$$+\left(\varepsilon_n-\varepsilon_d-\Lambda(\varepsilon_n)\right)\left(\partial_R\Phi_{n\alpha}\right)-\left(\varepsilon_k-\varepsilon_d-\Lambda(\varepsilon_k)\right)\left(\partial_R\Phi_{k\beta}\right)+\frac{\partial_R\Gamma_\alpha(\varepsilon_n)}{4\Gamma_\alpha(\varepsilon_n)}\Gamma(\varepsilon_n)-\frac{\partial_R\Gamma_\beta(\varepsilon_k)}{4\Gamma_\beta(\varepsilon_k)}\Gamma(\varepsilon_k)\Bigg\}$$

$$\times\Bigg\{-\frac{1}{4}\partial_R\partial_{\varepsilon_n}\Gamma(\varepsilon_n)+\partial_{\varepsilon_n}\Bigg\{\sum_{m\gamma}\mathrm{PP}\frac{|V_{m\gamma}|^2\left(\partial_R\Phi_{m\gamma}\right)}{\varepsilon_n-\varepsilon_m}\Bigg\}+\left(1-\partial_{\varepsilon_n}\Lambda(\varepsilon_n)\right)\left(\partial_R\Phi_{n\alpha}\right)+\left(\varepsilon_n-\varepsilon_d-\Lambda(\varepsilon_n)\right)\left(\partial_{\varepsilon_n}\partial_R\Phi_{n\alpha}\right)$$

$$+\partial_{\varepsilon_n}\left(\frac{\partial_R\Gamma_\alpha(\varepsilon_n)}{4\Gamma_\alpha(\varepsilon_n)}\Gamma(\varepsilon_n)\right)\Bigg\}\Bigg]$$

29(SG22)

Thus

$$\delta(\varepsilon_n - \varepsilon_k)\partial_{\varepsilon_n}\left|\langle\psi_{k\beta}|\partial_R\hat{V}|\psi_{n\alpha}\rangle\right|^2$$

$$= \delta(\varepsilon_n - \varepsilon_k)\left(\partial_{\varepsilon_n}\left(\frac{A_{dd}(\varepsilon_n)}{\Gamma(\varepsilon_n)}\right)\frac{A_{dd}(\varepsilon_n)}{\Gamma(\varepsilon_n)}|V_{n\alpha}V_{n\beta}|^2 + \left(\frac{A_{dd}(\varepsilon_n)}{\Gamma(\varepsilon_n)}\right)^2\left(\partial_{\varepsilon_n}|V_{n\alpha}|^2\right)|V_{n\beta}|^2\right)$$

$$\times\left[\left\{\partial_R(\Lambda(\varepsilon_n) + \varepsilon_d) + \frac{1}{2}\sum_\gamma\{\partial_R(2\Phi_{n\gamma} - \Phi_{n\alpha} - \Phi_{n\beta})\Gamma_\gamma(\varepsilon_n)\} + (\varepsilon_n - \varepsilon_d - \Lambda(\varepsilon_n))\left(\frac{\partial_R\Gamma_\alpha(\varepsilon_n)}{2\Gamma_\alpha(\varepsilon_n)} + \frac{\partial_R\Gamma_\beta(\varepsilon_n)}{2\Gamma_\beta(\varepsilon_n)}\right)\right\}^2$$

$$+ \left\{(\varepsilon_n - \varepsilon_d - \Lambda(\varepsilon_n))\partial_R(\Phi_{n\alpha} - \Phi_{n\beta}) + \Gamma(\varepsilon_n)\left(\frac{\partial_R\Gamma_\alpha(\varepsilon_n)}{4\Gamma_\alpha(\varepsilon_n)} - \frac{\partial_R\Gamma_\beta(\varepsilon_n)}{4\Gamma_\beta(\varepsilon_n)}\right)\right\}^2\right]$$

$$+ \left(\frac{A_{dd}(\varepsilon_n)}{\Gamma(\varepsilon_n)}\right)^2\delta(\varepsilon_n - \varepsilon_k)|V_{n\alpha}V_{n\beta}|^2$$

$$\times\left[2\left\{\partial_R(\Lambda(\varepsilon_n) + \varepsilon_d) + \frac{1}{2}\sum_\gamma\{\partial_R(2\Phi_{n\gamma} - \Phi_{n\alpha} - \Phi_{n\beta})\Gamma_\gamma(\varepsilon_n)\} + (\varepsilon_n - \varepsilon_d - \Lambda(\varepsilon_n))\left(\frac{\partial_R\Gamma_\alpha(\varepsilon_n)}{2\Gamma_\alpha(\varepsilon_n)} + \frac{\partial_R\Gamma_\beta(\varepsilon_n)}{2\Gamma_\beta(\varepsilon_n)}\right)\right\}\right.$$

$$\times\left\{\partial_R\partial_{\varepsilon_n}\left(\frac{\Lambda(\varepsilon_n)}{2}\right) + \frac{1}{2}\sum_\gamma\partial_{\varepsilon_n}\{\partial_R(\Phi_{n\gamma} - \Phi_{n\alpha})\Gamma_\gamma(\varepsilon_n)\} + (1 - \partial_{\varepsilon_n}\Lambda(\varepsilon_n))\frac{\partial_R\Gamma_\alpha(\varepsilon_n)}{2\Gamma_\alpha(\varepsilon_n)} + (\varepsilon_n - \varepsilon_d - \Lambda(\varepsilon_n))\partial_{\varepsilon_n}\left(\frac{\partial_R\Gamma_\alpha(\varepsilon_n)}{2\Gamma_\alpha(\varepsilon_n)}\right)\right\}$$

$$+ 2\left\{(\varepsilon_n - \varepsilon_d - \Lambda(\varepsilon_n))\partial_R(\Phi_{n\alpha} - \Phi_{n\beta}) + \Gamma(\varepsilon_n)\left(\frac{\partial_R\Gamma_\alpha(\varepsilon_n)}{4\Gamma_\alpha(\varepsilon_n)} - \frac{\partial_R\Gamma_\beta(\varepsilon_n)}{4\Gamma_\beta(\varepsilon_n)}\right)\right\}$$

$$\times\left\{-\frac{1}{4}\partial_R\partial_{\varepsilon_n}\Gamma(\varepsilon_n) + \frac{1}{2\pi}\partial_{\varepsilon_n}\left\{\int_{-\infty}^{\infty}\sum_{m\gamma}PP\frac{2\pi|V_{m\gamma}|^2\delta(\varepsilon_m - \varepsilon')(\partial_R\Phi_{m\gamma})}{\varepsilon_n - \varepsilon_m}d\varepsilon'\right\}\right.$$

$$\left.\left. + (1 - \partial_{\varepsilon_n}\Lambda(\varepsilon_n))(\partial_R\Phi_{n\alpha}) + (\varepsilon_n - \varepsilon_d - \Lambda(\varepsilon_n))(\partial_{\varepsilon_n}\partial_R\Phi_{n\alpha}) + \partial_{\varepsilon_n}\left(\frac{\partial_R\Gamma_\alpha(\varepsilon_n)}{4\Gamma_\alpha(\varepsilon_n)}\Gamma(\varepsilon_n)\right)\right\}\right]$$

(SG23)

From Eq. (SG23) it follows that

$$\sum_k \pi \delta(\varepsilon_n - \varepsilon_k) \left( \partial_{\varepsilon_n} \left| \langle \psi_{k\beta} | \partial_R \hat{V} | \psi_{n\alpha} \rangle \right|^2 - \partial_{\varepsilon_n} \left| \langle \psi_{k\alpha} | \partial_R \hat{V} | \psi_{n\beta} \rangle \right|^2 \right)$$

$$= \frac{1}{4\pi D_\alpha(\varepsilon_n)} \left( \frac{A_{dd}(\varepsilon_n)}{\Gamma(\varepsilon_n)} \right)^2 \left\{ \left( \partial_{\varepsilon_n} \Gamma_\alpha(\varepsilon_n) \right) \Gamma_\beta(\varepsilon_n) - \left( \partial_{\varepsilon_n} \Gamma_\beta(\varepsilon_n) \right) \Gamma_\alpha(\varepsilon_n) \right\}$$

$$\times \left[ \partial_R \left( \Lambda(\varepsilon_n) + \varepsilon_d \right) + \frac{1}{2} \sum_\gamma \left\{ \partial_R \left( 2\Phi_{n\gamma} - \Phi_{n\alpha} - \Phi_{n\beta} \right) \Gamma_\gamma(\varepsilon_n) \right\} \right.$$

$$\left. + \left( \varepsilon_n - \varepsilon_d - \Lambda(\varepsilon_n) \right) \left( \frac{\partial_R \Gamma_\alpha(\varepsilon_n)}{2\Gamma_\alpha(\varepsilon_n)} + \frac{\partial_R \Gamma_\beta(\varepsilon_n)}{2\Gamma_\beta(\varepsilon_n)} \right) \right]^2$$

$$+ \frac{1}{4\pi D_\alpha(\varepsilon_n)} \left( \frac{A_{dd}(\varepsilon_n)}{\Gamma(\varepsilon_n)} \right)^2 \Gamma_\alpha(\varepsilon_n) \Gamma_\beta(\varepsilon_n)$$

$$\times \left[ 2 \left\{ \partial_R \left( \Lambda(\varepsilon_n) + \varepsilon_d \right) + \frac{1}{2} \sum_\gamma \left\{ \partial_R \left( 2\Phi_{n\gamma} - \Phi_{n\alpha} - \Phi_{n\beta} \right) \Gamma_\gamma(\varepsilon_n) \right\} \right. \right.$$

$$\left. + \left( \varepsilon_n - \varepsilon_d - \Lambda(\varepsilon_n) \right) \left( \frac{\partial_R \Gamma_\alpha(\varepsilon_n)}{2\Gamma_\alpha(\varepsilon_n)} + \frac{\partial_R \Gamma_\beta(\varepsilon_n)}{2\Gamma_\beta(\varepsilon_n)} \right) \right\}$$

$$\times \left\{ \frac{1}{2} \sum_\gamma \partial_{\varepsilon_n} \left\{ \partial_R \left( \Phi_{n\beta} - \Phi_{n\alpha} \right) \Gamma_\gamma(\varepsilon_n) \right\} \right.$$

$$\left. + \left( 1 - \partial_{\varepsilon_n} \Lambda(\varepsilon_n) \right) \left( \frac{\partial_R \Gamma_\alpha(\varepsilon_n)}{2\Gamma_\alpha(\varepsilon_n)} - \frac{\partial_R \Gamma_\beta(\varepsilon_n)}{2\Gamma_\beta(\varepsilon_n)} \right) + \left( \varepsilon_n - \varepsilon_d - \Lambda(\varepsilon_n) \right) \partial_{\varepsilon_n} \left( \frac{\partial_R \Gamma_\alpha(\varepsilon_n)}{2\Gamma_\alpha(\varepsilon_n)} - \frac{\partial_R \Gamma_\beta(\varepsilon_n)}{2\Gamma_\beta(\varepsilon_n)} \right) \right\}$$

$$+ 2 \left\{ \left( \varepsilon_n - \varepsilon_d - \Lambda(\varepsilon_n) \right) \partial_R \left( \Phi_{n\alpha} - \Phi_{n\beta} \right) + \Gamma(\varepsilon_n) \left( \frac{\partial_R \Gamma_\alpha(\varepsilon_n)}{4\Gamma_\alpha(\varepsilon_n)} - \frac{\partial_R \Gamma_\beta(\varepsilon_n)}{4\Gamma_\beta(\varepsilon_n)} \right) \right\}$$

$$\times \left\{ -\frac{1}{2} \partial_R \partial_{\varepsilon_n} \Gamma(\varepsilon_n) + \partial_{\varepsilon_n} \left\{ \mathrm{PP} \int_{-\infty}^{\infty} d\varepsilon' \sum_\gamma \Gamma_\gamma(\varepsilon') \frac{\partial_R \Phi_\gamma(\varepsilon')}{\pi(\varepsilon_n - \varepsilon')} \right\} + \left( 1 - \partial_{\varepsilon_n} \Lambda(\varepsilon_n) \right) \partial_R \left( \Phi_{n\alpha} + \Phi_{n\beta} \right) \right.$$

$$\left. \left. + \left( \varepsilon_n - \varepsilon_d - \Lambda(\varepsilon_n) \right) \partial_{\varepsilon_n} \partial_R \left( \Phi_{n\alpha} + \Phi_{n\beta} \right) + \partial_{\varepsilon_n} \left( \left\{ \frac{\partial_R \Gamma_\alpha(\varepsilon_n)}{4\Gamma_\alpha(\varepsilon_n)} + \frac{\partial_R \Gamma_\beta(\varepsilon_n)}{4\Gamma_\beta(\varepsilon_n)} \right\} \Gamma(\varepsilon_n) \right) \right\} \right]$$

(SG24)

and

$$\sum_k \pi \delta(\varepsilon_n - \varepsilon_k) |\langle \psi_{k\beta} | \partial_R \hat{V} | \psi_{n\alpha} \rangle|^2$$

$$= \frac{1}{4\pi D_\alpha(\varepsilon_n)} \left( \frac{A_{dd}(\varepsilon_n)}{\Gamma(\varepsilon_n)} \right)^2 \Gamma_\alpha(\varepsilon_n) \Gamma_\beta(\varepsilon_n) \left\{ \partial_R \left( \Lambda(\varepsilon_n) + \varepsilon_d \right) + \frac{1}{2} \sum_\gamma \left\{ \partial_R \left( 2\Phi_{n\gamma} - \Phi_{n\alpha} - \Phi_{n\beta} \right) \Gamma_\gamma(\varepsilon_n) \right\} \right.$$

$$\left. + (\varepsilon_n - \varepsilon_d - \Lambda(\varepsilon_n)) \left( \frac{\partial_R \Gamma_\alpha(\varepsilon_n)}{2\Gamma_\alpha(\varepsilon_n)} + \frac{\partial_R \Gamma_\beta(\varepsilon_n)}{2\Gamma_\beta(\varepsilon_n)} \right) \right\}^2$$

(SG25)

With (SG24) and (SG25), the double sums in Eqs. (SG11) and (SG10) can be converted to the following integrals ($\varepsilon$-dependences are dropped to shorten the notation):

$$\dot{W}_{II}^{(2)} = -\frac{1}{8\pi} (\dot{R})^2 \int_{-\infty}^{\infty} d\varepsilon \left( \frac{A_{dd}}{\Gamma} \right)^2 \sum_\alpha \sum_\beta (f_\alpha - f_\beta)$$

$$\times \left[ \left\{ (\partial_\varepsilon \Gamma_\alpha) \Gamma_\beta - (\partial_\varepsilon \Gamma_\beta) \Gamma_\alpha \right\} \left[ \partial_R (\Lambda + \varepsilon_d) + \frac{1}{2} \sum_\gamma \left\{ \partial_R (2\Phi_\gamma - \Phi_\alpha - \Phi_\beta) \Gamma_\gamma \right\} + (\varepsilon - \varepsilon_d - \Lambda) \left( \frac{\partial_R \Gamma_\alpha}{2\Gamma_\alpha} + \frac{\partial_R \Gamma_\beta}{2\Gamma_\beta} \right) \right]^2 \right.$$

$$+ \Gamma_\alpha \Gamma_\beta \left[ \left\{ 2\partial_R (\Lambda + \varepsilon_d) + \sum_\gamma \left\{ \partial_R (2\Phi_\gamma - \Phi_\alpha - \Phi_\beta) \Gamma_\gamma \right\} + (\varepsilon - \varepsilon_d - \Lambda) \left( \frac{\partial_R \Gamma_\alpha}{\Gamma_\alpha} + \frac{\partial_R \Gamma_\beta}{\Gamma_\beta} \right) \right\} \right.$$

$$\times \left\{ \frac{1}{2} \sum_\gamma \partial_\varepsilon \left\{ \partial_R (\Phi_\beta - \Phi_\alpha) \Gamma_\gamma \right\} + (1 - \partial_\varepsilon \Lambda) \left( \frac{\partial_R \Gamma_\alpha}{2\Gamma_\alpha} - \frac{\partial_R \Gamma_\beta}{2\Gamma_\beta} \right) + (\varepsilon - \varepsilon_d - \Lambda) \partial_\varepsilon \left( \frac{\partial_R \Gamma_\alpha}{2\Gamma_\alpha} - \frac{\partial_R \Gamma_\beta}{2\Gamma_\beta} \right) \right\}$$

$$+ \left\{ 2(\varepsilon - \varepsilon_d - \Lambda) \partial_R (\Phi_\alpha - \Phi_\beta) + \Gamma \left( \frac{\partial_R \Gamma_\alpha}{2\Gamma_\alpha} - \frac{\partial_R \Gamma_\beta}{2\Gamma_\beta} \right) \right\}$$

$$\times \left\{ -\frac{1}{2} \partial_R \partial_\varepsilon \Gamma + \partial_\varepsilon \partial_R \left\{ PP \int_{-\infty}^{\infty} d\varepsilon' \sum_\gamma \Gamma_\gamma(\varepsilon') \frac{(2\Phi_\gamma(\varepsilon') - \Phi_\alpha - \Phi_\beta)}{2\pi(\varepsilon - \varepsilon')} \right\} + \partial_R (\Phi_\alpha + \Phi_\beta) \right.$$

$$\left. \left. + (\varepsilon - \varepsilon_d) \partial_\varepsilon \partial_R (\Phi_\alpha + \Phi_\beta) + \partial_\varepsilon \left( \left\{ \frac{\partial_R \Gamma_\alpha}{4\Gamma_\alpha} + \frac{\partial_R \Gamma_\beta}{4\Gamma_\beta} \right\} \Gamma \right) \right] \right]$$

(SG26)

and

$$\dot{W}_I^{(2)} = -\frac{(\dot{R})^2}{4\pi} \int_{-\infty}^{\infty} d\varepsilon \left(\frac{A_{dd}}{\Gamma}\right)^2 \sum_\alpha (\partial_\varepsilon f_\alpha) \Gamma_\alpha \sum_\beta \Gamma_\beta$$

$$\times \left\{ \partial_R(\Lambda + \varepsilon_d) + \frac{1}{2} \sum_\gamma \{\partial_R(2\Phi_\gamma - \Phi_\alpha - \Phi_\beta)\Gamma_\gamma\} + (\varepsilon - \varepsilon_d - \Lambda)\left(\frac{\partial_R \Gamma_\alpha}{2\Gamma_\alpha} + \frac{\partial_R \Gamma_\beta}{2\Gamma_\beta}\right) \right\}^2$$

(SG27)

Now consider a specific scenario when only the dot energy is driven. In this case, only the first term in the bracket $[\![...]\!]$ in Eq. (SG26) (the one with the factor $\{(\partial_\varepsilon \Gamma_\alpha)\Gamma_\beta - (\partial_\varepsilon \Gamma_\beta)\Gamma_\alpha\}$) is non-zero:

$$\dot{W}^{(2)} = \dot{W}_I^{(2)} + \dot{W}_{II}^{(2)}$$

$$= -\frac{(\dot{R})^2}{8\pi} \int_{-\infty}^{\infty} d\varepsilon \left(\frac{A_{dd}}{\Gamma}\right)^2 K_d^2 \sum_\alpha \sum_\beta \left[ (\partial_\varepsilon f_\alpha + \partial_\varepsilon f_\beta)\Gamma_\alpha \Gamma_\beta + (f_\alpha - f_\beta)\{(\partial_\varepsilon \Gamma_\alpha)\Gamma_\beta - (\partial_\varepsilon \Gamma_\beta)\Gamma_\alpha\} \right]$$

$$= -\frac{(\dot{R})^2}{4\pi} \int_{-\infty}^{\infty} d\varepsilon \left(\frac{A_{dd}}{\Gamma}\right)^2 K_d^2 \sum_\alpha \left[ (\partial_\varepsilon f_\alpha)\Gamma_\alpha \Gamma + f_\alpha \{(\partial_\varepsilon \Gamma_\alpha)\Gamma - (\partial_\varepsilon \Gamma)\Gamma_\alpha\} \right]$$

$$= -\frac{(\dot{R})^2}{4\pi} \int_{-\infty}^{\infty} d\varepsilon A_{dd}^2 K_d^2 \partial_\varepsilon \left( \sum_\alpha \frac{f_\alpha \Gamma_\alpha}{\Gamma} \right)$$

(SG28)

This expression coincides with Eq. (SF10) as expected.

In the case of a single bath when the both dot and couplings are driven only $\dot{W}_I^{(2)}$ contributes:

$$\dot{W}^{(2)} = \dot{W}_I^{(2)} = -\frac{(\dot{R})^2}{4\pi} \int_{-\infty}^{\infty} d\varepsilon \left(\frac{A_{dd}}{\Gamma}\right)^2 (\partial_\varepsilon f) \Gamma^2$$

$$\times \left\{ \partial_R(\Lambda + \varepsilon_d) + \frac{1}{2} \sum_\gamma \{\partial_R(2\Phi - \Phi - \Phi)\Gamma\} + (\varepsilon - \varepsilon_d - \Lambda)\left(\frac{\partial_R \Gamma}{2\Gamma} + \frac{\partial_R \Gamma}{2\Gamma}\right) \right\}^2$$

$$= -\frac{(\dot{R})^2}{4\pi} \int_{-\infty}^{\infty} d\varepsilon A_{dd}^2 (\partial_\varepsilon f) \Gamma^2 \left( \frac{\partial_R(\varepsilon - \varepsilon_d - \Lambda)\Gamma - (\varepsilon - \varepsilon_d - \Lambda)\partial_R \Gamma}{\Gamma^2} \right)^2$$

$$= -\frac{(\dot{R})^2}{4\pi} \int_{-\infty}^{\infty} d\varepsilon A_{dd}^2 (\partial_\varepsilon f) \Gamma^2 \left( \partial_R \left( \frac{\varepsilon - \varepsilon_d - \Lambda}{\Gamma} \right) \right)^2$$

(SG29)

**Section I. Calculations of the excess current and its relation to power**

The excess steady-state current operator is defined as follows:

$$\hat{J}_{ex} = \sum_{\alpha}\hat{J}_{\alpha} = \sum_{\alpha}i[\hat{H},\hat{N}_{\alpha}] = -i[\hat{H},\hat{d}^{\dagger}\hat{d}] = -i\sum_{n\alpha}\sum_{k\beta}\left\{V_{n\alpha}^{*}V_{k\beta}G_{dd}^{r}(\varepsilon_{n\alpha})G_{dd}^{a}(\varepsilon_{k\beta})\right\}\left[\hat{H},\hat{\psi}_{k\beta}^{\dagger}\hat{\psi}_{n\alpha}\right]$$
$$= -i\sum_{n\alpha}\sum_{k\beta}\left\{V_{n\alpha}^{*}V_{k\beta}G_{dd}^{r}(\varepsilon_{n\alpha})G_{dd}^{a}(\varepsilon_{k\beta})\right\}\hat{\psi}_{k\beta}^{\dagger}\hat{\psi}_{n\alpha}(\varepsilon_{k\beta}-\varepsilon_{n\alpha}) \quad \text{(SH1)}$$

At steady sate $J_{ex}^{(0)} = \mathrm{Tr}\left(\hat{\rho}_{ss}^{(0)}\hat{J}_{ex}\right)=0$. Let's compute non-adiabatic correction to the excess current when the dot and the couplings are a subject of slow driving.

Using (SH1), (50), (SG5) and (SG13), after evaluating limits $\eta_{1(2)}\to +0$ one gets

$$J_{ex}^{(1)} = \mathrm{Tr}\left(\hat{\rho}_{ss}^{(1)}\hat{J}_{ex}\right) = i(\dot{R})\sum_{n\alpha}\sum_{k\beta}\pi\delta(\varepsilon_{n}-\varepsilon_{k})\left\{V_{n\alpha}^{*}V_{k\beta}G_{dd}^{r}(\varepsilon_{n})G_{dd}^{a}(\varepsilon_{k})\right\}\left(f_{\alpha}(\varepsilon_{n})-f_{\beta}(\varepsilon_{k})\right)$$

$$\times G_{dd}^{r}(\varepsilon_{k})G_{dd}^{a}(\varepsilon_{n})V_{n\alpha}V_{k\beta}^{*}\left[\left(\frac{1}{2}\sum_{m\gamma}\left(\partial_{R}|V_{m\gamma}|^{2}\right)\right)\left\{\mathrm{PP}\left(\frac{1}{\varepsilon_{n}-\varepsilon_{m}}+\frac{1}{\varepsilon_{k}-\varepsilon_{m}}\right)+i\pi\left(\delta(\varepsilon_{n}-\varepsilon_{m})-\delta(\varepsilon_{k}-\varepsilon_{m})\right)\right\}+\left(\partial_{R}\varepsilon_{d}\right)\right]$$

$$-i\sum_{m\gamma}|V_{m\gamma}|^{2}\left(\partial_{R}\Phi_{m\gamma}\right)\left\{\mathrm{PP}\left(\frac{1}{\varepsilon_{n}-\varepsilon_{m}}-\frac{1}{\varepsilon_{k}-\varepsilon_{m}}\right)+i\pi\left(\delta(\varepsilon_{n}-\varepsilon_{m})+\delta(\varepsilon_{k}-\varepsilon_{m})\right)\right\}$$

$$+\left(\varepsilon_{n}-\varepsilon_{d}-\Lambda(\varepsilon_{n})-i\frac{\Gamma(\varepsilon_{n})}{2}\right)\left(-i\left(\partial_{R}\Phi_{n\alpha}\right)+\frac{\partial_{R}\Gamma_{\alpha}(\varepsilon_{n})}{2\Gamma_{\alpha}(\varepsilon_{n})}\right)+\left(\varepsilon_{k}-\varepsilon_{d}-\Lambda(\varepsilon_{k})+i\frac{\Gamma(\varepsilon_{k})}{2}\right)\left(i\left(\partial_{R}\Phi_{k\beta}\right)+\frac{\partial_{R}\Gamma_{\beta}(\varepsilon_{k})}{2\Gamma_{\beta}(\varepsilon_{k})}\right)\right]$$

(SH2)

Assume the wide-band limit and the driving frequency $\dot{\Phi}_{n\alpha}=(\dot{R})\partial_{R}\Phi=(\dot{R})K_{\Phi}$ is the same for all leads. Thus

$$J_{ex}^{(1)} = i(\dot{R})\sum_{n\alpha}\sum_{k\beta}\pi\delta(\varepsilon_{n}-\varepsilon_{k})\left\{|V_{n\alpha}V_{n\beta}|^{2}G_{dd}^{r}(\varepsilon_{n})G_{dd}^{a}(\varepsilon_{n})\right\}\left(f_{\alpha}(\varepsilon_{n})-f_{\beta}(\varepsilon_{n})\right)G_{dd}^{r}(\varepsilon_{n})G_{dd}^{a}(\varepsilon_{n})$$

$$\times\left[\left(\partial_{R}\varepsilon_{d}\right)+K_{\Phi}\Gamma+\left(\varepsilon_{n}-\varepsilon_{d}-i\frac{\Gamma}{2}\right)\left(-iK_{\Phi}+\frac{\partial_{R}\Gamma_{\alpha}}{2\Gamma_{\alpha}}\right)+\left(\varepsilon_{n}-\varepsilon_{d}+i\frac{\Gamma}{2}\right)\left(iK_{\Phi}+\frac{\partial_{R}\Gamma_{\beta}}{2\Gamma_{\beta}}\right)\right]$$

(SH3)

Swapping $\alpha$ and $\beta$ in (SH3) eliminates the anti-symmetric (imaginary) part which leads to

$$J_{ex}^{(1)} = (\dot{R}) \sum_{n\alpha} \sum_{k\beta} \pi \delta(\varepsilon_n - \varepsilon_k) \left\{ |V_{n\alpha} V_{n\beta}|^2 G_{dd}^r(\varepsilon_n) G_{dd}^a(\varepsilon_n) \right\} (f_\alpha(\varepsilon_n) - f_\beta(\varepsilon_n)) G_{dd}^r(\varepsilon_n) G_{dd}^a(\varepsilon_n) \frac{\Gamma}{2} \left[ \frac{\partial_R \Gamma_\alpha}{2\Gamma_\alpha} - \frac{\partial_R \Gamma_\beta}{2\Gamma_\beta} \right]$$

$$= \frac{(\dot{R})}{8\pi} \int_{-\infty}^{\infty} \left( \frac{A_{dd}}{\Gamma} \right)^2 \Gamma_\alpha \Gamma_\beta (f_\alpha - f_\beta) \left[ \frac{\partial_R \Gamma_\alpha}{2\Gamma_\alpha} - \frac{\partial_R \Gamma_\beta}{2\Gamma_\beta} \right] d\varepsilon$$

(SH4)

Now it is time to establish a connection between the current (SH1) and the excess power. The correction to power is obtained as follows:

$$\dot{W}^{(2)} = (\dot{R}) \mathrm{Tr}(\hat{\rho}_{ss}^{(1)} \partial_R \hat{V}) = (\dot{R}) \mathrm{Tr}\left( \hat{\rho}_{ss}^{(1)} \left\{ \sum_\alpha K_{\Phi\alpha}(\partial_{\Phi_\alpha} \hat{V}) + K_d(\partial_{\varepsilon_d} \hat{V}) + \sum_\alpha K_{\Gamma\alpha}(\partial_{\Gamma_\alpha} \hat{V}) \right\} \right) \quad \text{(SH5)}$$

where (see (SA1))

$$(\partial_{\Phi_\alpha} \hat{V}) = \sum_{m\gamma} (\partial_{\Phi_\alpha} V_{m\gamma}^*) \hat{d}^\dagger \hat{c}_{m\gamma} + (\partial_{\Phi_\alpha} V_{m\gamma}) \hat{c}_{m\gamma}^\dagger \hat{d} = i \sum_m (V_{m\alpha}^* \hat{d}^\dagger \hat{c}_{m\alpha} - V_{m\alpha} \hat{c}_{m\alpha}^\dagger \hat{d}) = -\hat{J}_\alpha \quad \text{(SH6)}$$

Since $\dot{\Phi}_{n\alpha} = (\dot{R}) K_\Phi$, Eq. (SH5) becomes

$$\dot{W}^{(2)} = (\dot{R}) \mathrm{Tr}\left( \hat{\rho}_{ss}^{(1)} \left\{ -K_\Phi \sum_\alpha \hat{J}_\alpha + K_d(\partial_{\varepsilon_d} \hat{V}) + \sum_\alpha K_{\Gamma\alpha}(\partial_{\Gamma_\alpha} \hat{V}) \right\} \right)$$

$$= (\dot{R}) \mathrm{Tr}\left( \hat{\rho}_{ss}^{(1)} \left\{ -K_\Phi \hat{J}_{ex} + K_d(\partial_{\varepsilon_d} \hat{V}) + \sum_\alpha K_{\Gamma\alpha}(\partial_{\Gamma_\alpha} \hat{V}) \right\} \right) = (\dot{R}) \mathrm{Tr}\left( \hat{\rho}_{ss}^{(1)} \left\{ K_d(\partial_{\varepsilon_d} \hat{V}) + \sum_\alpha K_{\Gamma\alpha}(\partial_{\Gamma_\alpha} \hat{V}) \right\} \right)$$

$$-K_\Phi (\dot{R}) J_{ex}^{(1)}$$

(SH7)

Thus, the contribution to the correction which comes from the phase driving is non-zero because of the excess current.

Now consider the case when $\partial_R \Gamma_\alpha = 0$ (tunneling rates are not driven and the wide-band limit is assumed). Then

$$J_{ex}^{(1)} = \text{Tr}\left(\hat{\rho}_{ss}^{(1)} \hat{J}_{ex}\right)$$
$$= i(\dot{R})\sum_{n\alpha}\sum_{k\beta}\pi\delta(\varepsilon_n - \varepsilon_k)\left\{V_{n\alpha}^* V_{k\beta} G_{dd}^r(\varepsilon_n) G_{dd}^a(\varepsilon_k)\right\}\left(f_\alpha(\varepsilon_n) - f_\beta(\varepsilon_k)\right)\left\{G_{dd}^r(\varepsilon_k) G_{dd}^a(\varepsilon_n) V_{n\alpha} V_{k\beta}^*\right\}$$
$$\times\left[\left(\varepsilon_n - \varepsilon_d - i\frac{\Gamma}{2}\right)\left(-i(\partial_R \Phi_{n\alpha})\right) + \left(\varepsilon_k - \varepsilon_d + i\frac{\Gamma(\varepsilon_k)}{2}\right)\left(i(\partial_R \Phi_{k\beta})\right) + \partial_R \varepsilon_d\right] \quad \text{(SH8)}$$
$$= (\dot{R})\sum_{n\alpha}\sum_{k\beta}\pi\delta(\varepsilon_n - \varepsilon_k)\left\{|V_{n\alpha}V_{n\beta}|^2 \frac{A_{dd}^2(\varepsilon_n)}{\Gamma^2}\right\}\left(f_\alpha(\varepsilon_n) - f_\beta(\varepsilon_n)\right)(\varepsilon_n - \varepsilon_d)\left(\partial_R(\Phi_{n\alpha} - \Phi_{n\beta})\right)$$
$$= \frac{(\dot{R})}{4\pi}\partial_R\left[\Phi_\alpha - \Phi_\beta\right]\Gamma_\alpha \Gamma_\beta \int_{-\infty}^{\infty}\left(\frac{A_{dd}}{\Gamma}\right)^2 (\varepsilon - \varepsilon_d)(f_\alpha - f_\beta) d\varepsilon$$

where the anti-symmetric imaginary part got eliminated by swapping α and β. This current arises from the interference of the waves coming from different baths.

**Section J. The first order correction to the outgoing distribution.**

As in Sections G and H, the system is a subject of slow driving with time-dependent Hamiltonian $\hat{H}(R(t))$.

From Eq. (SB23) it follows:

$$\phi_{\alpha\beta,out}^{(1)}(\varepsilon,t) = \frac{1}{2\pi}\int_{-\infty}^{\infty} d\omega \exp(i\omega t)\text{Tr}\left\{\hat{\rho}_{ss}(\hat{\chi}_{(\varepsilon-\omega/2)\alpha,-}^\dagger \hat{\chi}_{(\varepsilon+\omega/2)\beta,-})^{(1)} + \hat{\rho}_{ss}^{(1)}\hat{\chi}_{(\varepsilon-\omega/2)\alpha,-}^\dagger \hat{\chi}_{(\varepsilon+\omega/2)\beta,-}\right\} \text{(SI1)}$$

Eq. (SI1) can be split on two terms:

$$\phi_{\alpha\beta,I}^{(1)}(\varepsilon,t) = \frac{1}{2\pi}\int_{-\infty}^{\infty} d\omega \text{Tr}\left\{\hat{\rho}_{ss}(\hat{\chi}_{(\varepsilon-\omega/2)\alpha,-}^\dagger \hat{\chi}_{(\varepsilon+\omega/2)\beta,-})^{(1)} \exp(i\omega t)\right\} \quad \text{(SI2)a}$$

$$\phi_{\alpha\beta,II}^{(1)}(\varepsilon,t) = \frac{1}{2\pi}\int_{-\infty}^{\infty} d\omega \text{Tr}\left\{\hat{\rho}_{ss}^{(1)} \hat{\chi}_{(\varepsilon-\omega/2)\alpha,-}^\dagger \hat{\chi}_{(\varepsilon+\omega/2)\beta,-} \exp(i\omega t)\right\} \quad \text{(SI2)b}$$

$$\phi_{\alpha\beta,out}^{(1)}(\varepsilon,t) = \phi_{\alpha\beta,I}^{(1)}(\varepsilon,t) + \phi_{\alpha\beta,II}^{(1)}(\varepsilon,t) \quad \text{(SI2)c}$$

Since $\hat{\chi}_{\varepsilon\alpha,-}^\dagger$ is an outgoing solution, its time evolution is prescribed by the same Schrodinger as for the incoming solution but with the reverse time direction.

Indeed, the outgoing waves satisfy the time-dependent Schrodinger equation:

$$\partial_t \{\hat{\chi}^\dagger_{\varepsilon_k\alpha,-}(t)\hat{\chi}_{\varepsilon_n\beta,-}(t)\} = -i\left[\hat{H}(R(t)), \hat{\chi}^\dagger_{\varepsilon_k\alpha,-}(t)\hat{\chi}_{\varepsilon_n\beta,-}(t)\right] \tag{SI3}$$

To solve (SI3) the following ansatz can be used:

$$\hat{\chi}^\dagger_{\varepsilon_k\alpha,-}(t)\hat{\chi}_{\varepsilon_n\beta,-}(t) = \exp\left(-i\hat{H}(R(t))t\right)\left(\hat{\chi}^\dagger_{\varepsilon_k\alpha,-}(R(t))\hat{\chi}_{\varepsilon_n\beta,-}(R(t)) + \Delta\left(\hat{\chi}^\dagger_{\varepsilon_k\alpha,-}\hat{\chi}_{\varepsilon_n\beta,-}\right)(t)\right)\exp\left(i\hat{H}(R(t))t\right) \tag{SI4}$$

After substituting (SI4) into (SI3) (see also Eq. (47)) it follows that the (exact) correction:

$$\exp\left(-i\hat{H}(R(t))t\right)\Delta\left(\hat{\chi}^\dagger_{\varepsilon_k\alpha,-}\hat{\chi}_{\varepsilon_n\beta,-}\right)(t)\exp\left(i\hat{H}(R(t))t\right)$$
$$= \exp\left(-i\hat{H}(R(T))T\right)\Delta\left(\hat{\chi}^\dagger_{\varepsilon_k\alpha,-}\hat{\chi}_{\varepsilon_n\beta,-}\right)(T)\exp\left(i\hat{H}(R(T))T\right)$$
$$-\dot{R}\lim_{\eta\to+0}\int_T^t \exp(-\eta|(\tau-t)|)\hat{U}(t,\tau)\partial_R\left(\hat{\chi}^\dagger_{\varepsilon_k\alpha,-}(R(\tau))\hat{\chi}_{\varepsilon_n\beta,-}(R(\tau))\right)\hat{U}^\dagger(t,\tau)\exp(i(\varepsilon_n-\varepsilon_k)\tau)d\tau$$
(SI5)

The boundary condition is set in the future $T=\infty$: $\Delta\left(\hat{\chi}^\dagger_{\varepsilon_k\alpha,-}\hat{\chi}_{\varepsilon_n\beta,-}\right)(T) = 0$. Thus, performing the adiabatic approximation for the integrand $\hat{U}(t,\tau) \approx \exp\left(-i\hat{H}(R(t))(\tau-t)\right)$, $\hat{\chi}^\dagger_{\varepsilon_k\alpha,-}(R(\tau))\hat{\chi}_{\varepsilon_n\beta,-}(R(\tau)) \approx \hat{\chi}^\dagger_{\varepsilon_k\alpha,-}(R(t))\hat{\chi}_{\varepsilon_n\beta,-}(R(t))$ one gets the following expression for the correction:

$$\Delta\left(\hat{\chi}^\dagger_{\varepsilon_k\alpha,-}\hat{\chi}_{\varepsilon_n\beta,-}\right)(t) \approx \left(\hat{\chi}^\dagger_{\varepsilon_k\alpha,-}\hat{\chi}_{\varepsilon_n\beta,-}\right)^{(1)}$$
$$= \dot{R}\lim_{\eta\to+0}\int_0^\infty \exp\{(-\eta-i(\varepsilon_k-\varepsilon_n))\tau\}\exp\left(-i\hat{H}(R(t))(\tau-t)\right)\partial_R\{\hat{\chi}^\dagger_{\varepsilon_k\alpha,-}\hat{\chi}_{\varepsilon_n\beta,-}\}\exp\left(i\hat{H}(R(t))(\tau-t)\right)d\tau$$
(SI6)

Thus for the integrand in (SI2)a

$$\text{Tr}\left(\hat{\rho}_{ss}\left(\hat{\chi}^\dagger_{\varepsilon_k\alpha,-}\hat{\chi}_{\varepsilon_n\beta,-}\right)^{(1)}\right) = \dot{R}\lim_{\eta\to+0}\left(\frac{\eta}{(\varepsilon_k-\varepsilon_n)^2+\eta^2} - i\frac{\varepsilon_k-\varepsilon_n}{(\varepsilon_k-\varepsilon_n)^2+\eta^2}\right)\text{Tr}\left(\hat{\rho}_{ss}\partial_R\left(\hat{\chi}^\dagger_{\varepsilon_k\alpha,-}\hat{\chi}_{\varepsilon_n\beta,-}\right)\right) \tag{SI7}$$

Consider the case of the driven resonant level ($R=\varepsilon_d$) coupled to one bath.
Let's split (SI7) on two terms:

$$\text{Tr}\left(\hat{\rho}_{ss}\left(\hat{\chi}^\dagger_{\varepsilon_k,-}\hat{\chi}_{\varepsilon_n,-}\right)^{(1)}\right) = \dot{R}\lim_{\eta\to+0}\left(\frac{\eta}{(\varepsilon_k-\varepsilon_n)^2+\eta^2} - i\frac{\varepsilon_k-\varepsilon_n}{(\varepsilon_k-\varepsilon_n)^2+\eta^2}\right)$$
$$\times\left\{\text{Tr}\left(\hat{\rho}_{ss}\left(\partial_R\hat{\chi}^\dagger_{\varepsilon_k,-}\right)\hat{\chi}_{\varepsilon_n,-}\right) + \text{Tr}\left(\hat{\rho}_{ss}\hat{\chi}^\dagger_{\varepsilon_k,-}\partial_R\left(\hat{\chi}_{\varepsilon_n,-}\right)\right)\right\} \tag{SI8}$$

Thus, from (SE5) it follows:

$$\partial_R \hat{\chi}^\dagger_{\varepsilon_k,-} = -\sum_m V_k^* \frac{V_m G^a_{dd}(\varepsilon_k) G^r_{dd}(\varepsilon_m)}{\varepsilon_m - \varepsilon_k + i\eta_1} \hat{\chi}^\dagger_{\varepsilon_m,-} \qquad (SI9)$$

To proceed further, the outgoing states need to be expressed through the incoming ones:

$$\mathrm{Tr}\{\hat{\rho}_{ss} \hat{\chi}^\dagger_{\varepsilon_k,-} \hat{\chi}_{\varepsilon_n,-}\} = \mathrm{Tr}\left\{\hat{\rho}_{ss} \sum_m \sum_l S_{mk} \hat{\chi}^\dagger_{\varepsilon_m,+} \hat{\chi}_{\varepsilon_l,+} S^\dagger_{nl}\right\} = \mathrm{Tr}\left\{\hat{\rho}_{ss} \sum_m \sum_l S_{kk} \hat{\chi}^\dagger_{\varepsilon_m,+} \hat{\chi}_{\varepsilon_l,+} S^\dagger_{nn} \delta_{nl} \delta_{km}\right\}$$

$$= \mathrm{Tr}\{\hat{\rho}_{ss} S_{kk} \hat{\chi}^\dagger_{\varepsilon_k,+} \hat{\chi}_{\varepsilon_n,+} S^\dagger_{nn}\} = 2\pi f(\varepsilon_k) \delta(\varepsilon_k - \varepsilon_n) S_{kk} S^\dagger_{kk} = 2\pi f(\varepsilon_k) \delta(\varepsilon_k - \varepsilon_n) \qquad (SI10)$$

Substituting (SI9) into the first trace of (SI8) and using (SI10)

$$\mathrm{Tr}\{\hat{\rho}_{ss}(\partial_R \hat{\chi}^\dagger_{\varepsilon_k,-}) \hat{\chi}_{\varepsilon_n,-}\} = -\sum_m V_k^* \frac{V_m G^a_{dd}(\varepsilon_k) G^r_{dd}(\varepsilon_m)}{\varepsilon_m - \varepsilon_k + i\eta_1} \mathrm{Tr}\{\hat{\rho}_{ss} \hat{\chi}^\dagger_{\varepsilon_m,-} \hat{\chi}_{\varepsilon_n,-}\}$$

$$= -2\pi \sum_m V_k^* \frac{V_m G^a_{dd}(\varepsilon_k) G^r_{dd}(\varepsilon_m)}{\varepsilon_m - \varepsilon_k + i\eta_1} f(\varepsilon_n) \delta(\varepsilon_m - \varepsilon_n) \qquad (SI11)$$

By analogy for the second trace:

$$\mathrm{Tr}\{\hat{\rho}_{ss} \hat{\chi}^\dagger_{\varepsilon_k,-} \partial_R(\hat{\chi}_{\varepsilon_n,-})\} = 2\pi \sum_m V_m^* \frac{V_n G^a_{dd}(\varepsilon_m) G^r_{dd}(\varepsilon_n)}{\varepsilon_n - \varepsilon_m + i\eta_1} f(\varepsilon_k) \delta(\varepsilon_k - \varepsilon_m) \qquad (SI12)$$

Thus, the first term (SI2)a

$$\phi_I^{(1)}(\varepsilon, t) = \dot{\varepsilon}_d \int_{-\infty}^\infty d\omega \, \mathrm{Tr}\{\hat{\rho}_{ss}(\hat{\chi}^\dagger_{\varepsilon-\omega/2,-} \hat{\chi}_{\varepsilon+\omega/2,-})^{(1)} \exp(i\omega t)\}$$

$$= \dot{\varepsilon}_d \int_{-\infty}^\infty d\omega \lim_{\eta \to +0} \left(\frac{\eta}{\omega^2+\eta^2} + i\frac{\omega}{\omega^2+\eta^2}\right) \exp(i\omega t)$$

$$\times \left\{\left(-\sum_m V^*_{\varepsilon-\omega/2} \frac{V_{\varepsilon_m} G^a_{dd}(\varepsilon-\omega/2) G^r_{dd}(\varepsilon_m)}{\varepsilon_m - (\varepsilon-\omega/2) + i\eta_1} f(\varepsilon+\omega/2) \delta\{\varepsilon_m - (\varepsilon+\omega/2)\}\right)\right.$$

$$\left. + \left(\sum_m V^*_{\varepsilon_m} \frac{V_{\varepsilon+\omega/2} G^a_{dd}(\varepsilon_m) G^r_{dd}(\varepsilon+\omega/2)}{\varepsilon+\omega/2 - \varepsilon_m + i\eta_1} f(\varepsilon-\omega/2) \delta\{\varepsilon-\omega/2 - \varepsilon_m\}\right)\right\} \qquad (SI13)$$

where Eqs. (SI11) - (SI12) are used with $V_m \to V_{\varepsilon_m}$, $\varepsilon_n \to \varepsilon+\omega/2$ and $\varepsilon_k \to \varepsilon-\omega/2$.

To evaluate the second term $\phi_{II}^{(1)}(\varepsilon, t)$, recall (SI10) thus

$$\mathrm{Tr}\{(\partial_R \hat{\rho}_{ss})(\hat{\chi}^\dagger_{\varepsilon_k,-} \hat{\chi}_{\varepsilon_n,-})\} = -\mathrm{Tr}\{\hat{\rho}_{ss} \partial_R(\hat{\chi}^\dagger_{\varepsilon_k,-} \hat{\chi}_{\varepsilon_n,-})\} \text{ (see (SD3)) and Eq. (50)) can be employed:}$$

$$\mathrm{Tr}\left(\hat{\rho}_{ss}^{(1)}\hat{\chi}^{\dagger}_{\varepsilon_k,-}\hat{\chi}_{\varepsilon_n,-}\right)=\dot{R}\lim_{\eta\to+0}\left(\frac{\eta}{(\varepsilon_k-\varepsilon_n)^2+\eta^2}+i\frac{\varepsilon_k-\varepsilon_n}{(\varepsilon_k-\varepsilon_n)^2+\eta^2}\right)\mathrm{Tr}\left(\hat{\rho}_{ss}\partial_R\left(\hat{\chi}^{\dagger}_{\varepsilon_k,-}\hat{\chi}_{\varepsilon_n,-}\right)\right)\,(\mathrm{SI}14)$$

It is clear that (SI14) and (SI8) are only different in the sign before the principal part $i\dfrac{\varepsilon_k-\varepsilon_n}{(\varepsilon_k-\varepsilon_n)^2+\eta^2}$, thus they cancel out each other in the total correction. This is a consequence of the time reversal symmetry. With (SI13) the total correction takes the form:

$$\phi^{(1)}(\varepsilon,t)=\phi_I^{(1)}(\varepsilon,t)+\phi_{II}^{(1)}(\varepsilon,t)=\dot{\varepsilon}_d\int_{-\infty}^{\infty}d\omega\lim_{\eta\to+0}\left(\frac{2\eta}{\omega^2+\eta^2}\right)\exp(i\omega t)$$

$$\times\left\{\left(-\sum_m V^*_{\varepsilon-\omega/2}\frac{V_{\varepsilon_m}G^a_{dd}(\varepsilon-\omega/2)G^r_{dd}(\varepsilon_m)}{\varepsilon_m-(\varepsilon-\omega/2)+i\eta_1}f(\varepsilon+\omega/2)\delta\{\varepsilon_m-(\varepsilon+\omega/2)\}\right)\right.$$

$$\left.+\left(\sum_m V^*_{\varepsilon_m}\frac{V_{\varepsilon+\omega/2}G^a_{dd}(\varepsilon_m)G^r_{dd}(\varepsilon+\omega/2)}{\varepsilon+\omega/2-\varepsilon_m+i\eta_1}f(\varepsilon-\omega/2)\delta(\varepsilon-\omega/2-\varepsilon_m)\right)\right\} \quad (\mathrm{SI}15)$$

Now one can integrate (SI15) with respect to $\omega$: the integration will give a sum of two infinite series over $\varepsilon_m$:

$$\phi^{(1)}(\varepsilon,t)$$
$$=\dot{\varepsilon}_d\lim_{\eta\to+0}\left\{\left(-\sum_m\left(\frac{2\eta}{4(\varepsilon_m-\varepsilon)^2+\eta^2}\right)\exp(2i(\varepsilon_m-\varepsilon)t)V^*_{2\varepsilon-\varepsilon_m}\frac{V_{\varepsilon_m}G^a_{dd}(2\varepsilon-\varepsilon_m)G^r_{dd}(\varepsilon_m)}{2(\varepsilon_m-\varepsilon)+i\eta_1}f(\varepsilon_m)\right)\right.\quad(\mathrm{SI}16)$$
$$\left.+\left(\sum_m\left(\frac{2\eta}{4(\varepsilon_m-\varepsilon)^2+\eta^2}\right)\exp(-2i(\varepsilon_m-\varepsilon)t)V^*_{\varepsilon_m}\frac{V_{2\varepsilon-\varepsilon_m}G^a_{dd}(\varepsilon_m)G^r_{dd}(2\varepsilon-\varepsilon_m)}{-2(\varepsilon_m-\varepsilon)+i\eta_1}f(\varepsilon_m)\right)\right\}$$

Introducing new variable $\Delta\varepsilon_m=2(\varepsilon_m-\varepsilon)$ where $-\infty<\Delta\varepsilon_m<\infty$ we have for (SI16)

$$\phi^{(1)}(\varepsilon,t)$$
$$=\dot{\varepsilon}_d\lim_{\eta\to+0}\left\{\left(-\sum_m\left(\frac{2\eta}{\Delta\varepsilon_m^2+\eta^2}\right)\exp(i\Delta\varepsilon_m t)V^*_{\varepsilon-\Delta\varepsilon_m/2}\frac{V_{\varepsilon+\Delta\varepsilon_m/2}G^a_{dd}(\varepsilon-\Delta\varepsilon_m/2)G^r_{dd}(\varepsilon+\Delta\varepsilon_m/2)}{\Delta\varepsilon_m+i\eta_1}f(\varepsilon+\Delta\varepsilon_m/2)\right)\right.$$
$$\left.+\left(\sum_m\left(\frac{2\eta}{\Delta\varepsilon_m^2+\eta^2}\right)\exp(-i\Delta\varepsilon_m t)V^*_{\varepsilon+\Delta\varepsilon_m/2}\frac{V_{\varepsilon-\Delta\varepsilon_m/2}G^a_{dd}(\varepsilon+\Delta\varepsilon_m/2)G^r_{dd}(\varepsilon-\Delta\varepsilon_m/2)}{-\Delta\varepsilon_m+i\eta_1}f(\varepsilon+\Delta\varepsilon_m/2)\right)\right\}$$

$$(\mathrm{SI}17)$$

Reversing the sign $\Delta\varepsilon_m \to -\Delta\varepsilon_m$ in the second series in (SI17) one gets:

$$\phi^{(1)}(\varepsilon,t)$$
$$= \dot{\varepsilon}_d \lim_{\eta \to +0}\left\{\left(-\sum_m \left(\frac{2\eta}{\Delta\varepsilon_m^2+\eta^2}\right)\exp(i\Delta\varepsilon_m t)V^*_{\varepsilon-\Delta\varepsilon_m/2}\frac{V_{\varepsilon+\Delta\varepsilon_m/2}G^a_{dd}(\varepsilon+\Delta\varepsilon_m/2)G^r_{dd}(\varepsilon-\Delta\varepsilon_m/2)}{\Delta\varepsilon_m+i\eta_1}f(\varepsilon+\Delta\varepsilon_m/2)\right)\right.$$
$$\left.+\left(\sum_m\left(\frac{2\eta}{\Delta\varepsilon_m^2+\eta^2}\right)\exp(i\Delta\varepsilon_m t)V^*_{\varepsilon-\Delta\varepsilon_m/2}\frac{V_{\varepsilon+\Delta\varepsilon_m/2}G^a_{dd}(\varepsilon+\Delta\varepsilon_m/2)G^r_{dd}(\varepsilon-\Delta\varepsilon_m/2)}{\Delta\varepsilon_m+i\eta_1}f(\varepsilon-\Delta\varepsilon_m/2)\right)\right\}$$
$$= -\dot{\varepsilon}_d \lim_{\eta \to +0}\left(\sum_m\left(\frac{2\eta}{\Delta\varepsilon_m^2+\eta^2}\right)\exp(i\Delta\varepsilon_m t)V^*_{\varepsilon-\Delta\varepsilon_m/2}V_{\varepsilon+\Delta\varepsilon_m/2}G^a_{dd}(\varepsilon+\Delta\varepsilon_m/2)G^r_{dd}(\varepsilon-\Delta\varepsilon_m/2)\frac{f(\varepsilon+\Delta\varepsilon_m/2)-f(\varepsilon-\Delta\varepsilon_m/2)}{\Delta\varepsilon_m+i\eta_1}\right)$$

(SI18)

Note that in (SI18) limit $\eta_1 \to +0$ is implied and should be evaluated before $\eta \to +0$. Then, recalling (S 3), one gets:

$$\frac{f(\varepsilon+\Delta\varepsilon_m/2)-f(\varepsilon-\Delta\varepsilon_m/2)}{\Delta\varepsilon_m+i\eta_1} = \left(f(\varepsilon+\Delta\varepsilon_m/2)-f(\varepsilon-\Delta\varepsilon_m/2)\right)\times\left(-i\pi\delta(\Delta\varepsilon_m)+PP\frac{1}{\Delta\varepsilon_m}\right)$$
$$= \frac{f(\varepsilon+\Delta\varepsilon_m/2)-f(\varepsilon-\Delta\varepsilon_m/2)}{\Delta\varepsilon_m}$$

(SI19)

and

$$\lim_{\eta \to +0}\left(\frac{2\eta}{\Delta\varepsilon_m^2+\eta^2}\right)\frac{f(\varepsilon+\Delta\varepsilon_m/2)-f(\varepsilon-\Delta\varepsilon_m/2)}{\Delta\varepsilon_m} = 2\pi\delta(\Delta\varepsilon_m)\partial_\varepsilon f(\varepsilon) \qquad (SI20)$$

With (SI19) and (SI20), Eq. (SI18) becomes

$$\phi^{(1)}(\varepsilon,t) = -\dot{\varepsilon}_d\left(2\pi\delta(\Delta\varepsilon_m)\partial_\varepsilon f(\varepsilon)\exp(i\Delta\varepsilon_m t)V^*_{\varepsilon-\Delta\varepsilon_m/2}V_{\varepsilon+\Delta\varepsilon_m/2}G^a_{dd}(\varepsilon+\Delta\varepsilon_m/2)G^r_{dd}(\varepsilon-\Delta\varepsilon_m/2)\right)$$
$$= -\dot{\varepsilon}_d\left(\partial_\varepsilon f(\varepsilon)\right)A_{dd}(\varepsilon)$$

(SI21)

It is also possible to calculate the correction when the both dot energy and couplings are driven. From Eq. (SG2) it follows

$$\partial_R \hat{\chi}^\dagger_{\varepsilon_k,-} = \sum_m G^a_{mm}(\varepsilon_k)\langle\psi_{m,-}|\partial_R \hat{V}|\psi_{k,-}\rangle\hat{\chi}^\dagger_{\varepsilon_m,-} \qquad (SI22)$$

where (see Eq. (SG13))

$$\langle \psi_{m,-} | \partial_R \hat{V} | \psi_{k,-} \rangle$$
$$= G_{dd}^a(\varepsilon_k) G_{dd}^r(\varepsilon_m) \left\{ (\partial_R V_m^*) V_k (\varepsilon_m - \varepsilon_d - \Lambda(\varepsilon_m) + i\Gamma(\varepsilon_m)/2) + V_m^* (\partial_R V_k)(\varepsilon_k - \varepsilon_d - \Lambda(\varepsilon_k) - i\Gamma(\varepsilon_k)/2) \right.$$
$$\left. + V_m^* V_k \left( \tilde{\Sigma}_{dd}^r(\varepsilon_m) + \tilde{\Sigma}_{dd}^a(\varepsilon_k) \right) + (\partial_R \varepsilon_d) V_m^* V_k \right\}$$

(SI23)

with

$$\tilde{\Sigma}_{dd}^a(\varepsilon) = \sum_m \frac{V_m (\partial_R V_m^*)}{\varepsilon - \varepsilon_m - i\eta} = \sum_m \text{PP} \frac{V_m (\partial_R V_m^*)}{\varepsilon - \varepsilon_m} + i\pi D_\varepsilon V_\varepsilon (\partial_R V_\varepsilon^*) \quad \text{(SI24)a}$$

$$\tilde{\Sigma}_{dd}^r(\varepsilon) = \sum_m \frac{(\partial_R V_m) V_m^*}{\varepsilon - \varepsilon_m + i\eta} = \sum_m \text{PP} \frac{V_m^* (\partial_R V_m)}{\varepsilon - \varepsilon_m} - i\pi D_\varepsilon V_\varepsilon^* (\partial_R V_\varepsilon) \quad \text{(SI24)b}$$

From (SI7) and (SI14) one gets:

$$\text{Tr}\left( \hat{\rho}_{ss} \left( \hat{\chi}_{\varepsilon_k,-}^\dagger \hat{\chi}_{\varepsilon_n,-} \right)^{(1)} + \hat{\rho}_{ss}^{(1)} \hat{\chi}_{\varepsilon_k,-}^\dagger \hat{\chi}_{\varepsilon_n,-} \right) = \dot{R} \lim_{\eta \to +0} \left( \frac{2\eta}{(\varepsilon_k - \varepsilon_n)^2 + \eta^2} \right) \text{Tr}\left( \hat{\rho}_{ss} \partial_R \left( \hat{\chi}_{\varepsilon_k,-}^\dagger \hat{\chi}_{\varepsilon_n,-} \right) \right) \quad \text{(SI25)}$$

With (SI22) Eq. (SI25) becomes:

$$\text{Tr}\left( \hat{\rho}_{ss} \left( \hat{\chi}_{\varepsilon_k,-}^\dagger \hat{\chi}_{\varepsilon_n,-} \right)^{(1)} + \hat{\rho}_{ss}^{(1)} \hat{\chi}_{\varepsilon_k,-}^\dagger \hat{\chi}_{\varepsilon_n,-} \right) = \dot{R} \lim_{\eta \to +0} \left( \frac{2\eta}{(\varepsilon_k - \varepsilon_n)^2 + \eta^2} \right)$$
$$\times \text{Tr}\left( \hat{\rho}_{ss} \left( \sum_m G_{mm}^a(\varepsilon_k) \langle \psi_{m,-} | \partial_R \hat{V} | \psi_{k,-} \rangle \hat{\chi}_{\varepsilon_m,-}^\dagger \hat{\chi}_{\varepsilon_n,-} + \sum_m G_{mm}^r(\varepsilon_n) \langle \psi_{n,-} | \partial_R \hat{V} | \psi_{m,-} \rangle \hat{\chi}_{\varepsilon_k,-}^\dagger \hat{\chi}_{\varepsilon_m,-} \right) \right)$$
$$= \dot{R} \lim_{\eta \to +0} \left( \frac{2\eta}{(\varepsilon_k - \varepsilon_n)^2 + \eta^2} \right)$$
$$\times \left( \sum_m 2\pi f(\varepsilon_n) \delta(\varepsilon_n - \varepsilon_m) G_{mm}^a(\varepsilon_k) \langle \psi_{m,-} | \partial_R \hat{V} | \psi_{k,-} \rangle + \sum_m 2\pi f(\varepsilon_k) \delta(\varepsilon_k - \varepsilon_m) G_{mm}^r(\varepsilon_n) \langle \psi_{n,-} | \partial_R \hat{V} | \psi_{m,-} \rangle \right)$$

(SI26)

By denoting $V_R(\varepsilon_m, \varepsilon_k) = \langle \psi_{m,-} | \partial_R \hat{V} | \psi_{k,-} \rangle$, from (SI26) and (SI1) it follows:

$$\phi^{(1)}(\varepsilon,t) = \frac{1}{2\pi} \int_{-\infty}^{\infty} d\omega \exp(i\omega t) \operatorname{Tr}\left\{ \hat{\rho}_{ss,-}(\hat{\chi}^{\dagger}_{(\varepsilon-\omega/2),-}\hat{\chi}_{(\varepsilon+\omega/2),-})^{(1)} + \hat{\rho}^{(1)}_{ss,+}\hat{\chi}^{\dagger}_{(\varepsilon-\omega/2),-}\hat{\chi}_{(\varepsilon+\omega/2),-} \right\}$$

$$= \dot{R} \lim_{\eta \to +0} \int_{-\infty}^{\infty} d\omega \exp(i\omega t) \left( \frac{2\eta}{\omega^2 + \eta^2} \right)$$

$$\times \left( \sum_m f(\varepsilon+\omega/2)\delta(\varepsilon+\omega/2-\varepsilon_m) \frac{1}{\varepsilon-\omega/2-\varepsilon_m-i\eta_1} V_R(\varepsilon_m, \varepsilon-\omega/2) \right.$$

$$\left. + \sum_m f(\varepsilon-\omega/2)\delta(\varepsilon-\omega/2-\varepsilon_m) \frac{1}{\varepsilon+\omega/2-\varepsilon_m+i\eta_1} V_R(\varepsilon+\omega/2,\varepsilon_m) \right)$$

(SI27)

Integrating (SI27) over $\omega$ and introducing new variable $\Delta\varepsilon_m = 2(\varepsilon_m - \varepsilon)$ one obtains the following sum:

$$\phi^{(1)}(\varepsilon,t)$$

$$= \dot{R} \lim_{\eta \to +0} \left( \frac{2\eta}{(\Delta\varepsilon_m)^2 + \eta^2} \right)$$

$$\times \left( \sum_m f(\varepsilon+\Delta\varepsilon_m/2) \frac{1}{-\Delta\varepsilon_m - i\eta_1} V_R(\varepsilon+\Delta\varepsilon_m/2, \varepsilon-\Delta\varepsilon_m/2) \exp(i\Delta\varepsilon_m t) \right.$$

$$\left. + \sum_m f(\varepsilon+\Delta\varepsilon_m/2) \frac{1}{-\Delta\varepsilon_m + i\eta_1} V_R(\varepsilon-\Delta\varepsilon_m/2, \varepsilon+\Delta\varepsilon_m/2) \exp(-i\Delta\varepsilon_m t) \right)$$

(SI28)

Changing the sign of $\Delta\varepsilon_m$ in the second series in the expression above one gets (by analogy with (SI18)-(SI21)):

$$\phi^{(1)}(\varepsilon,t) = 2\pi\dot{R}D_\varepsilon \left( \partial_\varepsilon f(\varepsilon) \right) V_R(\varepsilon,\varepsilon)$$  (SI29)

where $D_\varepsilon$ is the density of states.

From (SI23) it follows:

$$V_R(\varepsilon_k,\varepsilon_k) = \langle \psi_{k,-}|\partial_R \hat{V}|\psi_{k,-}\rangle$$

$$= G^a_{dd}(\varepsilon_k)G^r_{dd}(\varepsilon_k)\{(\partial_R V_k^*)V_k(\varepsilon_k - \varepsilon_d - \Lambda(\varepsilon_k) + i\Gamma(\varepsilon_k)/2) + V_k^*(\partial_R V_k)(\varepsilon_k - \varepsilon_d - \Lambda(\varepsilon_k) - i\Gamma(\varepsilon_k)/2)$$

$$+ V_k^*V_k\left(\tilde{\Sigma}^r_{dd}(\varepsilon_k) + \tilde{\Sigma}^a_{dd}(\varepsilon_k)\right) + (\partial_R \varepsilon_d)V_k^*V_k\}$$

$$= G^a_{dd}(\varepsilon_k)G^r_{dd}(\varepsilon_k)$$

$$\times\left(\partial_R\left(V_k^*V_k\right)(\varepsilon_k - \varepsilon_d - \Lambda(\varepsilon_k)) + \partial_R(\varepsilon_d + \Lambda(\varepsilon_k))V_k^*V_k + i\Gamma(\varepsilon_k)/2\{(\partial_R V_k^*)V_k - V_k^*(\partial_R V_k)\}\right)$$

$$+ i\pi D_{\varepsilon_k}V_k^*V_k\left(V_k(\partial_R V_k^*) - V_k^*(\partial_R V_k)\right)$$

$$= G^a_{dd}(\varepsilon_k)G^r_{dd}(\varepsilon_k)$$

$$\times\left(\partial_R\left(V_k^*V_k\right)(\varepsilon_k - \varepsilon_d - \Lambda(\varepsilon_k)) + \partial_R(\varepsilon_d + \Lambda(\varepsilon_k))V_k^*V_k + i\Gamma(\varepsilon_k)/2\{(\partial_R V_k^*)V_k - V_k^*(\partial_R V_k)\}\right)$$

$$+ i\Gamma(\varepsilon_k)/2\left(V_k(\partial_R V_k^*) - V_k^*(\partial_R V_k)\right)$$

$$= G^a_{dd}(\varepsilon_k)G^r_{dd}(\varepsilon_k)\left(\partial_R\left(V_k^*V_k\right)(\varepsilon_k - \varepsilon_d - \Lambda(\varepsilon_k)) + \partial_R(\varepsilon_d + \Lambda(\varepsilon_k))V_k^*V_k\right)$$

(SI30)

Substituting (SI30) into (SI29) gives the final result for the correction:

$$\phi^{(1)}(\varepsilon,t) = 2\pi \dot{R} D_\varepsilon\left(\partial_\varepsilon f(\varepsilon)\right)G^a_{dd}(\varepsilon)G^r_{dd}(\varepsilon)\left(\partial_R\left(V_\varepsilon^*V_\varepsilon\right)(\varepsilon - \varepsilon_d - \Lambda(\varepsilon)) + \partial_R(\varepsilon_d + \Lambda(\varepsilon))V_\varepsilon^*V_\varepsilon\right)$$

$$= \dot{R}\left(\partial_\varepsilon f(\varepsilon)\right)A_{dd}(\varepsilon)\frac{\left(\partial_R(\Gamma(\varepsilon))\right)(\varepsilon - \varepsilon_d - \Lambda(\varepsilon)) + \partial_R(\varepsilon_d + \Lambda(\varepsilon))}{\Gamma(\varepsilon)}$$

$$= \dot{R}\left(\partial_\varepsilon f(\varepsilon)\right)A_{dd}(\varepsilon)\Gamma(\varepsilon)\partial_R\left(\frac{\varepsilon - \varepsilon_d - \Lambda(\varepsilon)}{\Gamma(\varepsilon)}\right)$$

(SI31)

**Section K. Evaluation of the thermodynamics quantities**

To find the correction to the Fermi distribution we define first order correction to the dot occupation thorough the distributions:

$$N^{(1)} = \sum_\beta \int d\varepsilon D_\beta(\varepsilon)\delta f_\beta(\varepsilon) \qquad (SJ1)$$

Comparing (SJ1) with (SF9) we see that

$$\delta f_\beta(\varepsilon) = -\frac{\dot{\varepsilon}_d}{2}A_{dd}(\varepsilon)\partial f_\beta(\varepsilon) \qquad (SJ2)$$

We also need to make sure that the Eq. (SJ2) is consistent with our previous definitions.

Specifically, we need to show that

$$E^{(1)} = \sum_\beta \int d\varepsilon\, \varepsilon D_\beta(\varepsilon) \delta f_\beta(\varepsilon) \tag{SJ3}$$

coincides with

$$\langle \hat{H}_{eff} \rangle^{(1)} = \mathrm{Tr}\{\rho^{(1)} \hat{H}_{eff}\} \tag{SJ4}$$

where $\hat{H}_{eff}$ is given by **Error! Reference source not found.**. Indeed, using (50) and (52) we have

$$\langle \hat{H}_{eff} \rangle^{(1)} = -\dot\varepsilon_d \sum_{k\alpha}\sum_{n\beta} |V_{k\alpha}|^2 |V_{n\beta}|^2\, G^r_{dd}(\varepsilon_k) G^a_{dd}(\varepsilon_k) G^r_{dd}(\varepsilon_n) G^a_{dd}(\varepsilon_n) \frac{(\varepsilon_n + \varepsilon_k)}{2} \frac{f_\beta(\varepsilon_n) - f_\alpha(\varepsilon_k)}{\varepsilon_n - \varepsilon_k - i\eta_1}$$

$$\times \left\{ \frac{\eta_2}{(\varepsilon_n - \varepsilon_k)^2 + \eta_2^2} + i\frac{\varepsilon_n - \varepsilon_k}{(\varepsilon_n - \varepsilon_k)^2 + \eta_2^2} \right\}$$

$$= -\dot\varepsilon_d \left(\frac{1}{2\pi}\right)^2 \int d\varepsilon \int d\varepsilon' \sum_{k\alpha}\sum_{n\beta} |V_{k\alpha}|^2 |V_{n\beta}|^2\, 2\pi\delta(\varepsilon - \varepsilon_n) 2\pi\delta(\varepsilon' - \varepsilon_k) \frac{(\varepsilon + \varepsilon')}{2}$$

$$\times G^r_{dd}(\varepsilon') G^a_{dd}(\varepsilon') G^r_{dd}(\varepsilon) G^a_{dd}(\varepsilon) \frac{f_\beta(\varepsilon) - f_\alpha(\varepsilon')}{\varepsilon - \varepsilon' - i\eta_1} \left\{ \frac{\eta_2}{(\varepsilon - \varepsilon')^2 + \eta_2^2} + i\frac{\varepsilon - \varepsilon'}{(\varepsilon - \varepsilon')^2 + \eta_2^2} \right\}$$

$$= -\dot\varepsilon_d \left(\frac{1}{2\pi}\right)^2 \int d\varepsilon \int d\varepsilon' \frac{(\varepsilon + \varepsilon')}{2} \frac{A_{dd}(\varepsilon) A_{dd}(\varepsilon')}{\Gamma(\varepsilon)\Gamma(\varepsilon')}$$

$$\times \sum_\alpha \sum_\beta \Gamma_\beta(\varepsilon) \Gamma_\alpha(\varepsilon') \frac{f_\beta(\varepsilon) - f_\alpha(\varepsilon')}{\varepsilon - \varepsilon' - i\eta_1} \left\{ \frac{\eta_2}{(\varepsilon - \varepsilon')^2 + \eta_2^2} + i\frac{\varepsilon - \varepsilon'}{(\varepsilon - \varepsilon')^2 + \eta_2^2} \right\}$$

$$\tag{SJ5}$$

Then, to evaluate (SJ5), we can apply the procedure form Section G, namely swapping the indexes α and β in the double sum and taking the limits $\eta_1 \to +0$ and $\eta_2 \to +0$. It will give us the following result:

$$\langle \hat{H}_{eff} \rangle^{(1)} = \frac{-\dot\varepsilon_d \hbar}{4\pi} \sum_\beta \int d\varepsilon\, \varepsilon A_{dd}^2(\varepsilon) \frac{\Gamma_\beta}{\Gamma} \partial_\varepsilon f_\beta \tag{SJ6}$$

On the other hand,

$$E^{(1)} = \sum_\beta \int d\varepsilon\, \varepsilon D_\beta(\varepsilon) \delta f_\beta(\varepsilon) = \frac{-\dot\varepsilon_d \hbar}{4\pi} \sum_\beta \int d\varepsilon\, \varepsilon A_{dd}^2(\varepsilon) \frac{\Gamma_\beta}{\Gamma} \left(\partial_\varepsilon f_\beta\right) \tag{SJ7}$$

thus $\left\langle \hat{H}_{eff} \right\rangle^{(1)} = E^{(1)}$

Let's evaluate heat:

$$\dot{Q}_\beta^{(1)} = (\dot{\varepsilon}_d)\left(\partial_{\varepsilon_d} E_\beta^{(0)} - \mu_\beta \partial_{\varepsilon_d} N_\beta^{(0)}\right) - \dot{W}_\beta^{(1)}$$

$$= (\dot{\varepsilon}_d)\int (\varepsilon - \mu_\beta)\frac{\Gamma_\beta}{2\pi\Gamma}\partial_{\varepsilon_d}\left(A_{dd}(\varepsilon)\right)f_\beta(\varepsilon)d\varepsilon - \frac{1}{2\pi}(\dot{\varepsilon}_d)\int A_{dd}(\varepsilon)\frac{\Gamma_\beta}{\Gamma}f_\beta(\varepsilon)d\varepsilon$$

$$= (\dot{\varepsilon}_d)\int \frac{\Gamma_\beta}{2\pi\Gamma}A_{dd}(\varepsilon)\partial_\varepsilon\left((\varepsilon-\mu_\beta)f_\beta(\varepsilon)\right)d\varepsilon - \frac{1}{2\pi}(\dot{\varepsilon}_d)\int A_{dd}(\varepsilon)\frac{\Gamma_\beta}{\Gamma}f_\beta(\varepsilon)d\varepsilon$$

$$= (\dot{\varepsilon}_d)\int \frac{\Gamma_\beta}{2\pi\Gamma}A_{dd}(\varepsilon)\partial_\varepsilon\left(\varepsilon-\mu_\beta\right)f_\beta(\varepsilon)d\varepsilon - \frac{1}{2\pi}(\dot{\varepsilon}_d)\int A_{dd}(\varepsilon)\frac{\Gamma_\beta}{\Gamma}f_\beta(\varepsilon)d\varepsilon$$

$$+ (\dot{\varepsilon}_d)\int \frac{\Gamma_\beta}{2\pi\Gamma}A_{dd}(\varepsilon)(\varepsilon-\mu_\beta)\partial_\varepsilon\left(f_\beta(\varepsilon)\right)d\varepsilon = (\dot{\varepsilon}_d)\int \frac{\Gamma_\beta}{2\pi\Gamma}A_{dd}(\varepsilon)(\varepsilon-\mu_\beta)\partial\left(f_\beta(\varepsilon)\right)d\varepsilon$$

(SJ8)

and for the second order correction:

$$\dot{Q}_\beta^{(2)} = (\dot{\varepsilon}_d)\left(\partial_{\varepsilon_d} E_\beta^{(1)} - \mu_\beta \partial_{\varepsilon_d} N_\beta^{(1)}\right) - \dot{W}_\beta^{(2)}$$

$$= -(\dot{\varepsilon}_d)^2 \hbar \int (\varepsilon-\mu_\beta)\frac{\Gamma_\beta}{4\pi\Gamma}\partial_{\varepsilon_d}\left(A_{dd}^2(\varepsilon)\right)\partial\left(f_\beta(\varepsilon)\right)d\varepsilon + \frac{1}{4\pi}(\dot{\varepsilon}_d)^2 \int A_{dd}^2(\varepsilon)\frac{\Gamma_\beta}{\Gamma}\partial\left(f_\beta(\varepsilon)\right)d\varepsilon$$

$$= -(\dot{\varepsilon}_d)^2 \hbar \int \frac{\Gamma_\beta}{4\pi\Gamma}A_{dd}^2(\varepsilon)\partial_\varepsilon\left((\varepsilon-\mu_\beta)\partial\left(f_\beta(\varepsilon)\right)\right)d\varepsilon + \frac{1}{4\pi}(\dot{\varepsilon}_d)^2 \int A_{dd}^2(\varepsilon)\frac{\Gamma_\beta}{\Gamma}\partial\left(f_\beta(\varepsilon)\right)d\varepsilon$$

$$= -(\dot{\varepsilon}_d)^2 \hbar \int \frac{\Gamma_\beta}{4\pi\Gamma}A_{dd}^2(\varepsilon)\partial_\varepsilon\left((\varepsilon-\mu_\beta)\right)\partial\left(f_\beta(\varepsilon)\right)d\varepsilon + \frac{1}{4\pi}(\dot{\varepsilon}_d)^2 \int A_{dd}^2(\varepsilon)\frac{\Gamma_\beta}{\Gamma}\partial\left(f_\beta(\varepsilon)\right)d\varepsilon$$

$$-(\dot{\varepsilon}_d)^2 \hbar \int \frac{\Gamma_\beta}{4\pi\Gamma}A_{dd}^2(\varepsilon)(\varepsilon-\mu_\beta)\partial^2\left(f_\beta(\varepsilon)\right)d\varepsilon$$

$$= -(\dot{\varepsilon}_d)^2 \hbar \int \frac{\Gamma_\beta}{4\pi\Gamma}A_{dd}^2(\varepsilon)(\varepsilon-\mu_\beta)\partial^2\left(f_\beta(\varepsilon)\right)d\varepsilon$$

(SJ9)

The first order in the driving speed correction to the entropy:

$$\dot{S}^{(1)} = \dot{\varepsilon}_d \partial_{\varepsilon_d}\left(S^{(0)}\right) = \dot{\varepsilon}_d k_B \sum_\beta \int D_\beta(\varepsilon)\frac{\partial\sigma(x)}{\partial x}\bigg|_{x=f_\beta(\varepsilon)}\partial\left(f_\beta(\varepsilon)\right)d\varepsilon$$

$$= (\dot{\varepsilon}_d)\sum_\beta \int \frac{\Gamma_\beta}{2\pi\Gamma}A_{dd}(\varepsilon)\frac{(\varepsilon-\mu_\beta)}{T_\beta}\partial\left(f_\beta(\varepsilon)\right)d\varepsilon = \sum_\beta \frac{\dot{Q}_\beta^{(1)}}{T_\beta}$$

(SJ10)

Eqs. (75), (SJ9) and (55) give for the second order correction to the entropy the following

expression

$$\begin{aligned}\dot{S}^{(2)} &= \frac{(\dot{\varepsilon}_d)^2}{2}\sum_\beta \int \frac{\Gamma_\beta}{2\pi\Gamma}\frac{\varepsilon-\mu_\beta}{T_\beta}\partial\left(A_{dd}^2(\varepsilon)\right)\partial\left(f_\beta(\varepsilon)\right)d\varepsilon \\ &= -\frac{(\dot{\varepsilon}_d)^2}{2}\sum_\beta \int \frac{\Gamma_\beta}{2\pi\Gamma}A_{dd}^2(\varepsilon)\partial\left\{\frac{\varepsilon-\mu_\beta}{T_\beta}\partial\left(f_\beta(\varepsilon)\right)\right\}d\varepsilon = -\frac{(\dot{\varepsilon}_d)^2}{2}\sum_\beta \int \frac{\Gamma_\beta}{2\pi\Gamma}A_{dd}^2(\varepsilon)\left(\frac{\varepsilon-\mu_\beta}{T_\beta}\right)\partial^2\left(f_\beta(\varepsilon)\right)d\varepsilon \\ &-\frac{(\dot{\varepsilon}_d)^2}{2}\sum_\beta \int \frac{\Gamma_\beta}{2\pi\Gamma}A_{dd}^2(\varepsilon)\frac{\partial\left(f_\beta(\varepsilon)\right)}{T_\beta}d\varepsilon = \sum_\beta \frac{\dot{Q}_\beta^{(2)}}{T_\beta}+\sum_\beta \frac{\dot{W}_\beta^{(2)}}{T_\beta}\end{aligned}$$

(SJ11)

References.